\documentclass[aps,prd,twocolumn,times,graphicx,amssymb,tighten,showpacs,groupedaddress,superscriptaddress,floatfix]{revtex4}
\pdfoutput=1

\usepackage{graphicx,amssymb,amsmath,pdftricks,soul}
\newcommand{\be}{\begin{equation}}
\newcommand{\ee}{\end{equation}}

\newcommand{\ignore}[1]{}
\begin{document}
\title{Meson-exchange currents and quasielastic predictions for charged-current neutrino-$^{12}$C scattering in the superscaling approach}

\author{G.~D. Megias\footnote{Corresponding author: megias@us.es}}
\affiliation{Departamento de F\'{i}sica At\'omica, Molecular y Nuclear, Universidad de Sevilla, 41080 Sevilla, SPAIN}
\author{T.~W. Donnelly}
\affiliation{Center for Theoretical Physics, Laboratory for Nuclear Science and Department of Physics,
Massachusetts Institute of Technology, Cambridge, Massachusetts 02139, USA}
\author{O. Moreno}
\affiliation{Center for Theoretical Physics, Laboratory for Nuclear Science and Department of Physics,
Massachusetts Institute of Technology, Cambridge, Massachusetts 02139, USA}
\author{C.~F. Williamson}
\affiliation{Center for Theoretical Physics, Laboratory for Nuclear Science and Department of Physics,
Massachusetts Institute of Technology, Cambridge, Massachusetts 02139, USA}
\author{J.~A. Caballero}
\affiliation{Departamento de F\'{i}sica At\'omica, Molecular y Nuclear, Universidad de Sevilla, 41080 Sevilla, SPAIN}
\author{R. Gonz\'alez-Jim\'enez}
\affiliation{Departamento de F\'{i}sica At\'omica, Molecular y Nuclear, Universidad de Sevilla, 41080 Sevilla, SPAIN}
\author{A. De Pace}
\affiliation{Istituto Nazionale di Fisica Nucleare, Sezione di Torino, Via P. Giuria 1, 10125 Torino, ITALY}
\author{M.~B. Barbaro}
\affiliation{Dipartimento di Fisica, Universit\`a di Torino, Sezione di Torino, Via P. Giuria 1, 10125 Torino, ITALY}
\affiliation{Istituto Nazionale di Fisica Nucleare, Sezione di Torino, Via P. Giuria 1, 10125 Torino, ITALY}
\author{W.~M. Alberico}
\affiliation{Dipartimento di Fisica, Universit\`a di Torino, Sezione di Torino, Via P. Giuria 1, 10125 Torino, ITALY}
\affiliation{Istituto Nazionale di Fisica Nucleare, Sezione di Torino, Via P. Giuria 1, 10125 Torino, ITALY}
\author{M. Nardi}
\affiliation{Istituto Nazionale di Fisica Nucleare, Sezione di Torino, Via P. Giuria 1, 10125 Torino, ITALY}
\author{J.~E. Amaro}
\affiliation{Departamento de F\'isica At\'omica, Molecular y Nuclear and Instituto Carlos I de F\'isica Te\'orica y Computacional, Universidad de Granada, 18071 Granada, SPAIN}
\begin{abstract}
We evaluate and discuss the impact of meson-exchange currents (MEC) on charged-current quasielastic (QE) neutrino cross sections. We consider the nuclear transverse response arising from 2p-2h states excited by the action of electromagnetic, purely isovector meson-exchange currents in a fully relativistic framework, based on the work by the Torino collaboration \cite{DePace03}. An accurate parametrization of this MEC response as a function of the momentum and energy transfers involved is presented. Results of neutrino-nucleus cross sections using this MEC parametrization together with a recent scaling approach for the 1p-1h contributions (SuSAv2) are compared with experimental data.
\end{abstract}


\maketitle
\section{Introduction}\label{introduction}

A correct interpretation of atmospheric and accelerator-based neutrino oscillation experiments strongly relies on our understanding of neutrino-nucleus scattering at intermediate energies (from 0.5 to 10 GeV) and in particular of the nuclear-structure effects involved. One of the simplest descriptions of the nucleus, the relativistic Fermi gas (RFG) model, which is known to be inadequate for inclusive electron scattering in the QE regime \cite{footnote1}, also fails to reproduce recent measurements of QE neutrino and antineutrino scattering cross sections \cite{MiniBooNECC10, MiniBooNECC13, MINERVAnub13, MINERVAnu13, MINOS14, NOMAD09}. This supports the need for considering mechanisms such as final-state interactions, nuclear correlations or MEC, in particular through their contribution to multinucleon knock-out around and beyond the QE peak as suggested by explicit modeling \cite{Martini09, Nieves11, Amaro11}.

In particular, the recent muon neutrino charged-current quasielastic (CCQE)
cross sections measured by the MiniBooNE Collaboration \cite{MiniBooNECC10, MiniBooNECC13} show discrepancies with a RFG description of the nuclear target. This simple model, widely used in experimental analyses, underestimates the total cross section, unless {\it ad hoc} assumptions are made such as a larger mass parameter in the nucleon axial form factor ($M_A =$ 1.35 GeV/c$^2$ versus $M_A =$ 1.032 GeV/c$^2$). Relativistic effects cannot be neglected for the kinematics of experiments such as MiniBooNE, with neutrino energies as high as 3 GeV. Although the RFG model has the merit of accounting properly for relativistic effects, it is too crude to account for detailed nuclear dynamics, as is well known from comparisons with QE electron scattering data \cite{future_ee}. More sophisticated 
relativistic nuclear models have been applied in recent years to neutrino reactions. In addition, phenomenological techniques have been proposed, such as the superscaling approach (SuSA) \cite{Amaro05a} 
which assumes the 
existence of universal scaling functions for the electromagnetic and weak interactions. Analyses of inclusive ($e, e'$) data have shown that at energy transfers below the QE peak, superscaling is fulfilled rather well \cite{Day90, Donnelly99a, Donnelly99b}, which implies that the reduced cross section is largely independent of the momentum transfer (first-kind scaling) and of the nuclear target (second-kind scaling) when expressed as a function of the appropriate scaling variable. From these analyses a phenomenological scaling function was extracted from the longitudinal QE electron scattering responses. It was subsequently used to predict neutrino-nucleus cross sections by multiplying it by the single-nucleon weak cross sections, assuming that the single universal scaling function was appropriate for all of the various responses involved, namely CC, CL, LL, T(VV), T(AA) and T$^{\prime}$(VA). In this work we will use a recently developed improved version of the superscaling model, called SuSAv2 
\cite{SuSAv2}, that incorporates relativistic mean field (RMF) effects \cite{Caballero05, Caballero06, Caballero07} in the longitudinal and transverse nuclear responses, as well as in the isovector and isoscalar channels independently. Three reference scaling functions are provided to describe in a consistent way both electron- and (anti)neutrino-nucleus reactions in the QE region: transverse ($\tilde{f}_T$), longitudinal isovector ($\tilde{f}_L^{T=1}$) and longitudinal isoscalar ($\tilde{f}_L^{T=0}$). This model also includes in a natural way an enhancement of the transverse response through RMF effects without resorting to inelastic processes or two-particle emission via MEC. 

Strictly speaking only the longitudinal part of the response appears to superscale; in the scaling region some degree of scaling violation is found which can be attributed to the transverse part of the response. The assumption that the various types of response (CC, CL, LL, T(VV), T(AA) and T$^{\prime}$(VA)) scale the same way has been denoted zeroth-kind scaling; the most recent SuSAv2 approach builds in the degree of violation of zeroth-kind scaling demanded by the RMF results. Specifically, the longitudinal contributions, apparently being essentially impulsive at high energies, are usually used to determine the basic nuclear physics of QE scattering, notably, including any correlations present in that sector, since the results are obtained by fitting electron scattering data. Beyond the QE region it is natural to have scaling violations, since the reaction mechanism there is not solely the impulsive knockout of a nucleon, but may proceed via meson production including baryon resonances such 
as the $\Delta$. It is known that the latter contributions are much more prominent in the transverse than in the longitudinal responses \cite{Maieron09, Amaro05a}. However, it is also known that even with only the 1p-1h contributions there are expected to be violations of zeroth-kind scaling arising from purely dynamical relativistic effects (see the discussions of how the SuSAv2 approach is constructed).

However, even below the meson production threshold there are scaling violations in the transverse response \cite{Donnelly99b}, one source of which could be the MEC contributions, again predominantly transverse. The MEC are two-body currents that can excite both one-particle one-hole (1p-1h) and two-particle two-hole (2p-2h) states. Most studies of electromagnetic ($e,e'$) processes performed for low-to-intermediate momentum transfers with MEC in the 1p-1h sector (see, {\it e.g.,} \cite{Amaro02, Amaro03, Alberico98, Amaro10}) have shown a small reduction of the total response at the QE peak, mainly due to diagrams involving the electroexcitation of the $\Delta$ resonance; they are roughly compensated by the positive contributions of correlation diagrams, where the virtual photon couples to a correlated pair of nucleons. In the present work we shall therefore neglect them and restrict our attention to 2p-2h final states, computed in a fully relativistic way. It has been found \cite{Martini09, Nieves11, Amaro11,
 AmaroMEC14, AmaroMEC14b} that the MEC give a significant positive contribution to the cross section, which helps to account for the discrepancy observed in ($e,e'$) processes between theory and experiment in the ``dip'' region between the QE peak and $\Delta$-resonance as well as for the discrepancies between some recent neutrino CCQE measurements ({\it e.g.,} MiniBooNE, NOMAD, MINER$\nu$A). In particular, in \cite{Amaro11b, Amaro12} we used a parametrization of the results of \cite{DePace03} to evaluate the contribution of MEC to the vector transverse (anti)neutrino response at MiniBooNE kinematics.


The presence of nucleon-nucleon correlation interactions
involving the one-nucleon current may lead to the excitation of 2p-2h
final states, and interference between these processes and those
involving MEC must also be taken into account. Results of
calculations carried out within the Green's Function Monte Carlo
approach \cite{Benhar2013} suggest that these interference contributions
may in fact be quite large. This is in agreement with
our preliminary calculation of the correlation current plus MEC
effects in the response functions within the scheme of the relativistic Fermi gas model \cite{Amaro10b}. These effects, also taken into account in the RFG-based descriptions
of 2p-2h provided by Nieves {\it et al.}~\cite{Nieves11} and Martini~\cite{Martini09},
are not included explicitly in our RFG MEC model, that relies on a hybrid description where the one-particle
emission already contains contributions of nuclear ejections due to
nuclear correlations --- through the experimental scaling
function. Explicit calculations of the correlation-MEC interference terms are still in progress and
their contributions will be presented in a forthcoming publication.

This paper is organized as follows. In Sect. \ref{MEC} we briefly describe the computation of the MEC considered in this work and show for the first time the corresponding responses of $^{12}$C for several momentum transfers as a function of the QE scaling variable. We also show a new parametrization of these responses and compare it with the one used in \cite{Amaro11, Amaro11b, Amaro12}. In Sect. \ref{cross-sections} we apply the new MEC parametrization and the SuSAv2 model to the computation of neutrino-$^{12}$C CCQE cross sections and compare the results with MiniBooNE, NOMAD and MINER$\nu$A data. Finally, in Sect. \ref{conclusions} we show the conclusions of our analysis.

\section{Results for MEC responses} \label{MEC}

We consider in this work the purely isovector pion-exchange currents involving virtual $\Delta$ resonances as well as the seagull (contact) and pion-in-flight 
currents obtained in previous work~\cite{DePace03, DePace04}. The evaluation was performed within the RFG model in which a fully Lorentz and transitionally 
invariant calculation of the MEC can be developed. Deviations from the Fermi gas model 2p-2h responses produced by ingredients such as final-state interactions, 
finite nuclear effects or nuclear correlations are expected to be moderate, which would result in small corrections in the impulsive cross section as the MEC 
contributions are also moderate.

The previous statement is not insubstantial, and it requires further explanation. We expect 
the finite-size effects to be moderate on the 2p-2h responses. This is in accordance with the
calculations performed by one of the authors and presented in a series of papers, 
[see for instance ~\cite{Amaro92,Amaro94}]. 
These are the only calculations up to date concerning the inclusive 2p-2h transverse response function
at low-to-intermediate momentum transfers for $^{12}$C and $^{40}$Ca within the framework of the continuum shell model.
The results were similar to those found in nuclear matter by Van Orden and Donnelly~\cite{VanOrden81}, Alberico, Ericson and Molinari~\cite{Alberico84}, and Dekker, Brussaard and Tjon ~\cite{Dekker94}. The non-relativistic 2p-2h response function is a rather smooth function. Its general behavior is clearly dominated by the 2p-phase space and by the
nucleon and pion electromagnetic form factors, whereas it is rather insensitive to details of the finite size nucleus.

The previous works, together with ~\cite{Gil97,DePace03}, are the only calculations
available for the 2p-2h electromagnetic responses for medium nuclei. The
studies presented in \cite{Dekker94,DePace03} clearly showed that the relativistic effects,
mainly in the delta MEC, dominate the 2p-2h transverse response.

It has been known for a long time that ground-state correlations deplete the occupation
numbers of the hole states, the values of which drop from unity to
$\sim0.8$. 
The main effect of such depletion is known to be a
redistribution of the strength to higher energies. In the case of the
longitudinal response, dominated by the impulse approximation, this is
translated into a hardening of the response function with respect to
an uncorrelated model, like the Fermi Gas or the semirelativistic
shell model~\cite{Amaro05b}, with the
appearance of a long tail at high energy. This is precisely the shape of the scaling function we
are using. Being a phenomenological
observable, the scaling function already contains all the physics
embodied in the nuclear structure details, including correlations,
depletions and final state interactions.

In the case of the 2p-2h contributions, one expects the depletion of the occupation
numbers also to produce a redistribution of the strength to higher
energies. Although this could modify the position of the peak in the 2p-2h
response function, the resulting redistribution is expected to 
keep some resemblance with the behavior already shown in the 1p-1h channel.


As mentioned above, the kinematical regions contained under the integral over the neutrino fluxes considered here extend to relativistic domains, so that a relativistic treatment of the process is required. As was discussed in the previous work~\cite{DePace03, DePace04}, relativistic effects are important to describe the nuclear transverse response function for momentum transfers above 500 MeV/c.

All possible 2p-2h many-body diagrams containing two pionic lines and the virtual boson attached to the pion (pion-in-flight term), to the $NN\pi$ vertex (seagull or contact term) or involving the virtual $\Delta$ resonance are taken into account to compute the vector-vector transverse MEC response, $R_{T,\:VV}^{MEC}$, of $^{12}$C \cite{DePace03}. These responses can be given as a function of the energy transfer $\omega'$ or of the the scaling variable $\Psi'$, related through:
\begin{eqnarray}
\Psi'=\frac{1}{\sqrt{\xi_F}}\frac{\lambda'-\tau'}{\sqrt{(1+\lambda')\:\tau' + \kappa\:\sqrt{\tau'\:(1+\tau')}}} \:,
\end{eqnarray}
where $\xi_F$ is the dimensionless Fermi kinetic energy and the following dimensionless transfer variables have been defined: $\lambda=\omega/2m_N$, $\kappa=q/2m_N$, $\tau=\kappa^2-\lambda^2$. Primed variables contain an energy transfer shift, $\omega'=\omega-E_s$, which accounts (at least) for the binding energy of the ejected nucleon, but is usually determined phenomenologically; for $^{12}$C we use $E_s =$ 20 MeV. The scaling variable considerably distorts the $\omega$ dependence, but it has the advantage of allowing us to easily locate the QE peak at $\Psi'=0$, from which the peaks of the MEC responses are shifted.
Over 100,000 terms are involved in the calculation, with subsequent seven-dimensional integrations, which make it a highly non-trivial computational procedure. In order to include these results in the neutrino generators used in the analysis of neutrino experiments a parametrization of the MEC responses is essential to reduce the computational burden of performing the calculation for a large number of kinematic conditions (momentum and energy transfers).

\begin{figure}[htbp]
\begin{center}
\includegraphics[scale=0.3, bb=0 -09 794 529, clip]{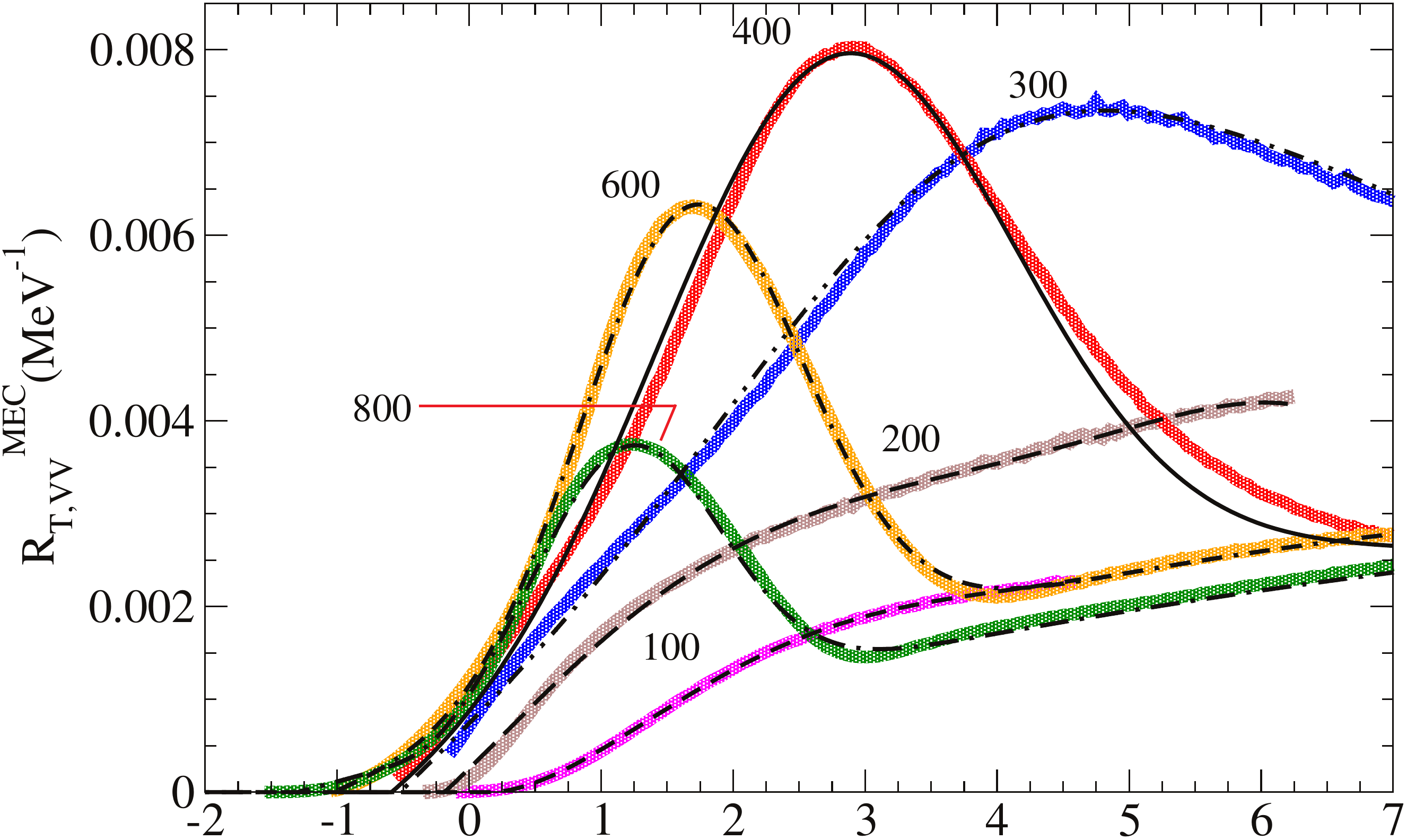} \\
\includegraphics[scale=0.3, bb=0 -09 859 529, clip]{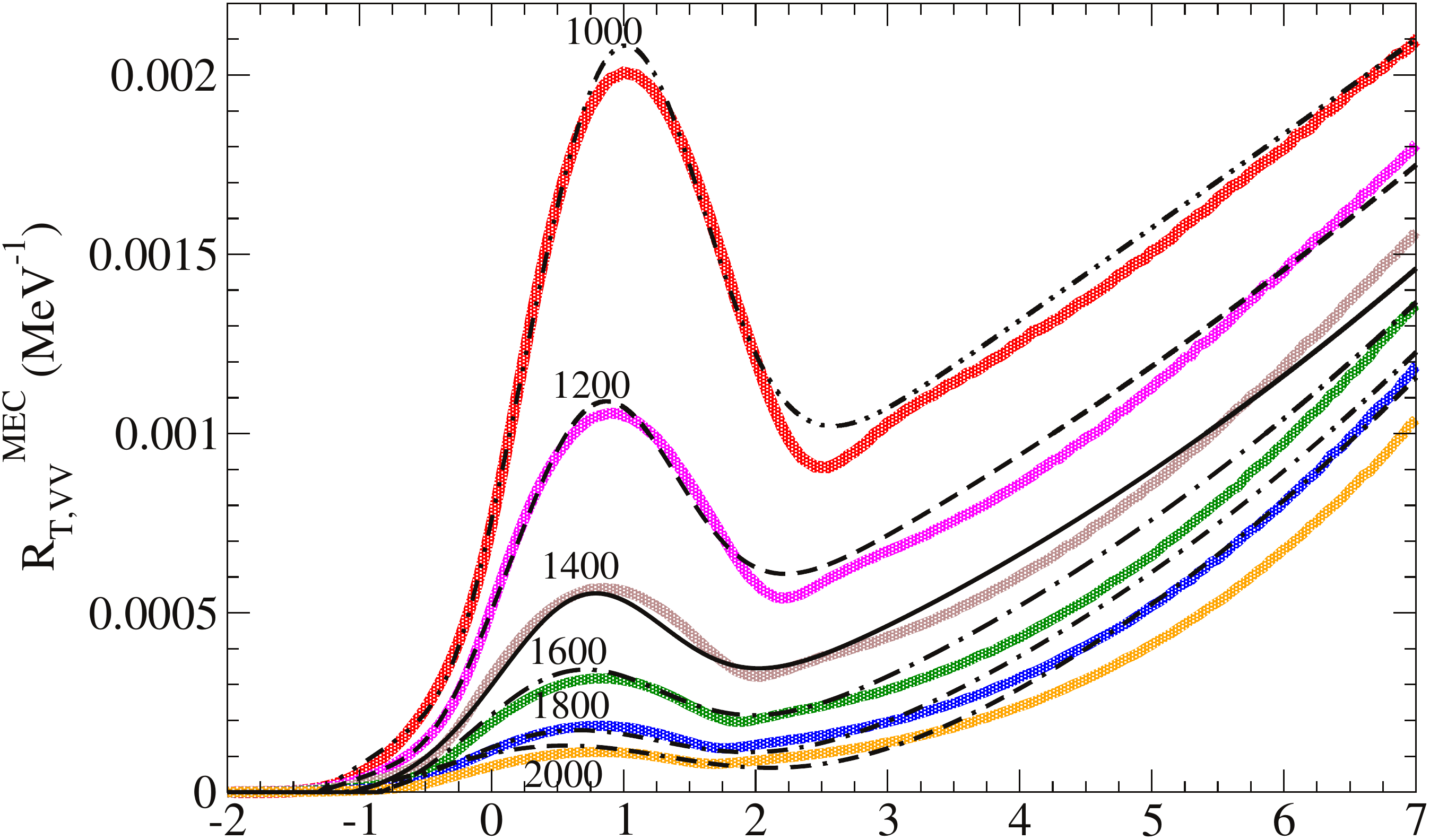} \\
\includegraphics[scale=0.3, bb=0 0 794 529, clip]{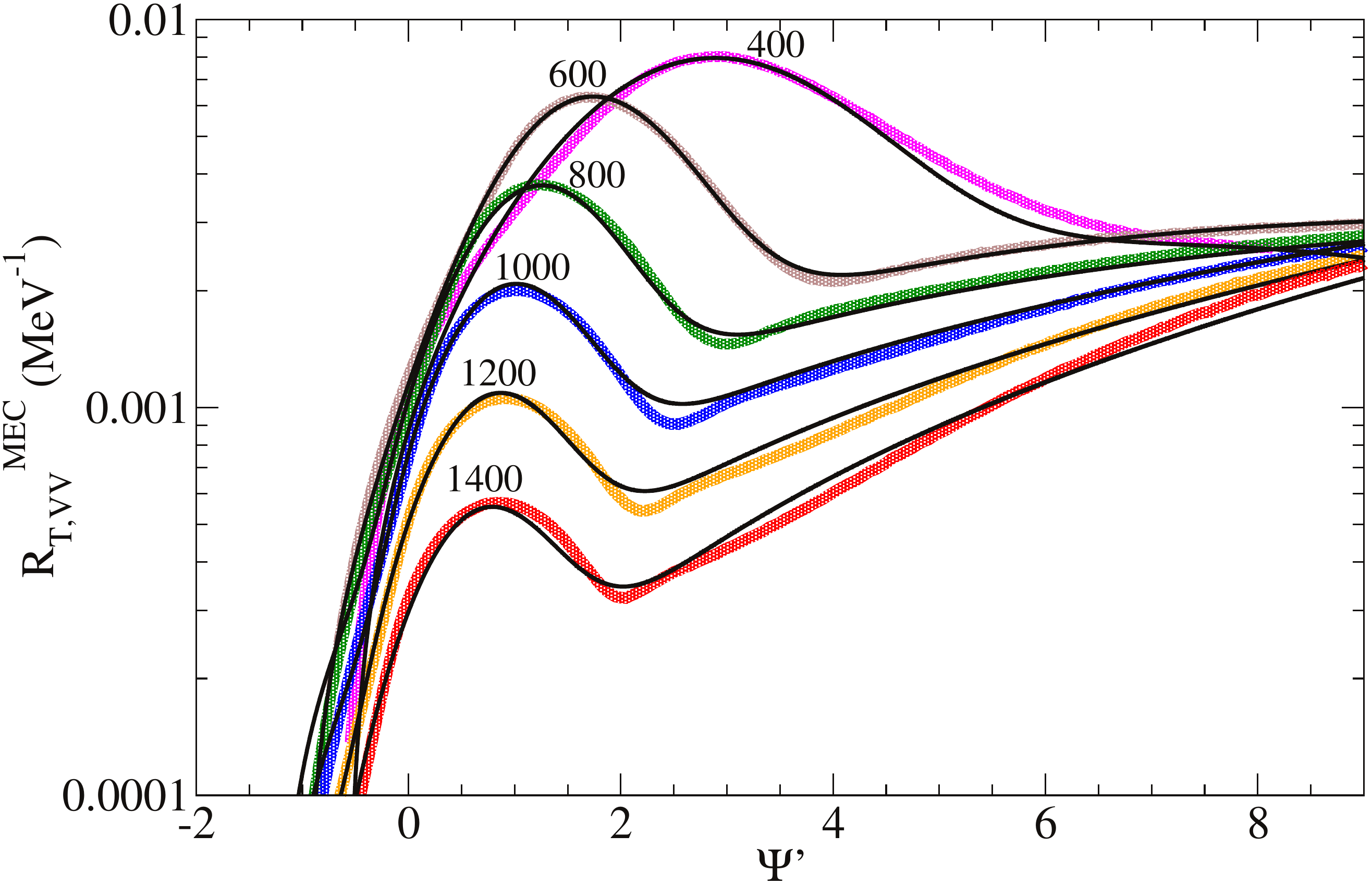} \\
\vspace{0.178cm}
\includegraphics[scale=0.3, bb=0 1 794 529, clip]{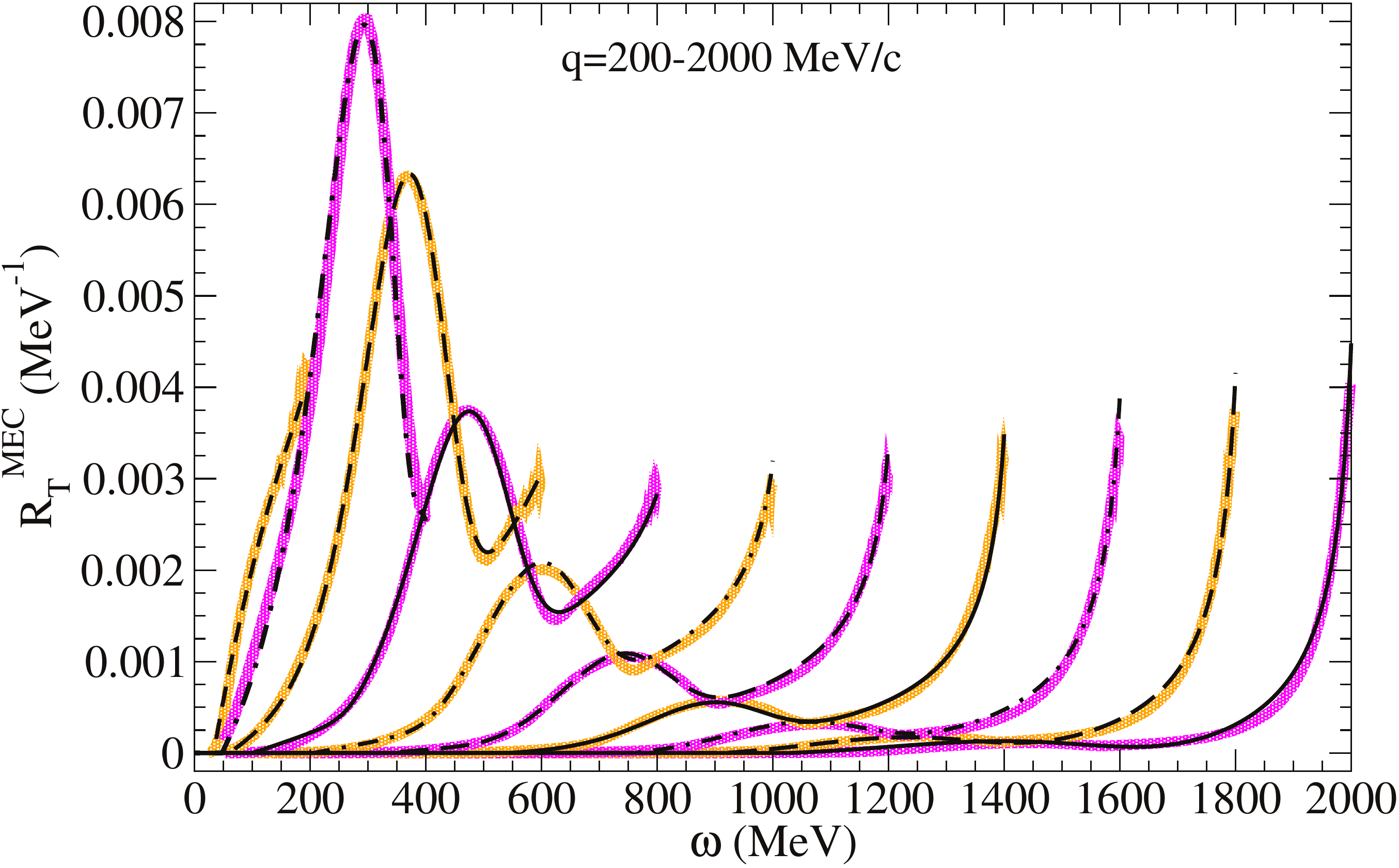} 
\caption{(Color online) $R_{T,\:VV}^{MEC,param}$ (MeV$^{-1}$) versus $\Psi'$ (first three panels) and versus $\omega$ (bottom panel), where for the last one the curves are displayed from left to right in steps of $q=$200 MeV/c. The parameterized responses are shown as black lines. Comparisons with Torino results (coloured thick lines) are also displayed. Note that the y-axis in the third panel is shown as a logarithmic scale. \label{fig2}}
\end{center}
\end{figure}

\begin{figure}[htbp]
\begin{center}
\includegraphics[scale=0.3, bb=-57 1 794 529, clip]{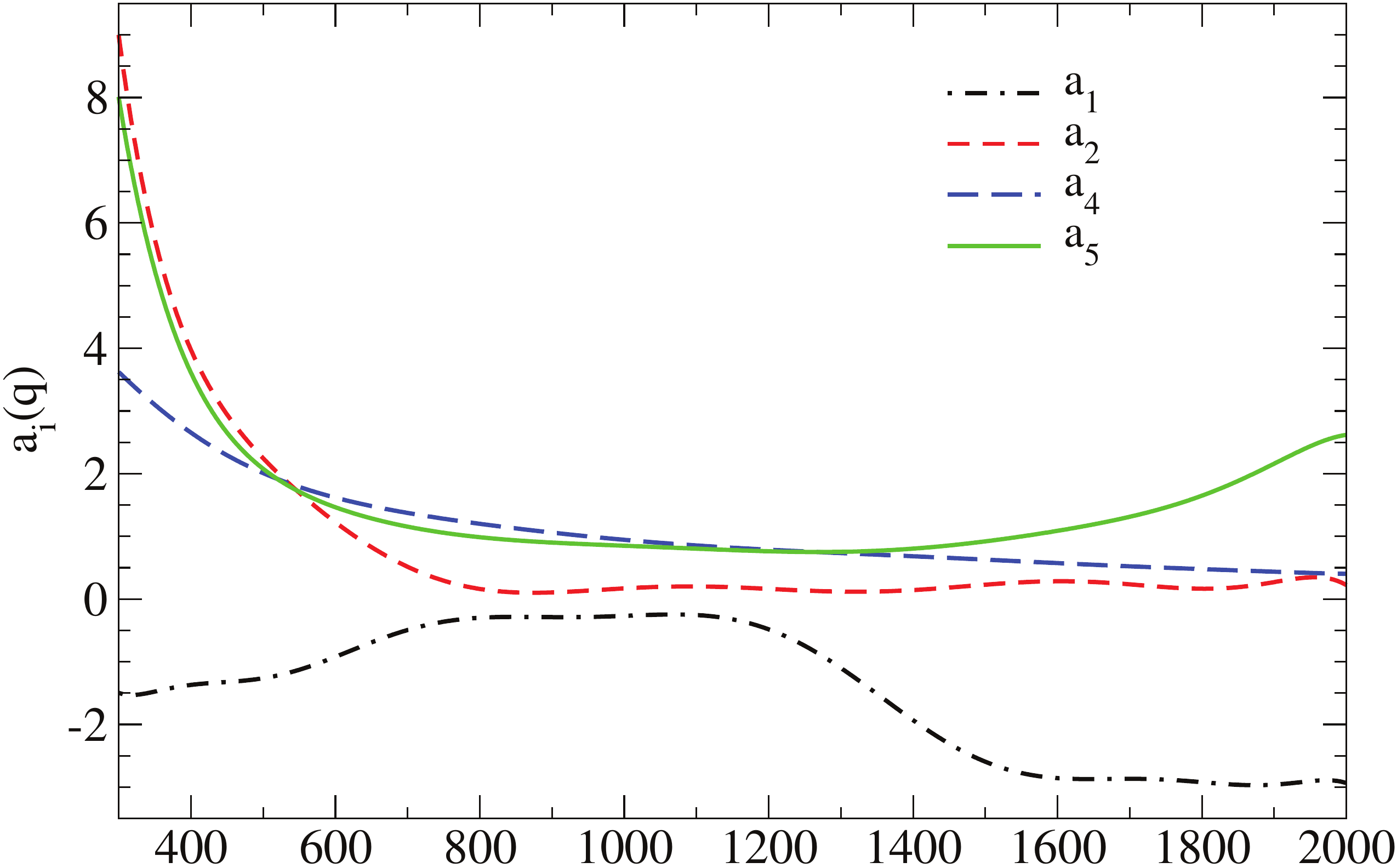} \\
\vspace{0.1cm}
\includegraphics[scale=0.3, bb=0 1 794 529, clip]{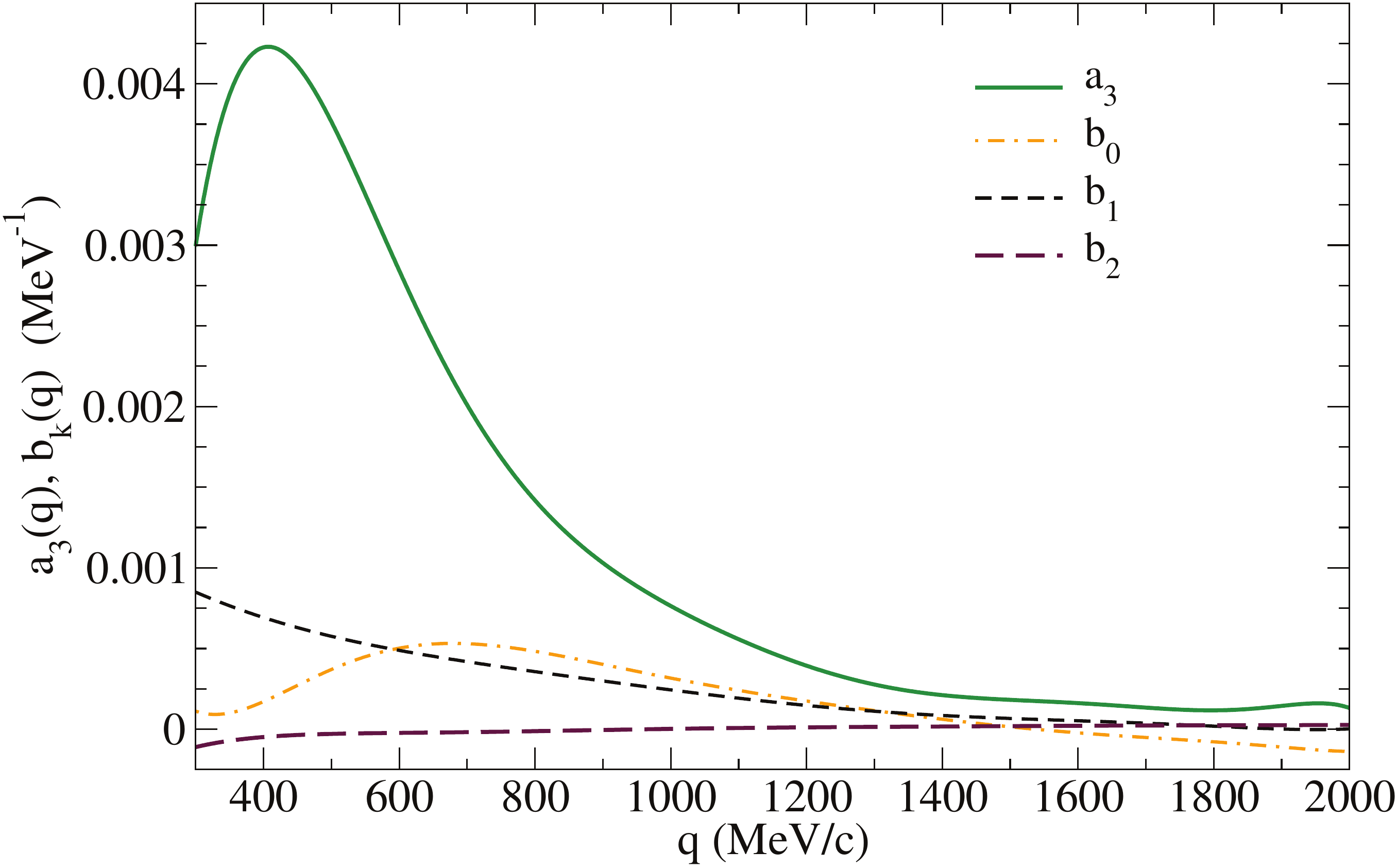}
\caption{(Color online) Dependence on q of the fitting parameters $\{a_i,b_k\}$.\label{fig1}}
\end{center}
\end{figure}

The MEC response functions for $q\geq400$ MeV/c exhibit a peak that decreases with $q$ together with a tail that rises with $\Psi'$ and $q$. In order to parameterize these functions we applied an expression with two terms, the first one mainly fitting the peak of the response and the second fitting the tail at larger $\Psi'$:
\begin{equation}\label{R_param}
 R_{T,\:VV}^{MEC}(\Psi')=\displaystyle\frac{2a_3 e^{-\frac{(\Psi'-a_4)^2}{a_5}}}{1+e^{-\frac{(\Psi'-a_1)^2}{a_2}}}+\sum_{k=0}^2 b_{k}\:(\Psi')^k \,.
\end{equation}
In this expression the parameters $a_i$, $b_k$ are $q$-dependent, and they are used to fit the original $R_{T,\:VV}^{MEC}$ responses shown in Fig. \ref{fig2}. We first fit each response for a given $q$ to get the values of the $a_i$, $b_k$ parameters for that specific $q$-value, ensuring a smooth dependence on $q$ for each of them. The $q$-dependent values of the fitting parameters are shown in Fig. \ref{fig1}. We then parametrize the $q$-dependence of the parameters themselves using a polynomial in $q$. The response in Eq.~(\ref{R_param}) then becomes explicitly dependent on the momentum transfer, $R_{T,\:VV}^{MEC}(\Psi',q)$, through the dependence in the parameters, $a_i(q)$, $b_k(q)$. 

\begin{figure}[htbp]
\begin{center}
\includegraphics[scale=0.3, bb=0 11 794 529]{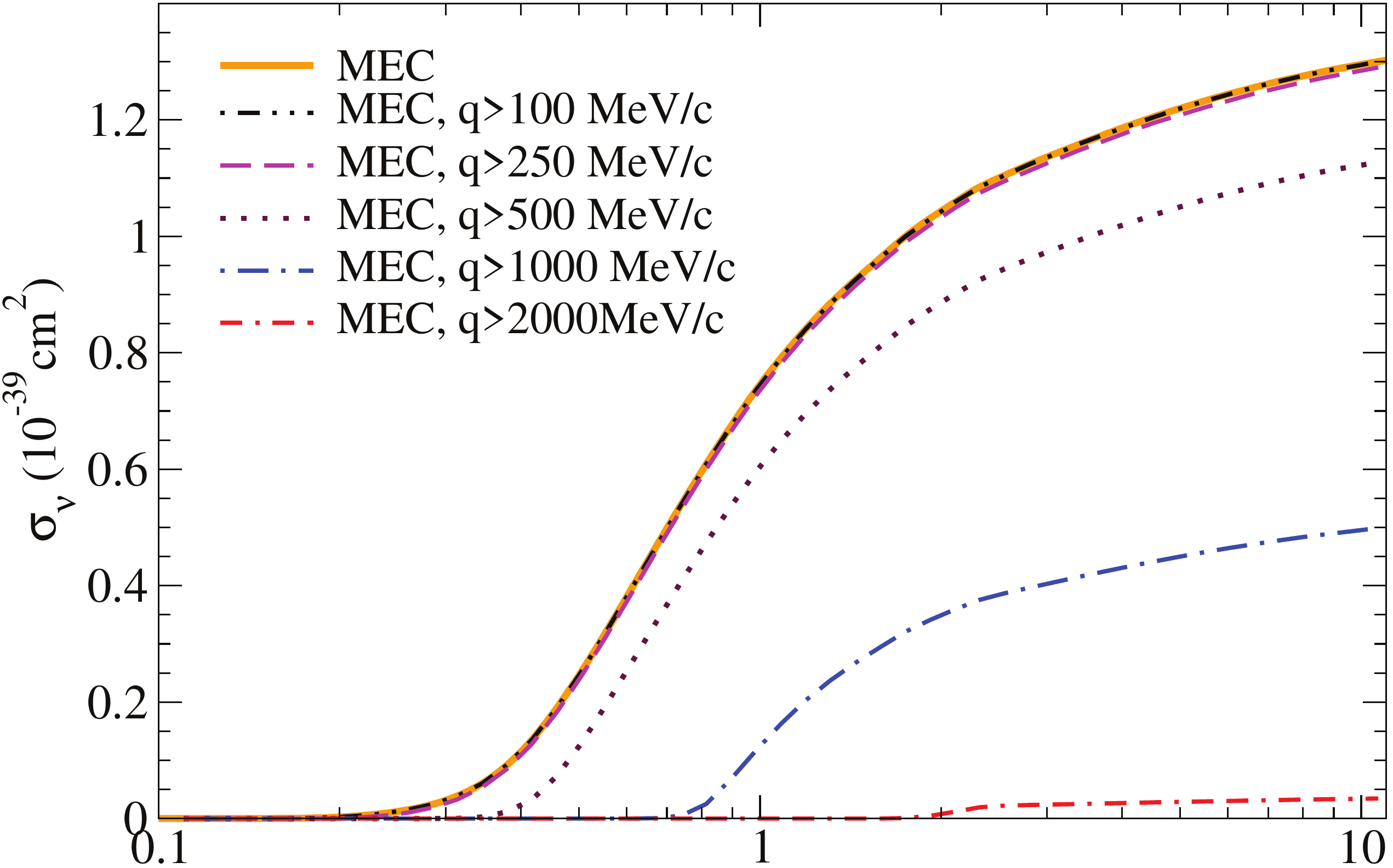} \\
\includegraphics[scale=0.3, bb=0 11 794 529]{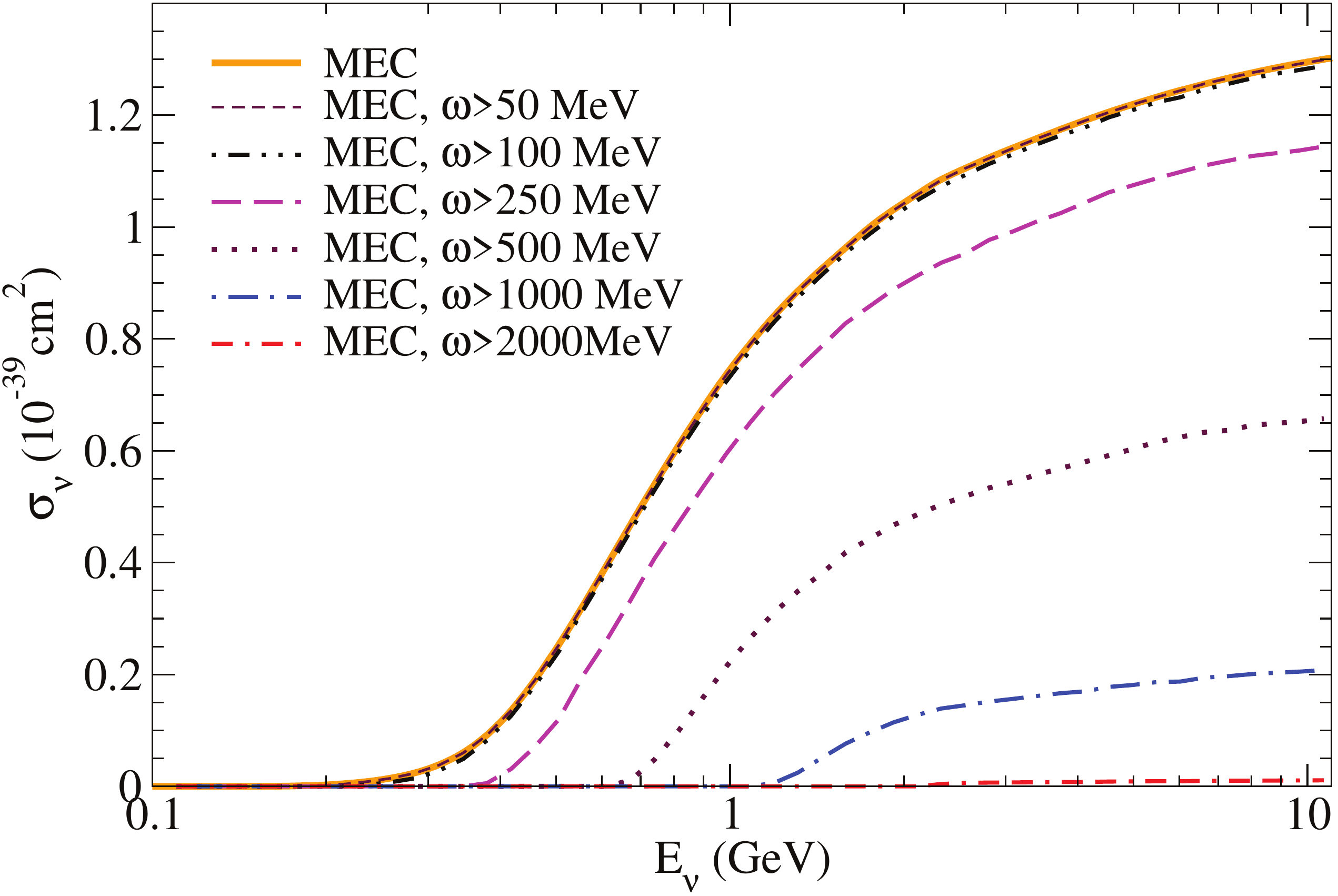}
\caption{(Color online) Total MEC neutrino cross section per target nucleon evaluated excluding all contributions coming from transferred momentum (upper panel) and energy (lower panel) below some selected values, as indicated in the figure. \label{fig11}}
\end{center}
\end{figure}

For the fitting of the responses above $q=2000$ MeV/c, which show almost no peak but a tail-like shape, we keep only the second term in Eq.~(\ref{R_param}), namely $a_3=0$; since these responses are very similar in the large-$q$ region under consideration (up to 3500 MeV/c), we use the same parametrization for all of them, namely $b_k(q>2000) = b_k(q=2000)$. In any case, as we can observe in Fig. \ref{fig11}, there are no significant MEC contributions for $q>$2000 MeV/c and the same is true for large $\omega >$ 1000 MeV. For the responses below $q=300$ MeV/c we use again a polynomial to fit the results, 
\begin{equation}
 \displaystyle R_{T,\:VV}^{MEC}(\Psi',q_{<300})=\sum_{k=0}^3 c_{k}(q) \:(\Psi')^k \,.
\end{equation}

The results of the above parametrization of the MEC responses are presented as a function of the scaling variable $\Psi'$ in Fig. \ref{fig2} where it is shown that it gives an excellent representation of the exact results in the full region of $q$ and $\Psi'$ explored.

\begin{figure}[htbp]
\begin{center}
\includegraphics[scale=0.3, bb=0 11 794 529]{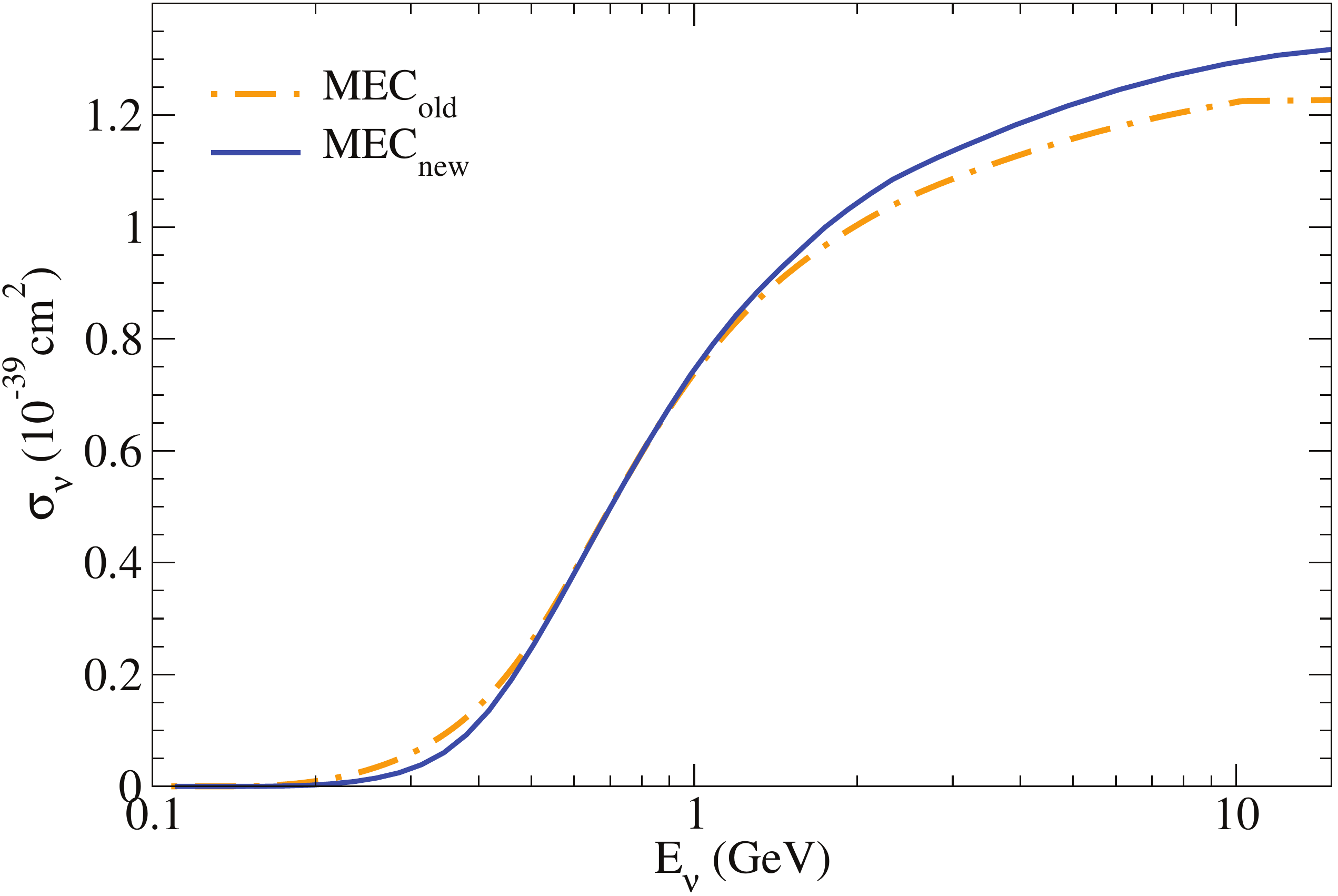}
\caption{(Color online) Comparison between the total MEC cross section in the present and past parametrizations. \label{fig8} }
\end{center}
\end{figure}

As already mentioned, in previous work \cite{Amaro11, Amaro11b, Amaro12} a simple parametrization of the exact MEC calculation was used in order to evaluate the MiniBooNE (anti)neutrino cross sections.  The present fit of the MEC responses improves the previous one in two respects: it uses data in a wider $q$ range and includes the tail of the responses at high $\Psi'$ or $\omega$ values. The previous parametrization was initially developed with electron scattering in mind and, since $(e,e')$ data are rarely available when $q \to \omega$, the high-$\omega$ region was ignored. Accordingly the old parametrization missed the high energy tails arising in the exact results and yielded lower peaks asymmetrically broadened towards higher $\Psi'$ values. In contrast, for CCQE reactions one must integrate over a broad neutrino spectrum and hence, potentially, the high-$\omega$ region may be relevant, and this motivated the re-evaluation of the MEC contributions. In Fig. \ref{fig2}, we also show the $R_{T,\:VV}^{MEC}$ 
results versus $\omega$ where it is noticed the negligible contribution below $q<300$ MeV/c as well as the relevance of the tail in the response at $q>800$ MeV/c. On the other hand, the tail of the MEC responses at high $q$ ($q>1000$ MeV/c) which appears at $\omega\gtrsim1000$ MeV does not contribute significantly to the cross section, as can be deduced from Fig. \ref{fig11}, and in fact the old and new parametrizations are observed to be very similar except at low neutrino energy where minor differences occur and at very high neutrino energy where the new parametrization yields somewhat larger contributions, as seen in Fig. \ref{fig8}.

\begin{figure}[htbp]
\begin{center}
\includegraphics[scale=0.3, bb=0 11 794 529]{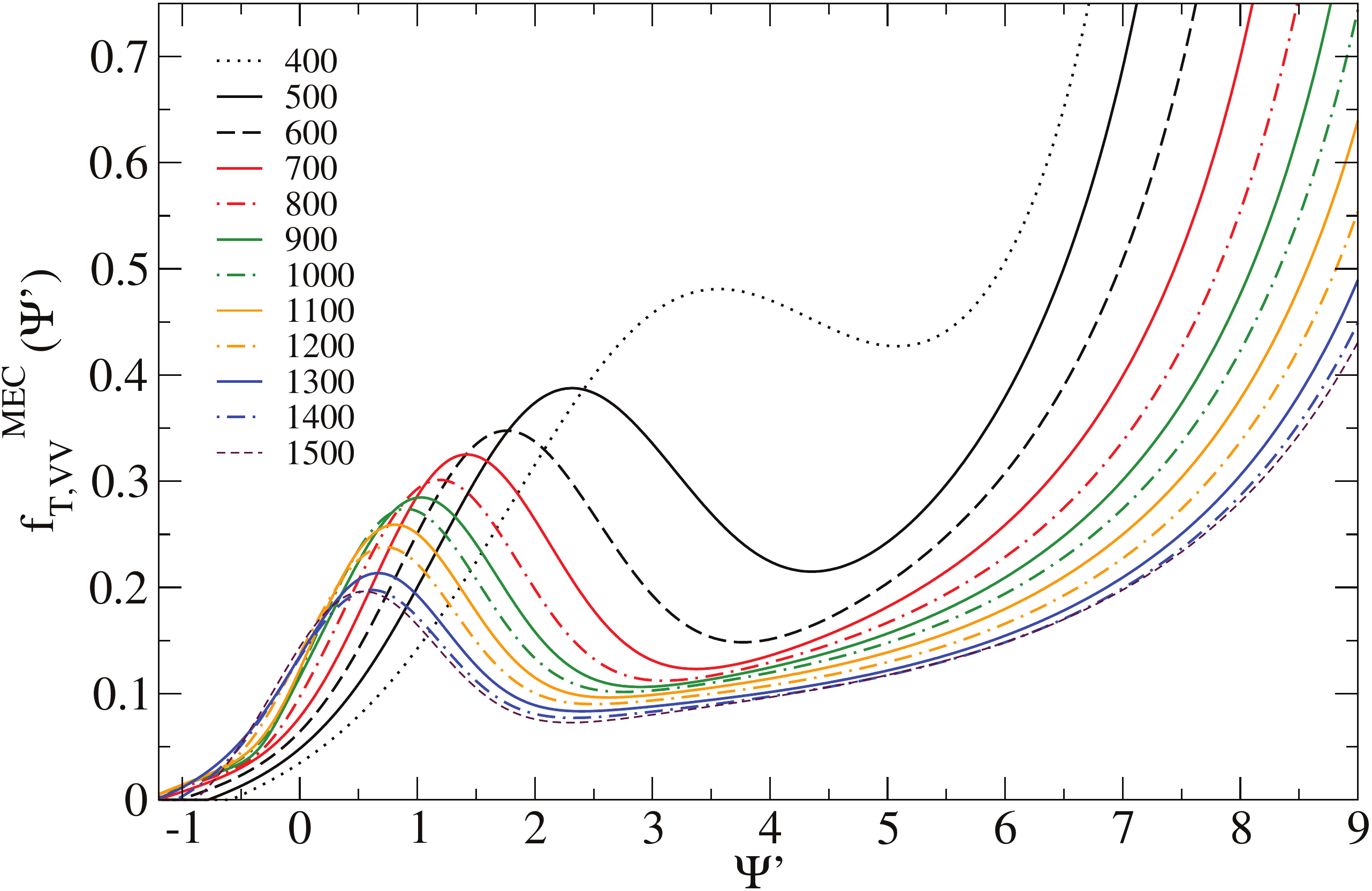} 
\caption{(Color online) Transverse 2p-2h MEC isovector scaling functions $f_{T,VV}^{MEC}$ versus the scaling variable $\Psi'$ from $q=$400 MeV/c to 1500 MeV/c.  \label{fig3}}
\end{center}
\end{figure}

In order to subtract some of the nucleonic and nuclear properties from the 2p-2h MEC parametrization, we can introduce a 2p-2h MEC isovector scaling function, $f_{T,\:VV}^{MEC}$, defined analogously to the transverse scaling function coming from the transverse one-body response:
\begin{eqnarray}
f_{T,VV}^{MEC}(\kappa,\lambda)=k_F\cdot\frac{R_{T,VV}^{MEC}(\kappa,\lambda)}{G_T(\kappa,\lambda)} \:,
\end{eqnarray}
where the $G_T$ factor depends on the momentum and energy transferred as well as on the isovector magnetic nucleon form factors and $k_F$ 
is the Fermi momentum of the nucleus. A detailed expression for $G_T$, including higher-order relativistic corrections, can be found 
in \cite{Amaro05b} and has been used in the calculation of $f_{T,\:VV}^{MEC}$ shown in Fig. \ref{fig3}.  The remaining dependence on $q$ 
of the scaling function seen in Fig.~\ref{fig3} is consistent with the violation of first-kind scaling exhibited by the MEC~\cite{DePace04}. The study of second-kind scaling violation, 
related to the dependence on the nuclear species, would require an in-depth study of the MEC contributions in other nuclei; some such studies were presented in~\cite{DePace04}. 

\begin{figure}[htbp]
\begin{center}
\includegraphics[scale=0.4, bb=0 11 794 288]{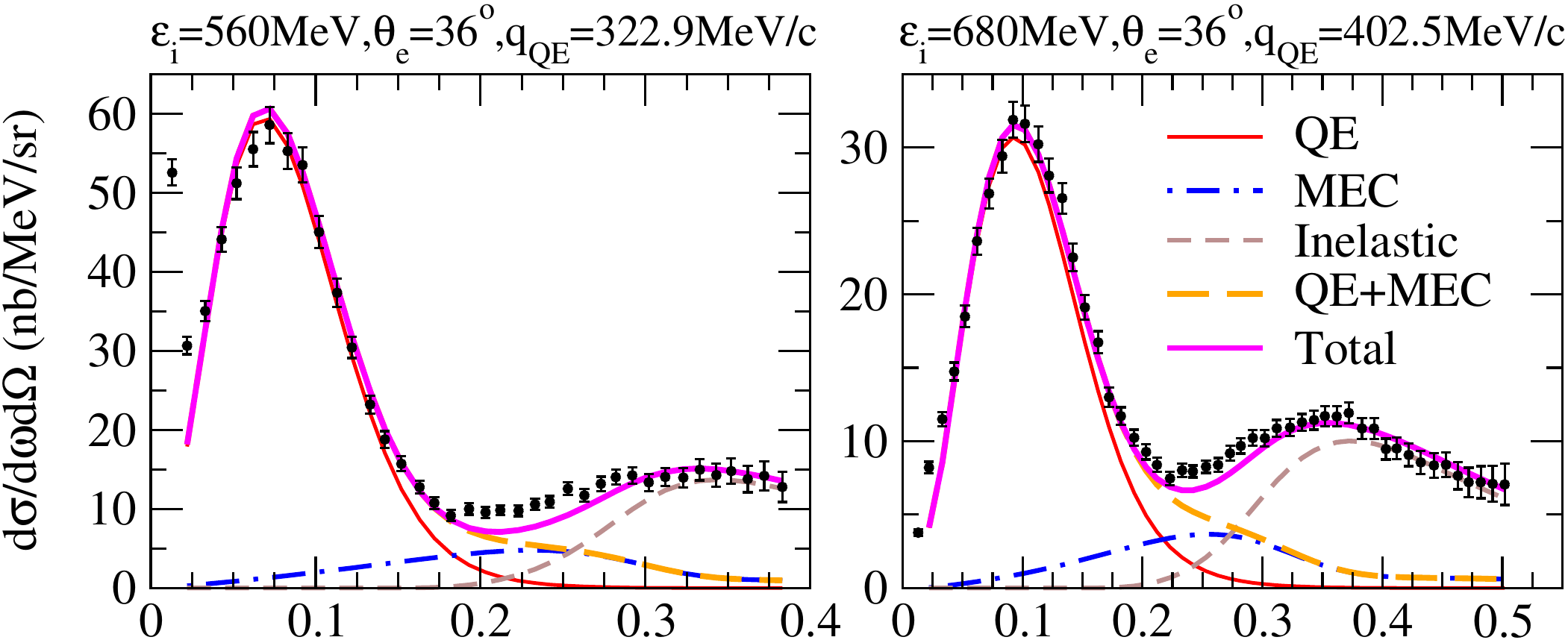} \\
\includegraphics[scale=0.4, bb=0 11 794 288]{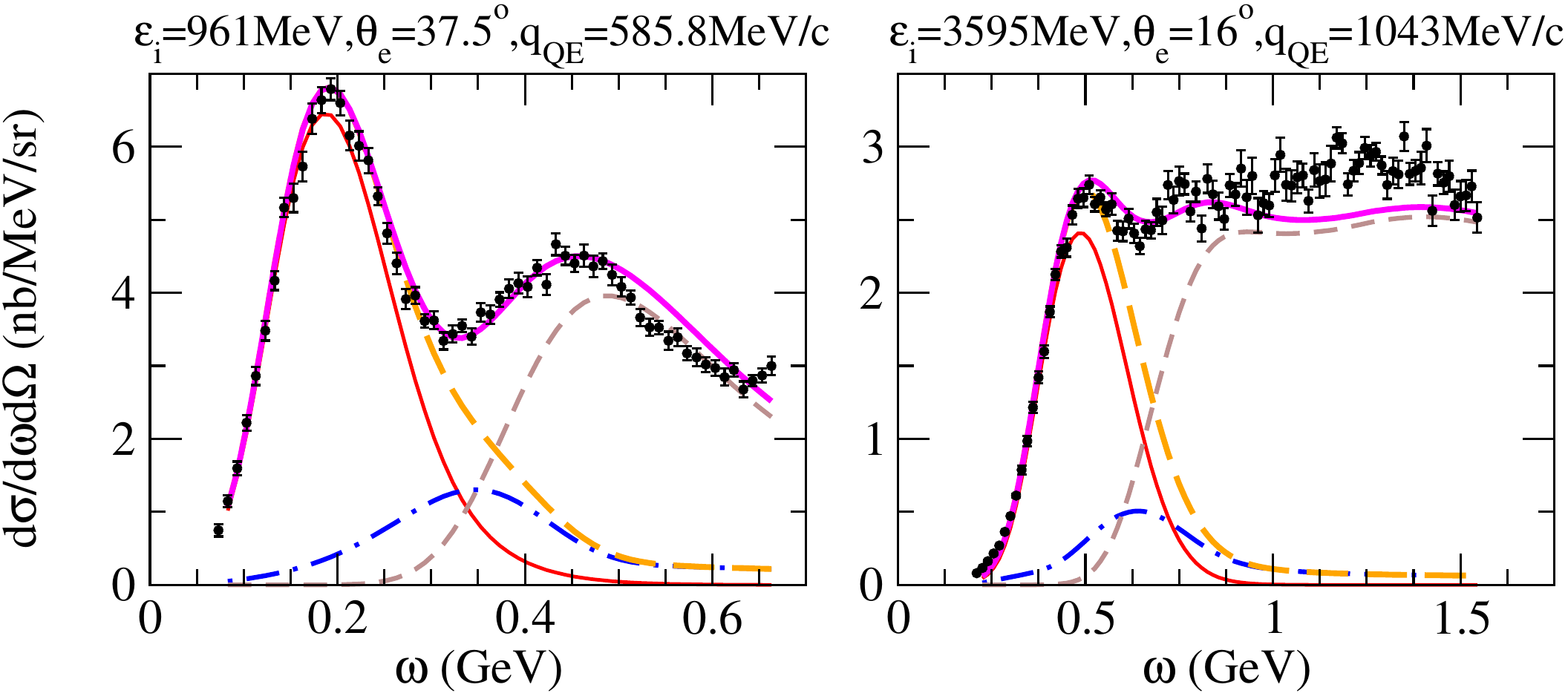}
\caption{(Color online) Comparison of inclusive $^{12}$C($e, e'$) cross
sections and predictions of the QE(SuSAv2), MEC and Inelastic(SuSAv2) models at different set values of the position of the QE peak ($q_{QE}$), incident electron energy ($\varepsilon_i$) and the scattering angle ($\theta_e$). Data taken from \cite{Benhar:2006wy}.\label{eeMEC}}
\end{center}
\end{figure}

For completeness, a comparison between our theoretical predictions and electron scattering data \cite{Benhar:2006wy} at kinematics 
where MEC contributions are relevant, extending from the non-relativistic to the highly-inelastic regime, is also presented in 
Fig.~\ref{eeMEC}. As shown, a model based solely on impulsive
response function is not able to reproduce the (e,e') data. Contributions beyond the impulse approximation such as 2p-2h MEC
could provide part of the missing strength in the transverse channel. Moreover, the addition of the impulsive inelastic contributions is shown 
to be essential to analyze the (e,e') data at high kinematics.

In general the inelastic contributions
can have a significant effect on the $(e,e')$ cross section even in
the QE regime, since the different domains can overlap. This agrees with the emerging pattern in Fig.~\ref{eeMEC} 
that suggests that the inclusion of inelastic processes --- the
contribution of which clearly extends into the region dominated by
quasielastic scattering---may lead to an enhancement of the theoretical results.
The inelastic part of the cross section is dominated by the delta
peak (mainly transverse) that contributes to the transverse response function. 
At low electron scattering angles the
longitudinal response function dominates the cross section and the
inelastic contribution is smaller. The opposite holds at large
scattering angles, where the delta peak contribution is important. On the other hand, 
for increasing values of the transferred momentum the peaks corresponding to the Delta and QE domains
become closer, and their overlap increases significantly. This general behaviour is clearly shown by our predictions
compared with data. In those kinematical situations where inelastic processes are expected to be important, our results
for the QE peak are clearly below the data. On the contrary, when the inelastic contributions are
expected to be small, the QE theoretical predictions get closer to data. It is important to point out that the description
presented in this work corresponds to a semi-phenomenological model where the
scaling function is fitted to the longitudinal ($e,e'$) scattering data (and extended to the transverse response via the RMF theory).
Thus, it does not encode the inelasticities that dominate the transverse response.

However, for completeness we also show in Fig.~\ref{eeMEC} some 
results for the inelastic contributions. As observed,  
the inclusion of the inelastic processes does not necessarily imply a {\sl ``significant''} enhancement of the cross
section in the region close to the QE peak. In fact, at the particular kinematics considered in Fig.~\ref{eeMEC} 
the overlap between the QE and inelastic regions is small and therefore the agreement with the data in the QE
region is not spoiled. However, more detailed results are needed before more definitive conclusions can be reached.
In this sense, a new analysis of the inelastic channel based on the use of the recent SuSAv2 and 
MEC models will be presented in a forthcoming paper \cite{future_ee}.

\section{Evaluation of neutrino cross sections} \label{cross-sections}

In this section, we evaluate the CCQE double-differential and total cross sections of (anti)neutrino scattering off $^{12}$C using our latest SuSAv2 
results and the new 2p-2h MEC parametrization. We compare the results with experimental data of MiniBooNE, NOMAD and MINER$\nu$A.

\begin{figure}[htbp]
\begin{center}

\includegraphics[scale=0.21, bb=0 0 784 557, clip]{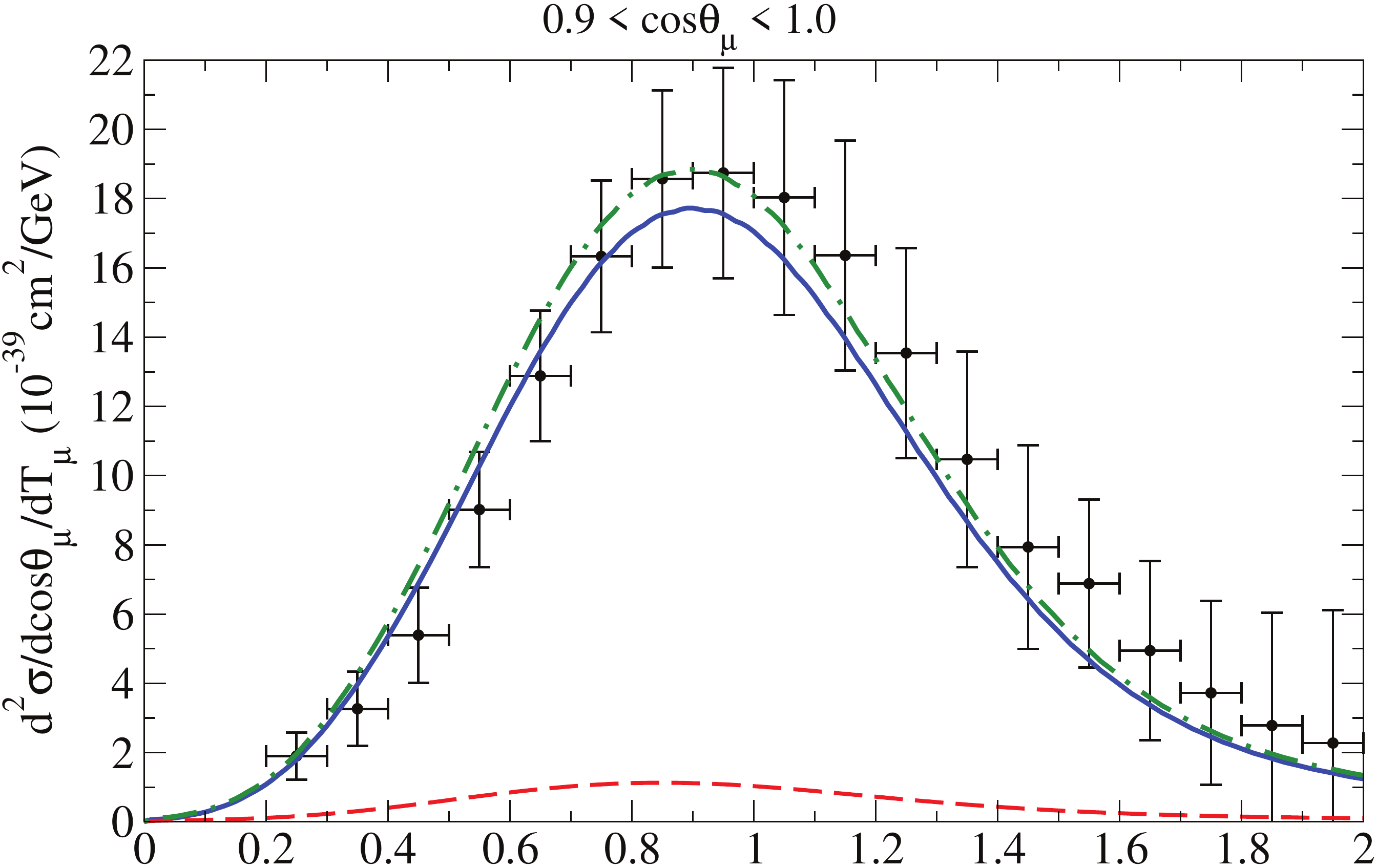}
\includegraphics[scale=0.21, bb=-20 0 784 557, clip]{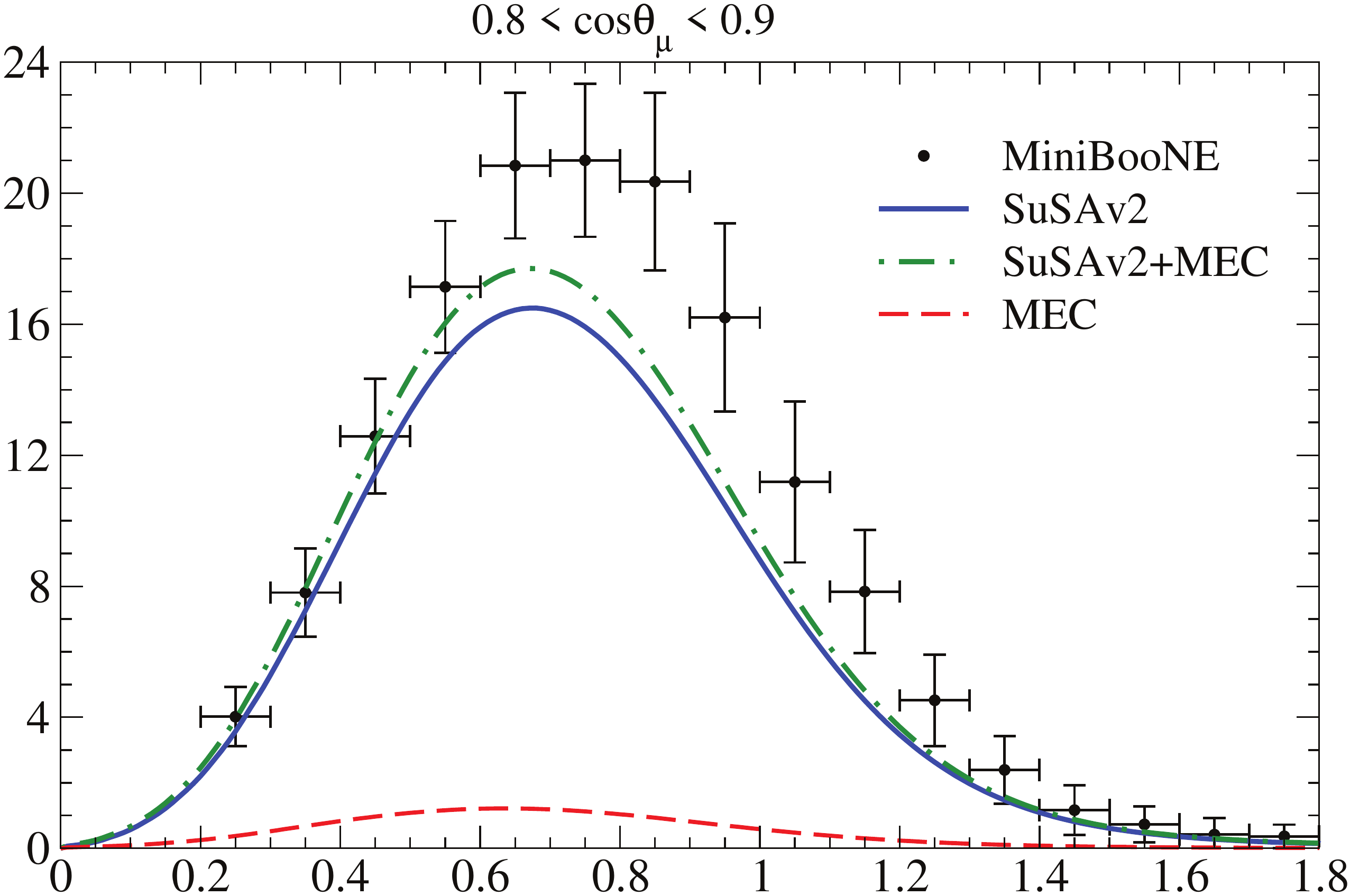} \hspace{-0.25cm}
\includegraphics[scale=0.21, bb=0 0 784 557, clip]{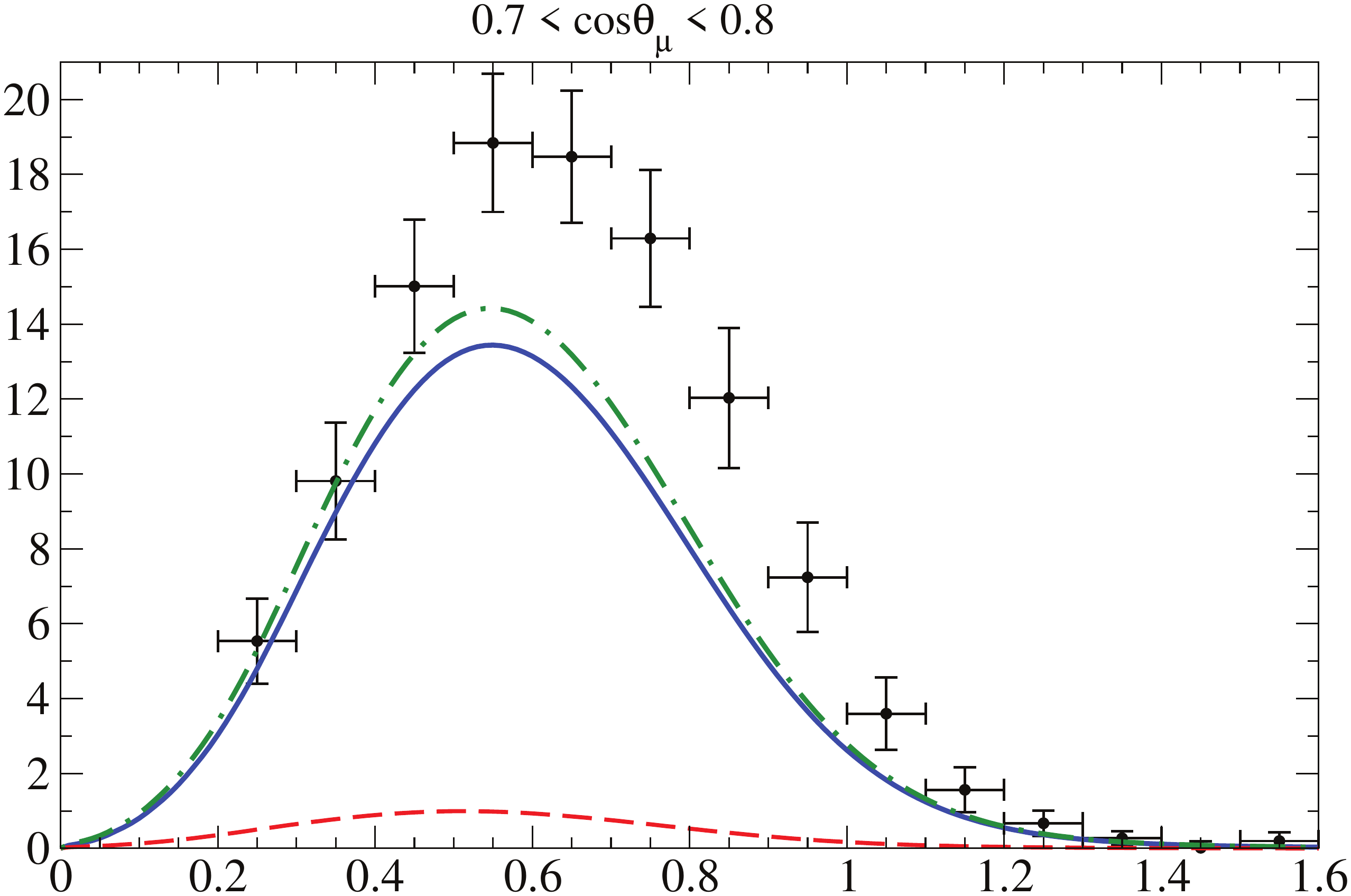} \\
\includegraphics[scale=0.21, bb=0 0 784 557, clip]{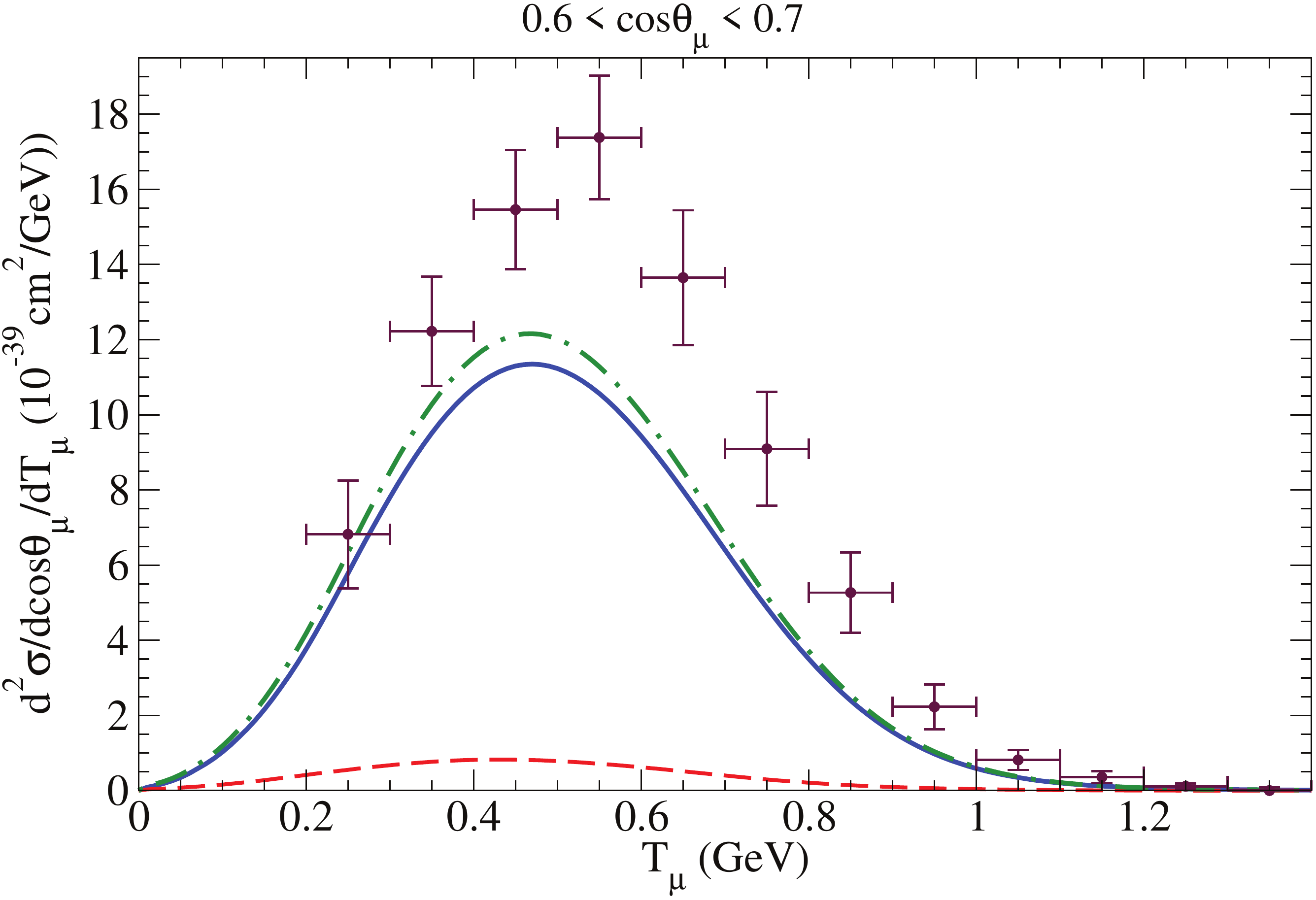}
\includegraphics[scale=0.21, bb=-20 0 784 557, clip]{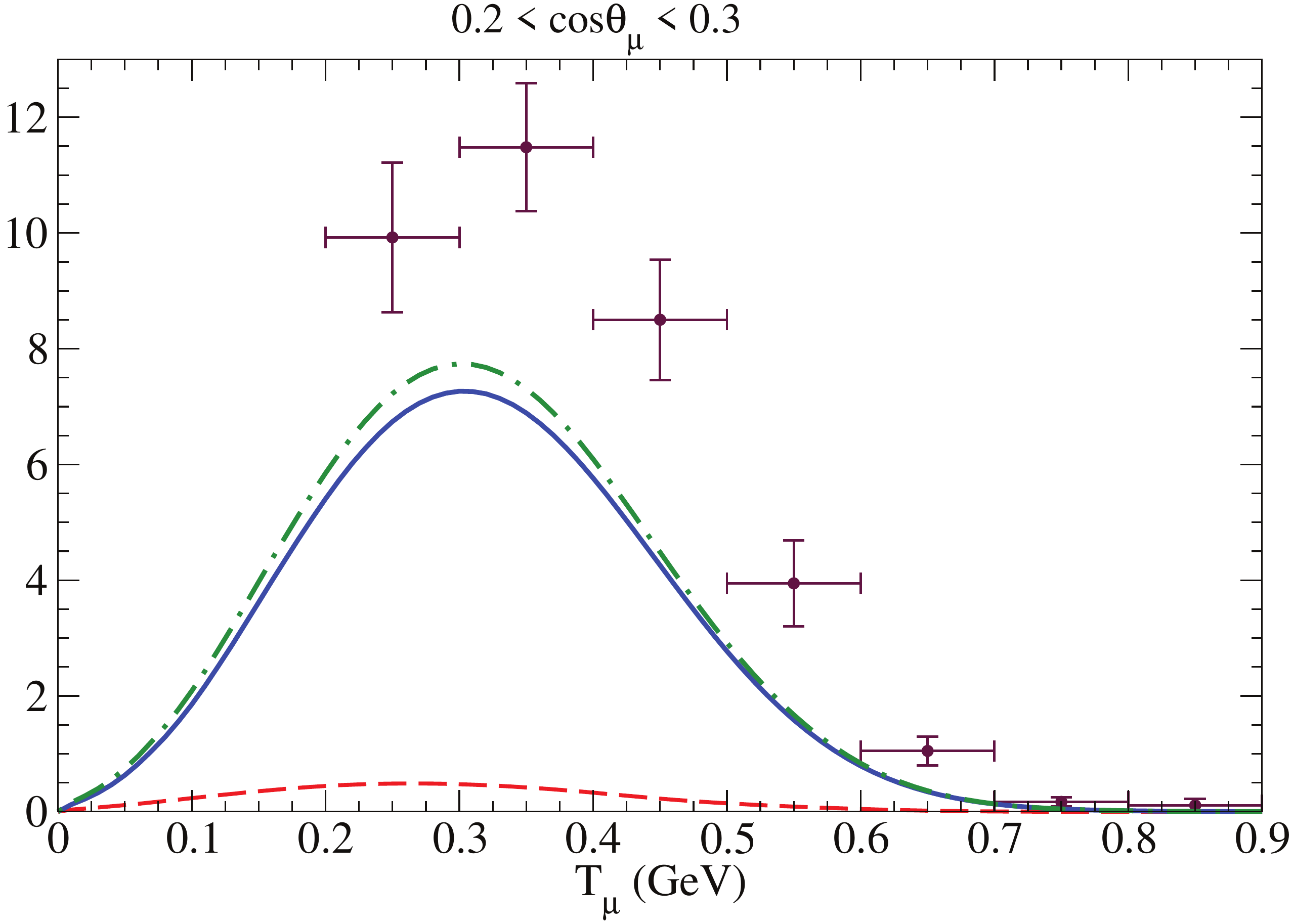} \hspace{-0.25cm}
\includegraphics[scale=0.21, bb=0 0 784 557, clip]{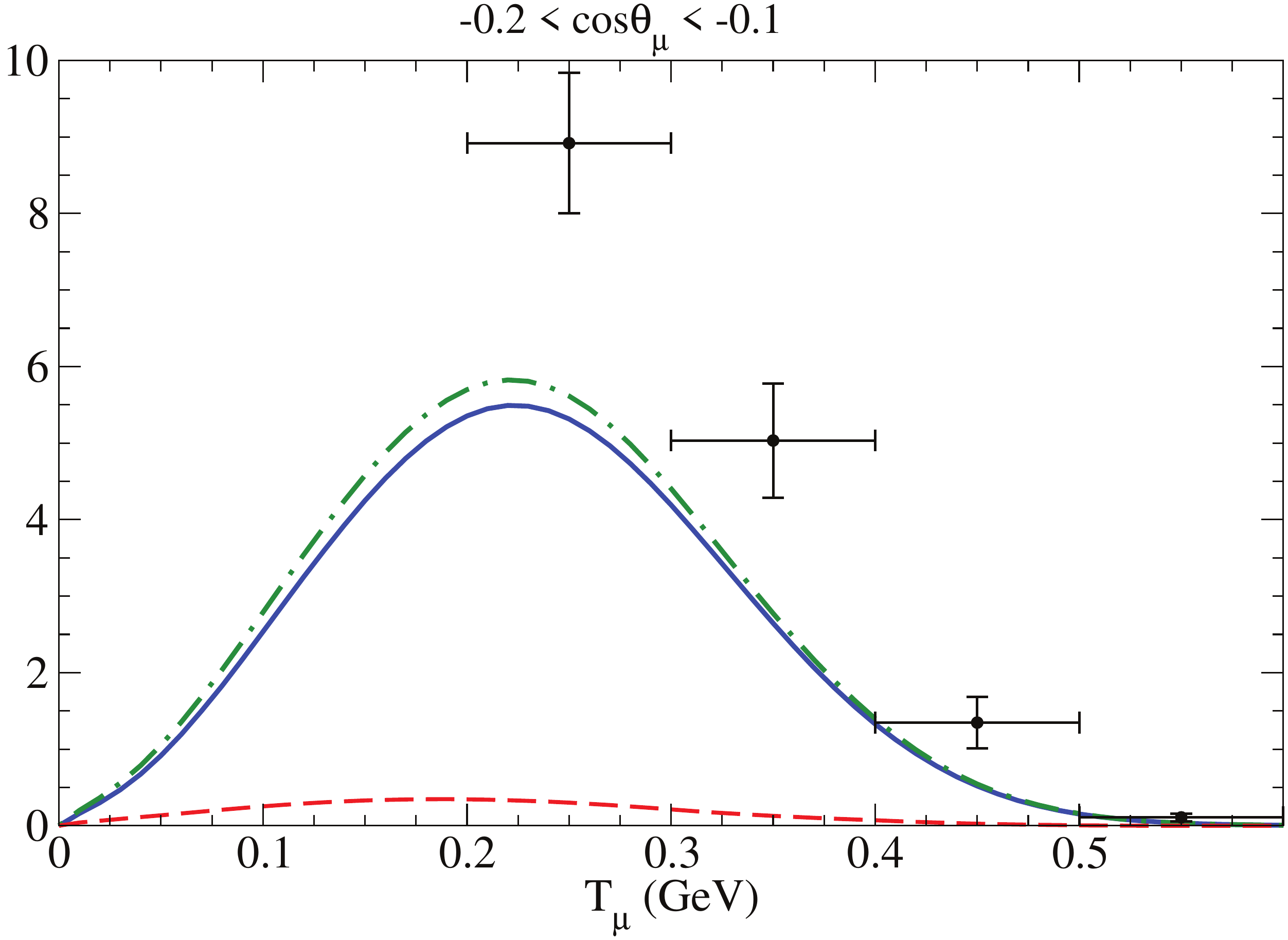}
\caption{(Color online) Flux-integrated double-differential cross section per target nucleon for the $\nu_\mu$ CCQE process on $^{12}$C displayed versus the $\mu^-$ kinetic energy $T_\mu$ for various bins of $\cos\theta_\mu$ obtained within the SuSAv2 and SuSAv2+MEC approaches. MEC results are also shown. The data are from Ref. \cite{MiniBooNECC10}.  \label{fig4}}
\end{center}
\end{figure}

\begin{figure}[htbp]
\begin{center}
\includegraphics[scale=0.21, bb=0 0 784 557, clip]{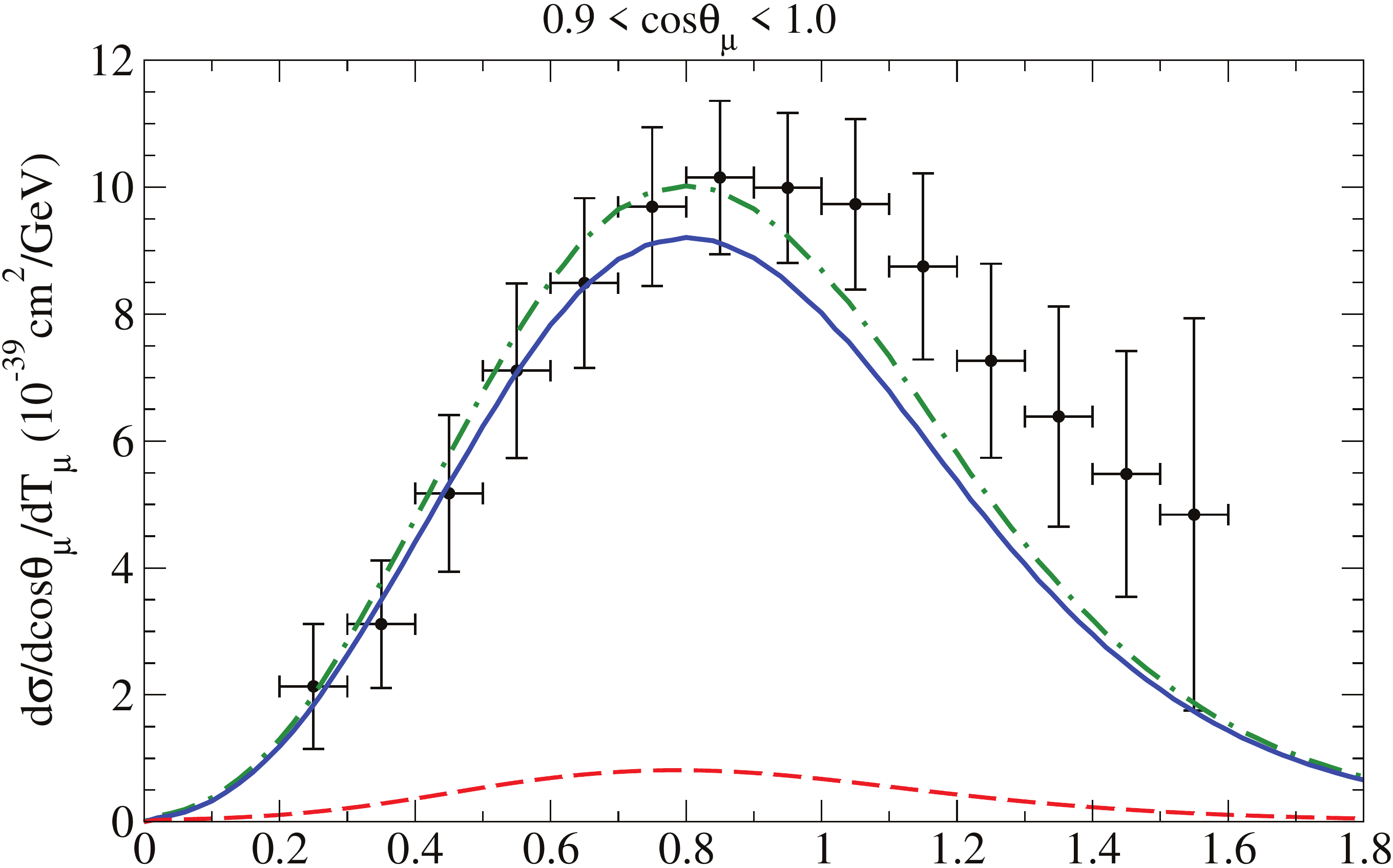}
\includegraphics[scale=0.21, bb=-20 0 784 557, clip]{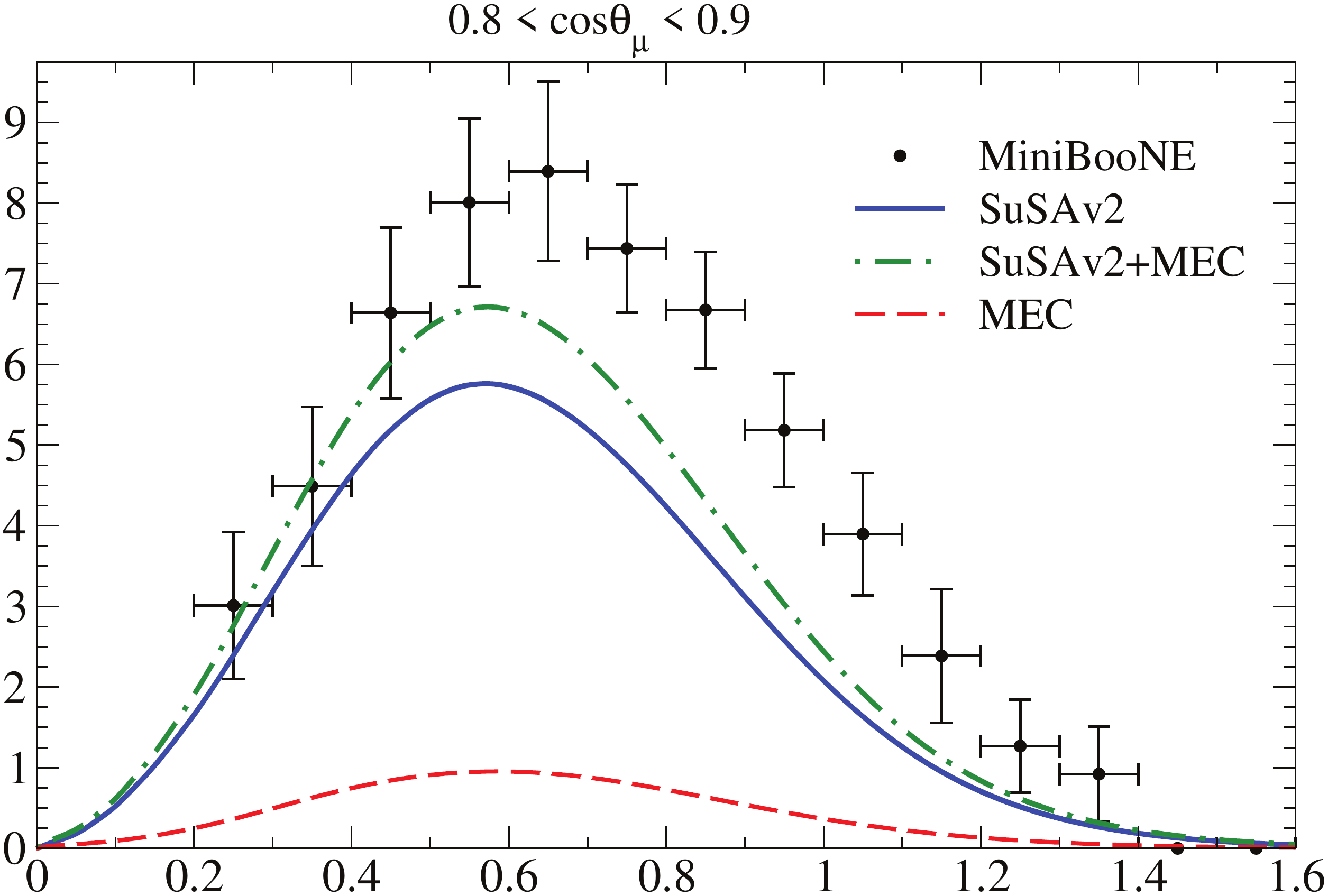} \hspace{-0.25cm}
\includegraphics[scale=0.21, bb=0 0 784 557, clip]{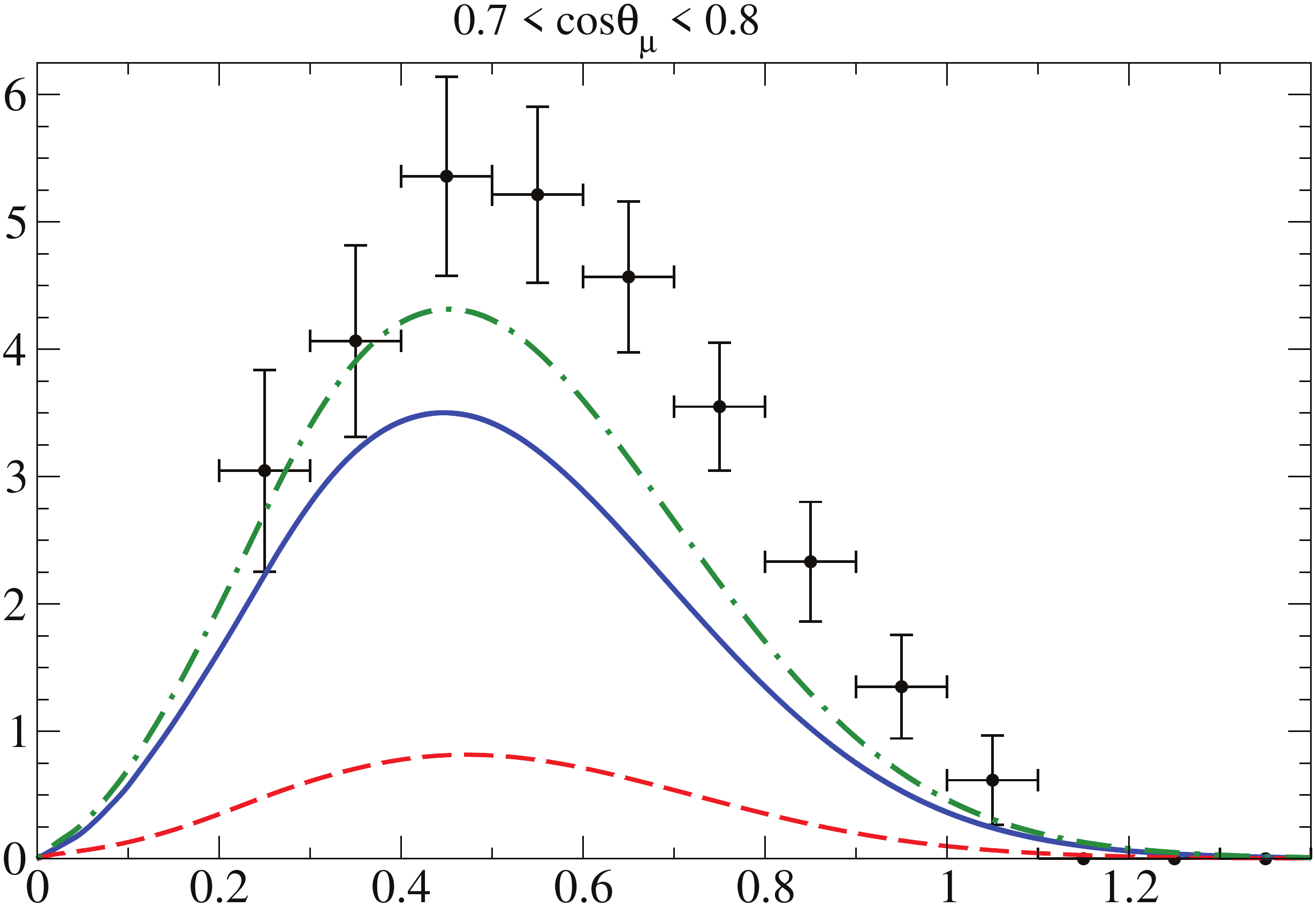} \\
\includegraphics[scale=0.21, bb=0 0 784 557, clip]{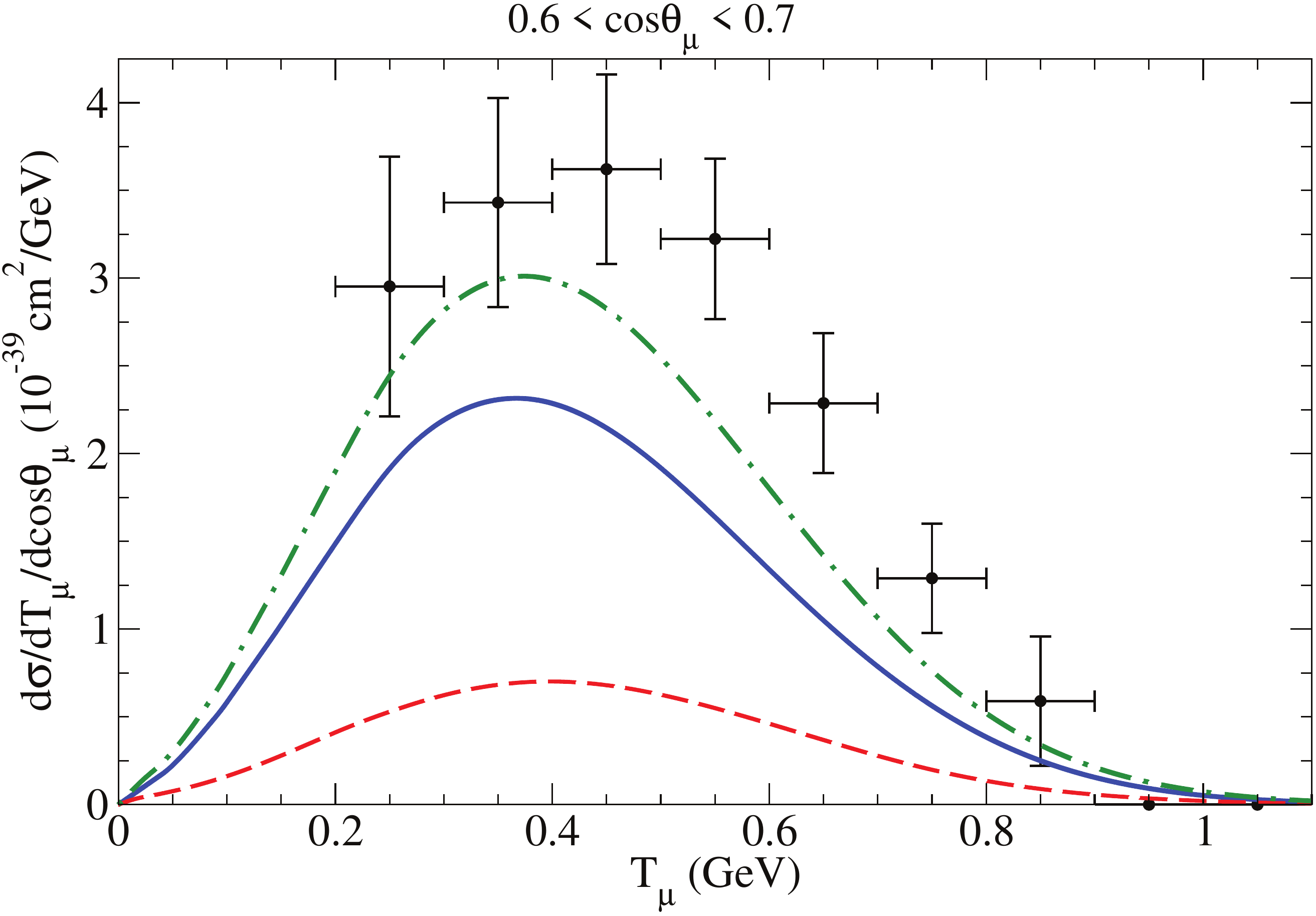}
\includegraphics[scale=0.21, bb=10 0 784 557, clip]{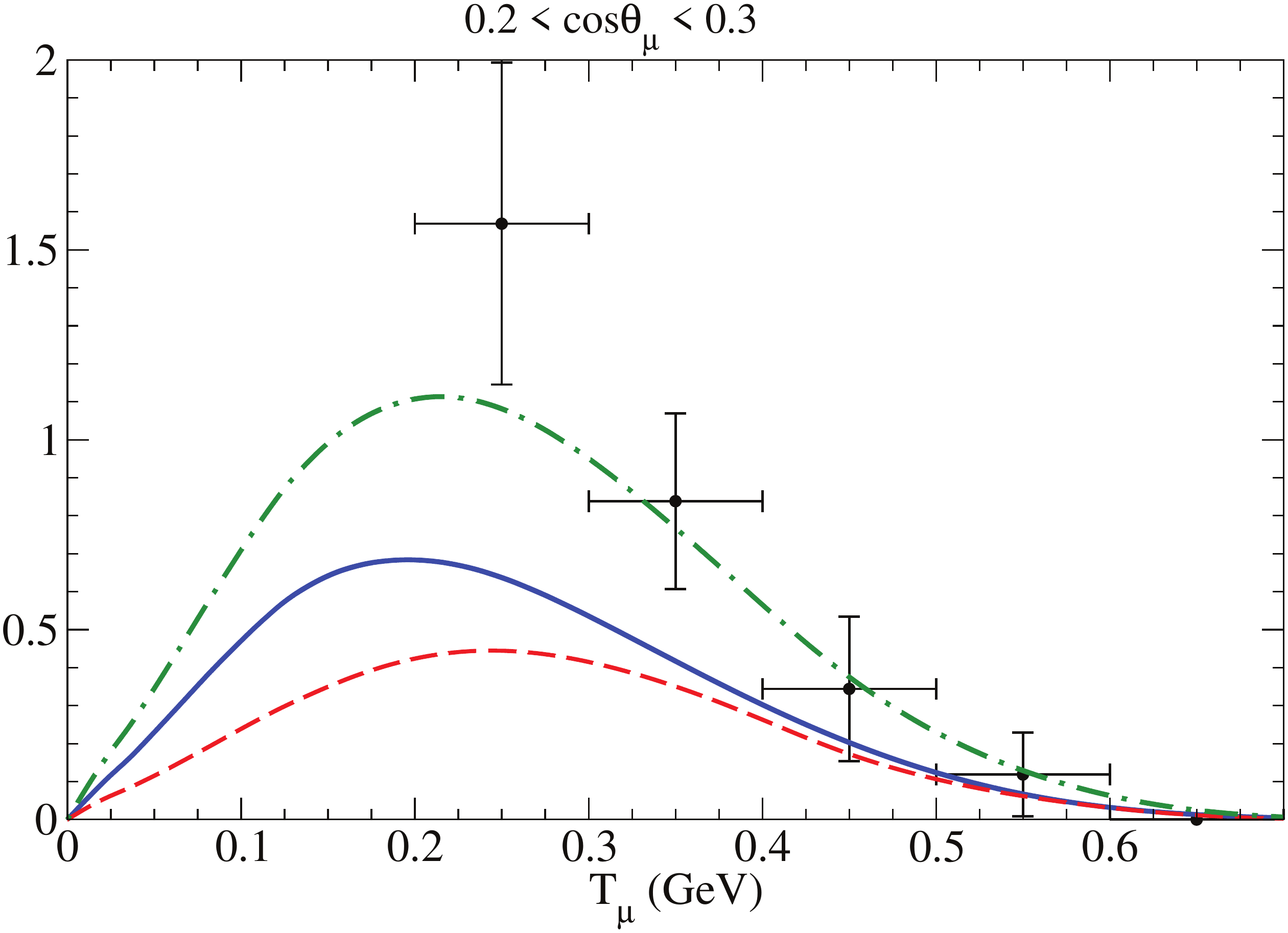}  \hspace{-0.25cm}
\includegraphics[scale=0.21, bb=0 0 784 557, clip]{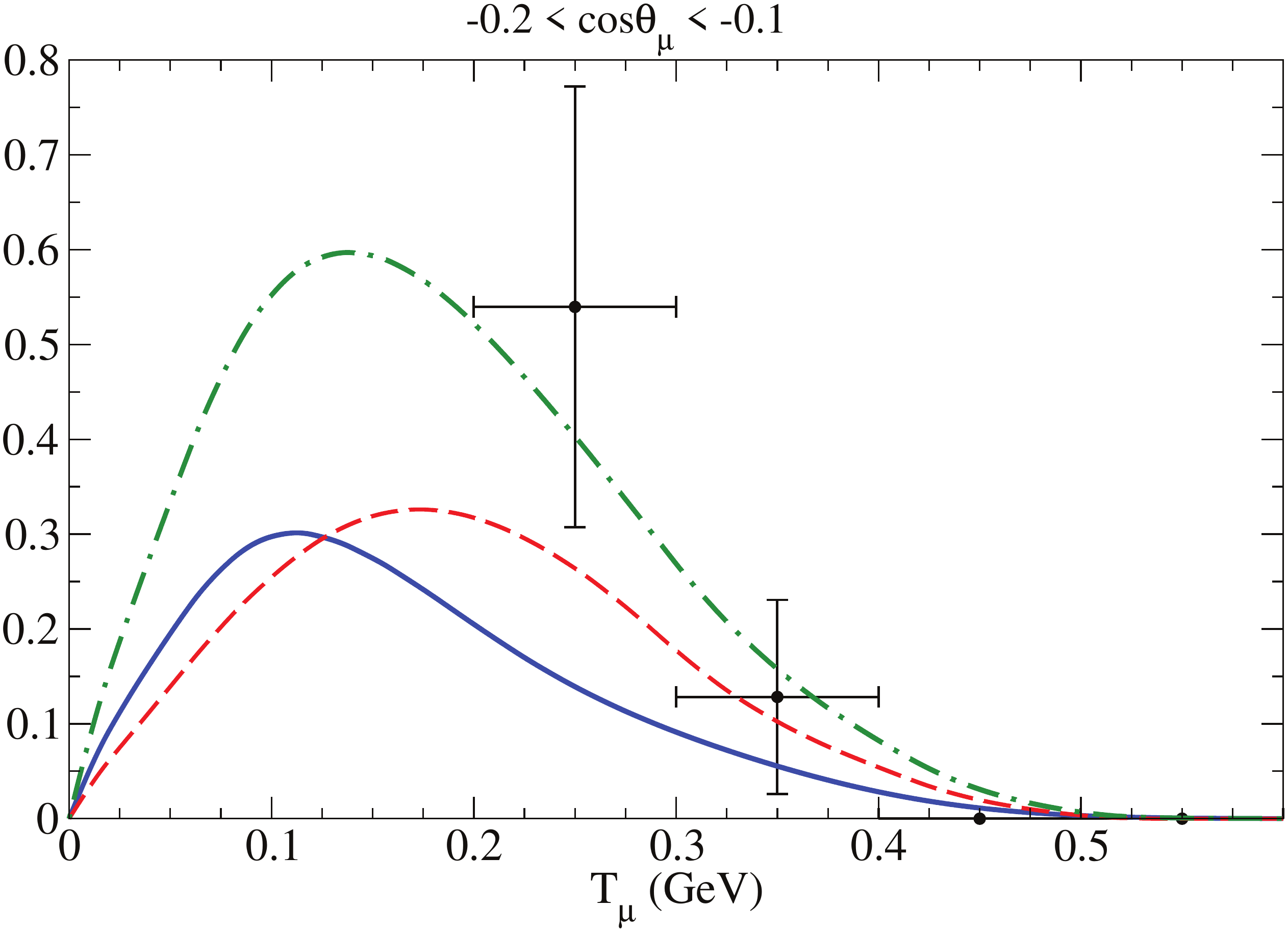}
\caption{(Color online) As for Fig. \ref{fig4}, but for $\bar\nu_\mu$ scattering versus $\mu^+$ kinetic energy $T_\mu$. The data are from Ref. \cite{MiniBooNECC13}.  \label{fig5}}
\end{center}
\end{figure}

\begin{figure}[htbp]
\begin{center}

\includegraphics[scale=0.21, bb=0 0 784 557, clip]{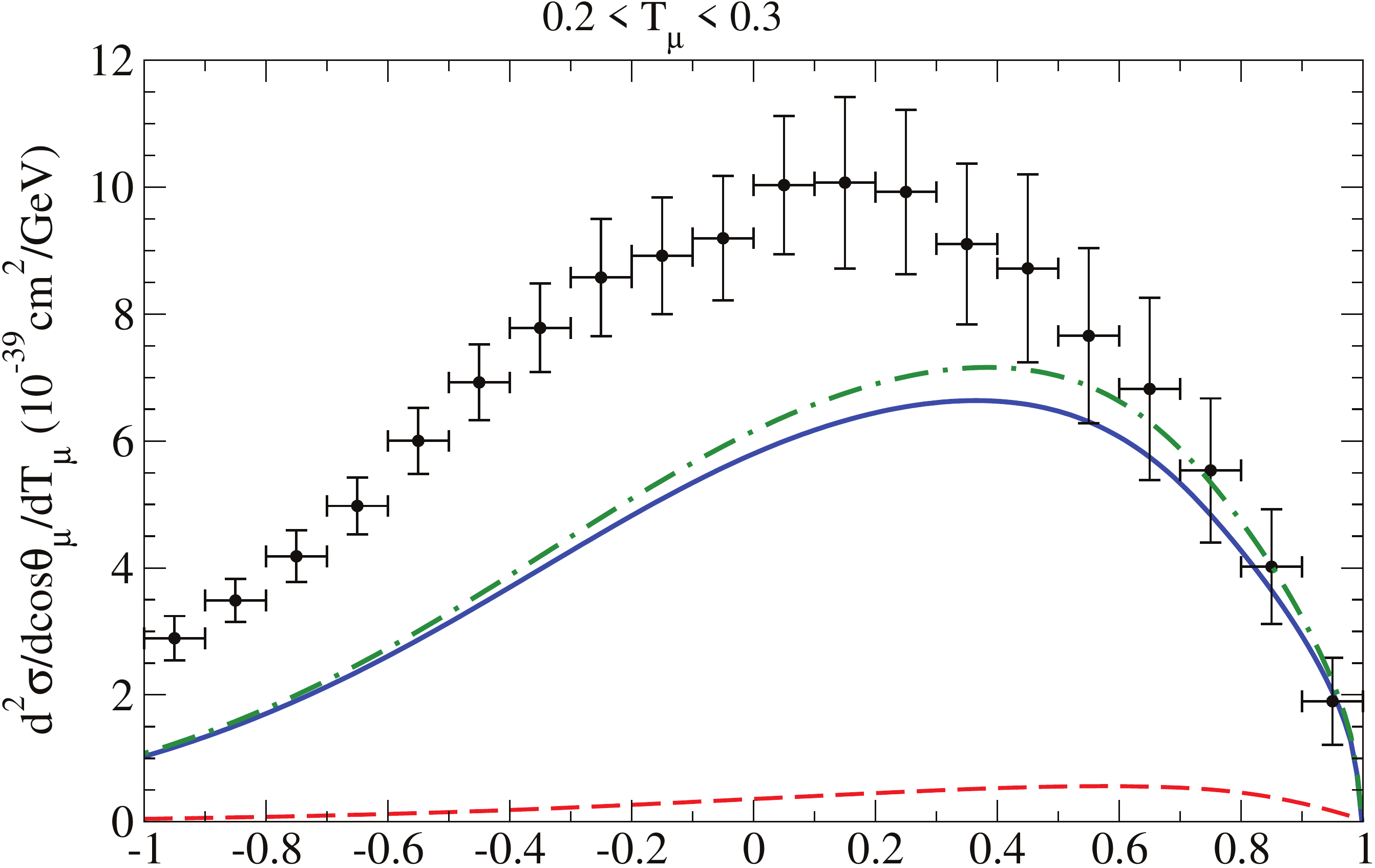}
\includegraphics[scale=0.21, bb=-20 0 784 557, clip]{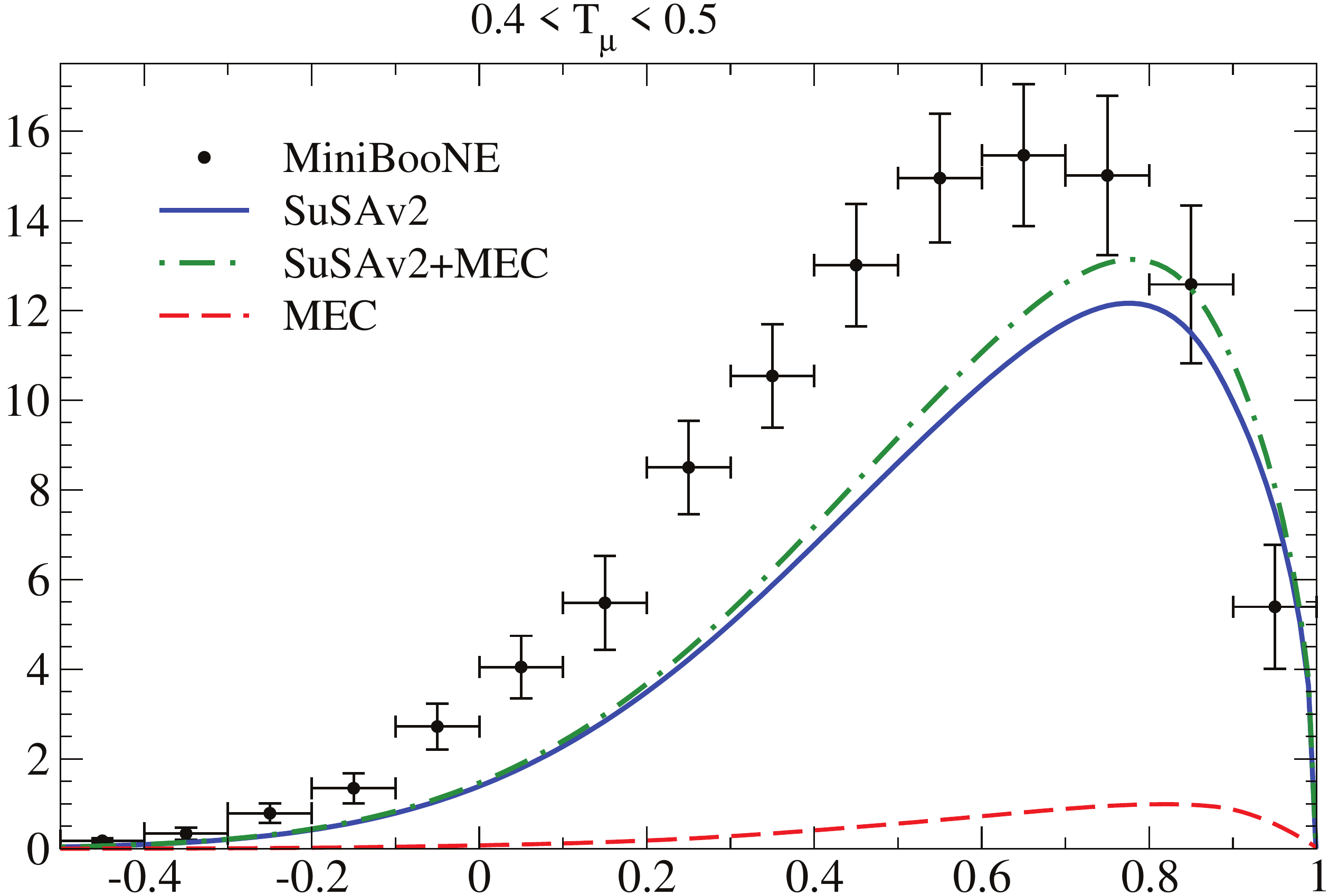} \hspace{-0.25cm}
\includegraphics[scale=0.21, bb=0 0 784 557, clip]{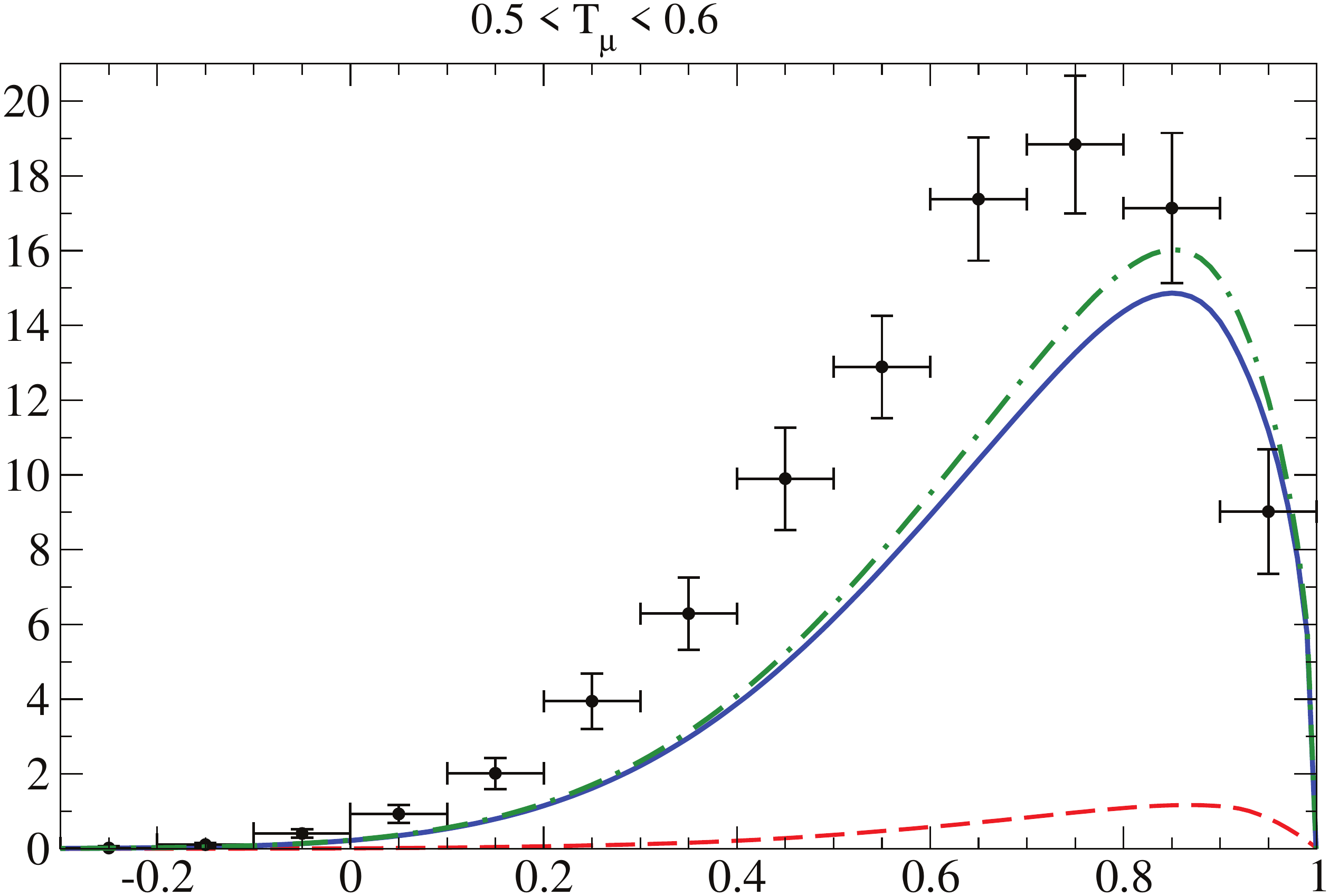} \\
\includegraphics[scale=0.21, bb=0 0 784 557, clip]{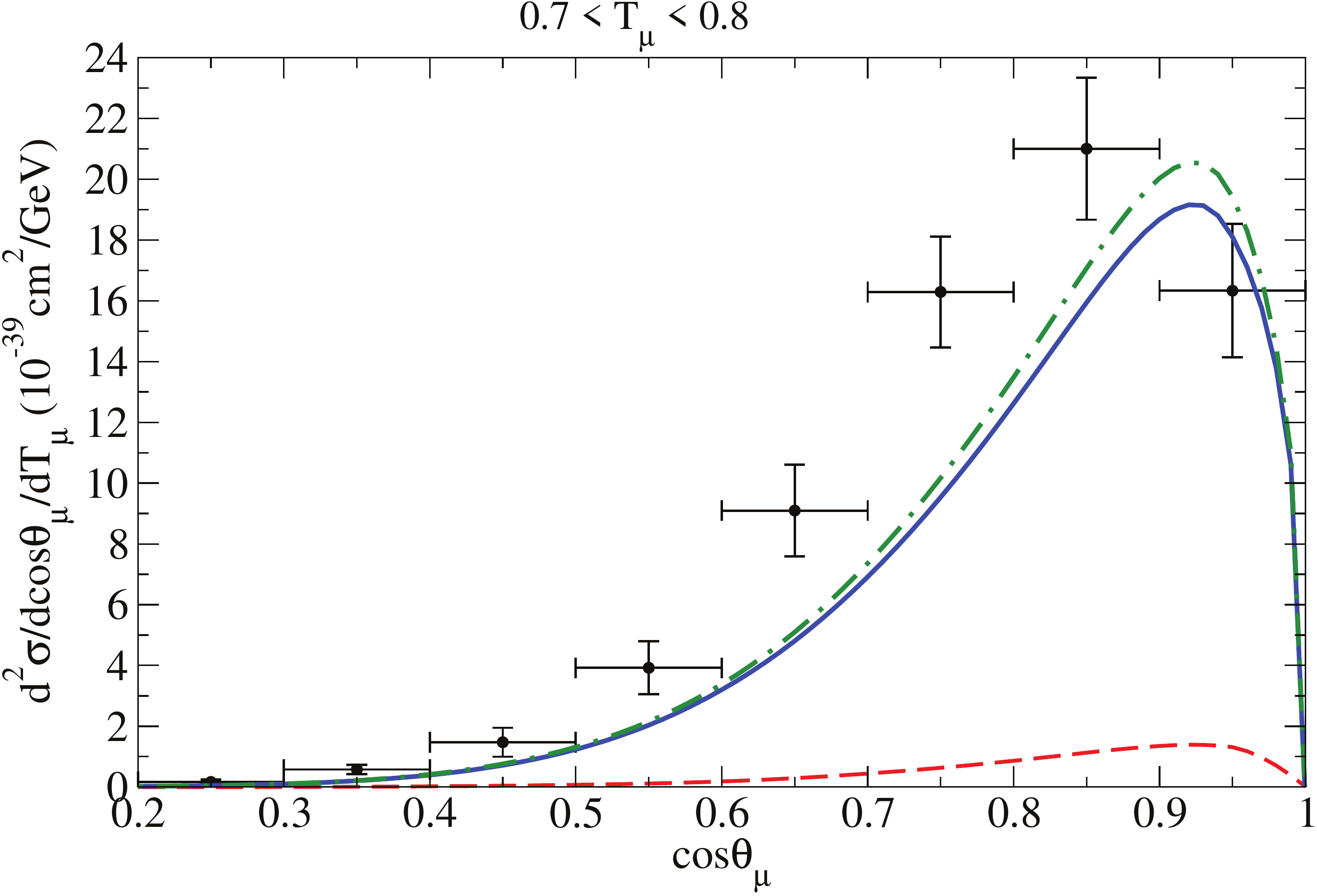}
\includegraphics[scale=0.21, bb=-20 0 784 557, clip]{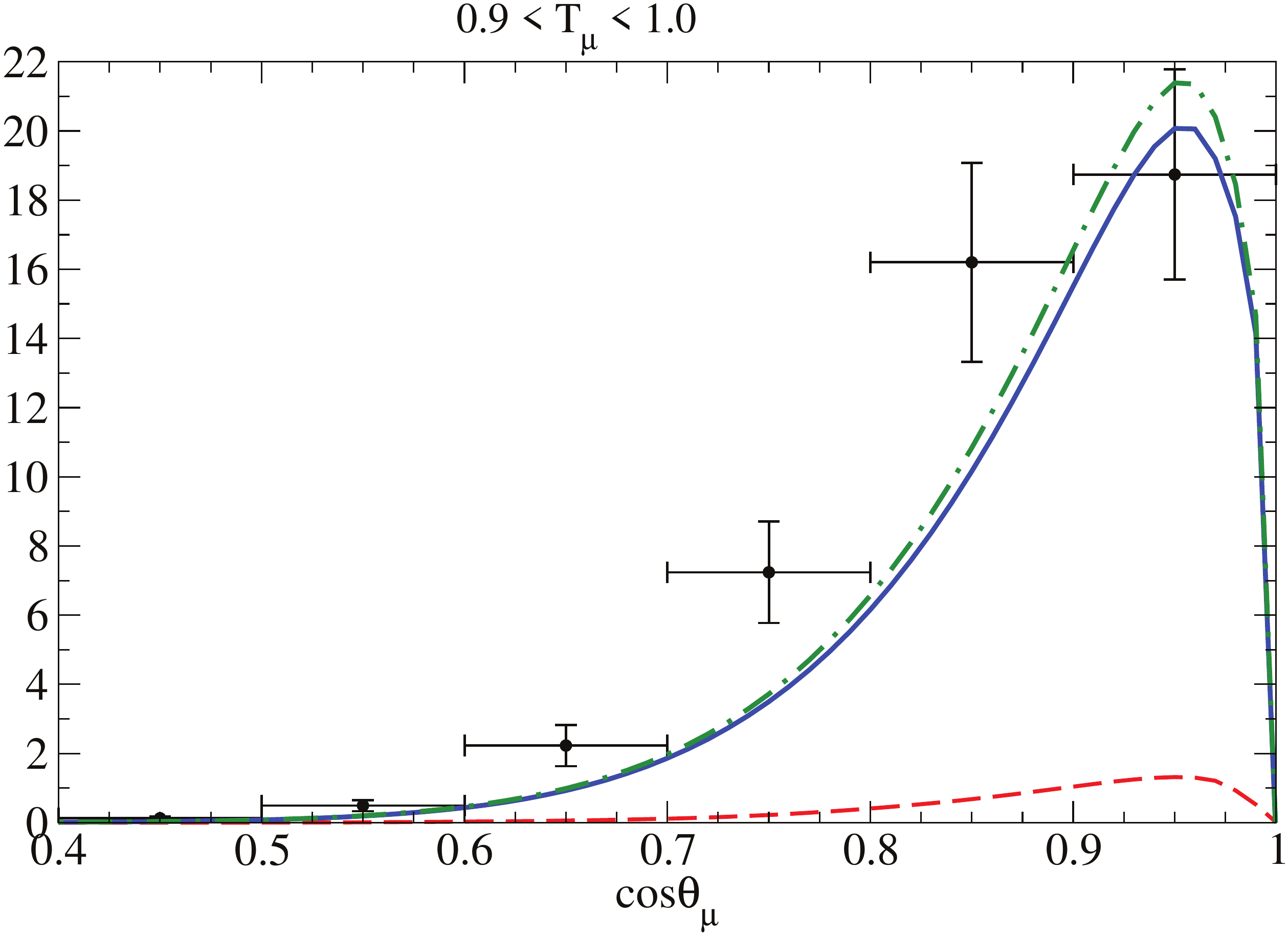} \hspace{-0.25cm}
\includegraphics[scale=0.21, bb=0 0 784 557, clip]{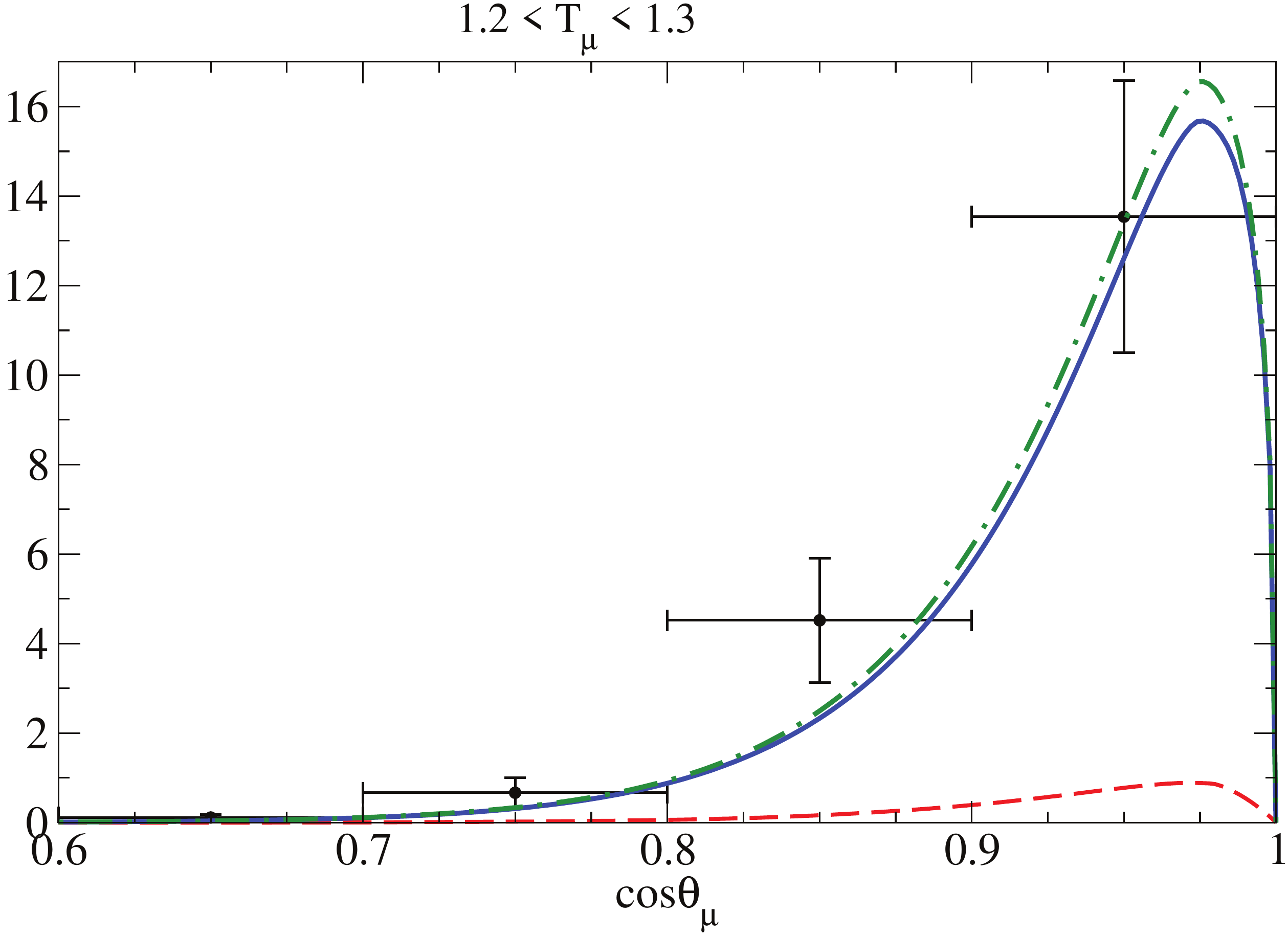}
\caption{(Color online) Flux-integrated double-differential cross section per target nucleon for the $\nu_\mu$ CCQE process on $^{12}$C displayed versus $\cos\theta_\mu$ for various bins of $T_\mu$ obtained within the SuSAv2 and SuSAv2+MEC approaches. MEC results are also shown. The data are from Ref. \cite{MiniBooNECC10}. \label{fig6} }
\end{center}
\end{figure}

\begin{figure}[htbp]
\begin{center}
\includegraphics[scale=0.21, bb=0 0 784 557, clip]{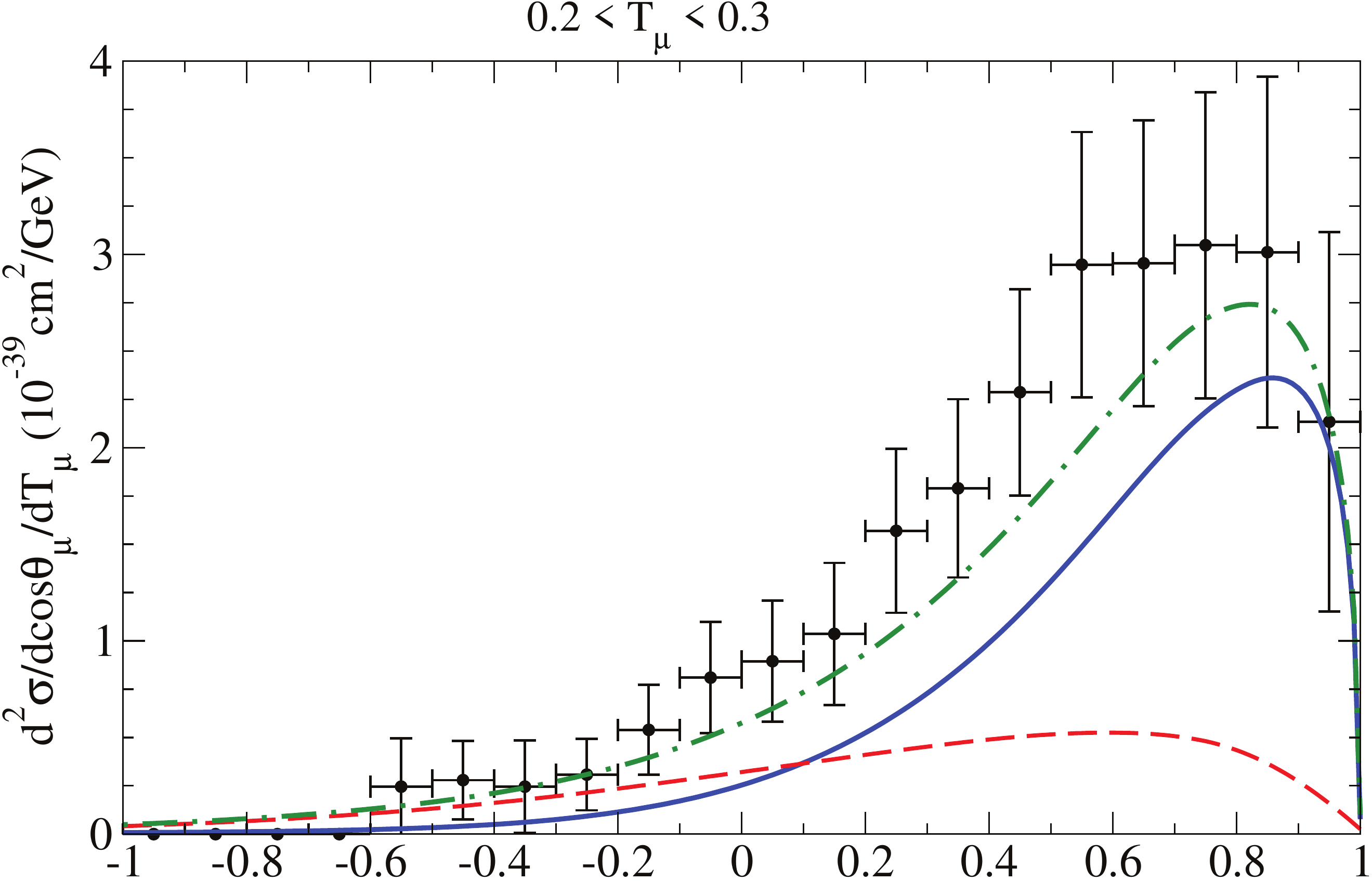}
\includegraphics[scale=0.21, bb=-20 0 784 557, clip]{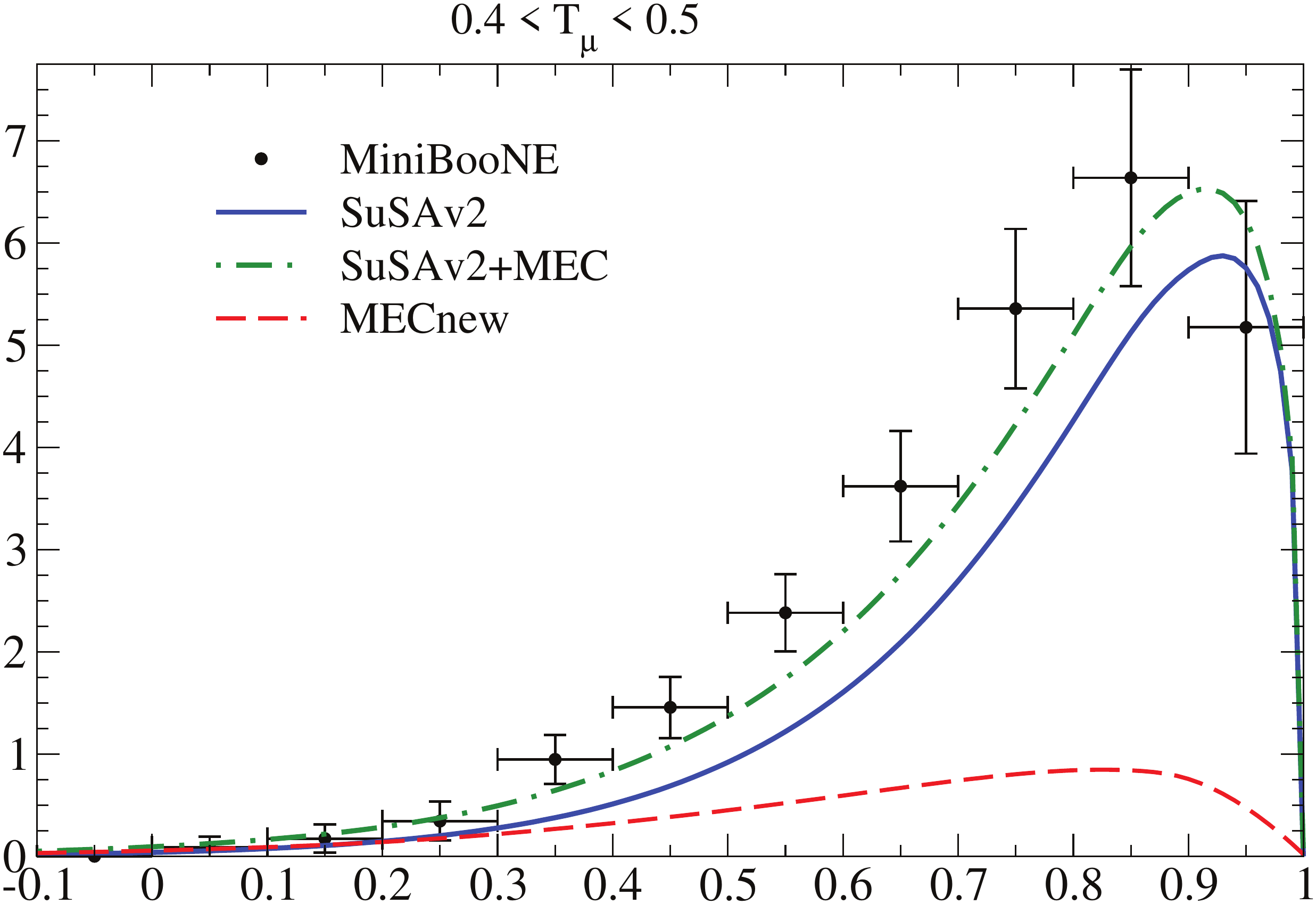} \hspace{-0.25cm}
\includegraphics[scale=0.21, bb=0 0 784 557, clip]{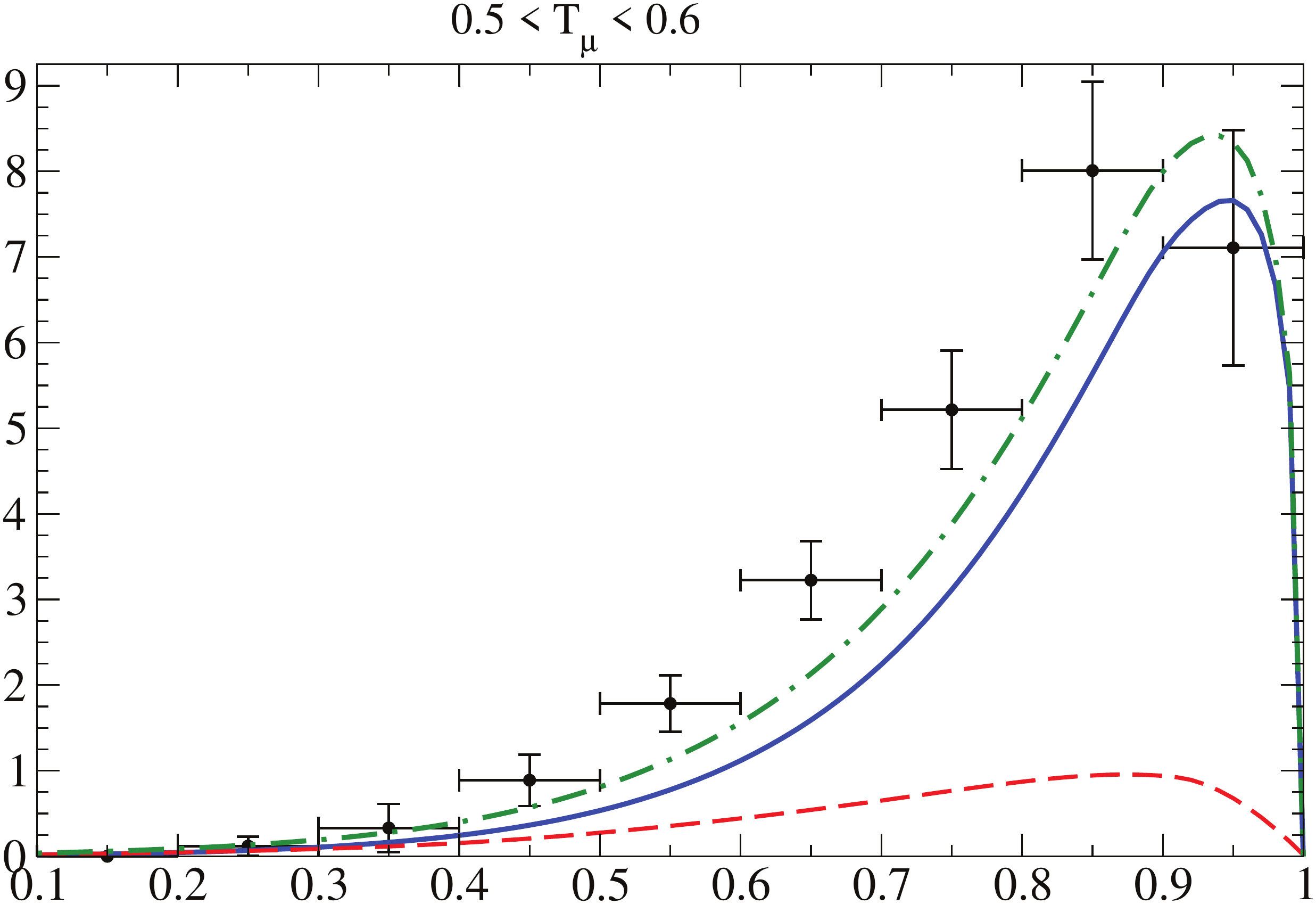} \\
\includegraphics[scale=0.21, bb=0 0 784 557, clip]{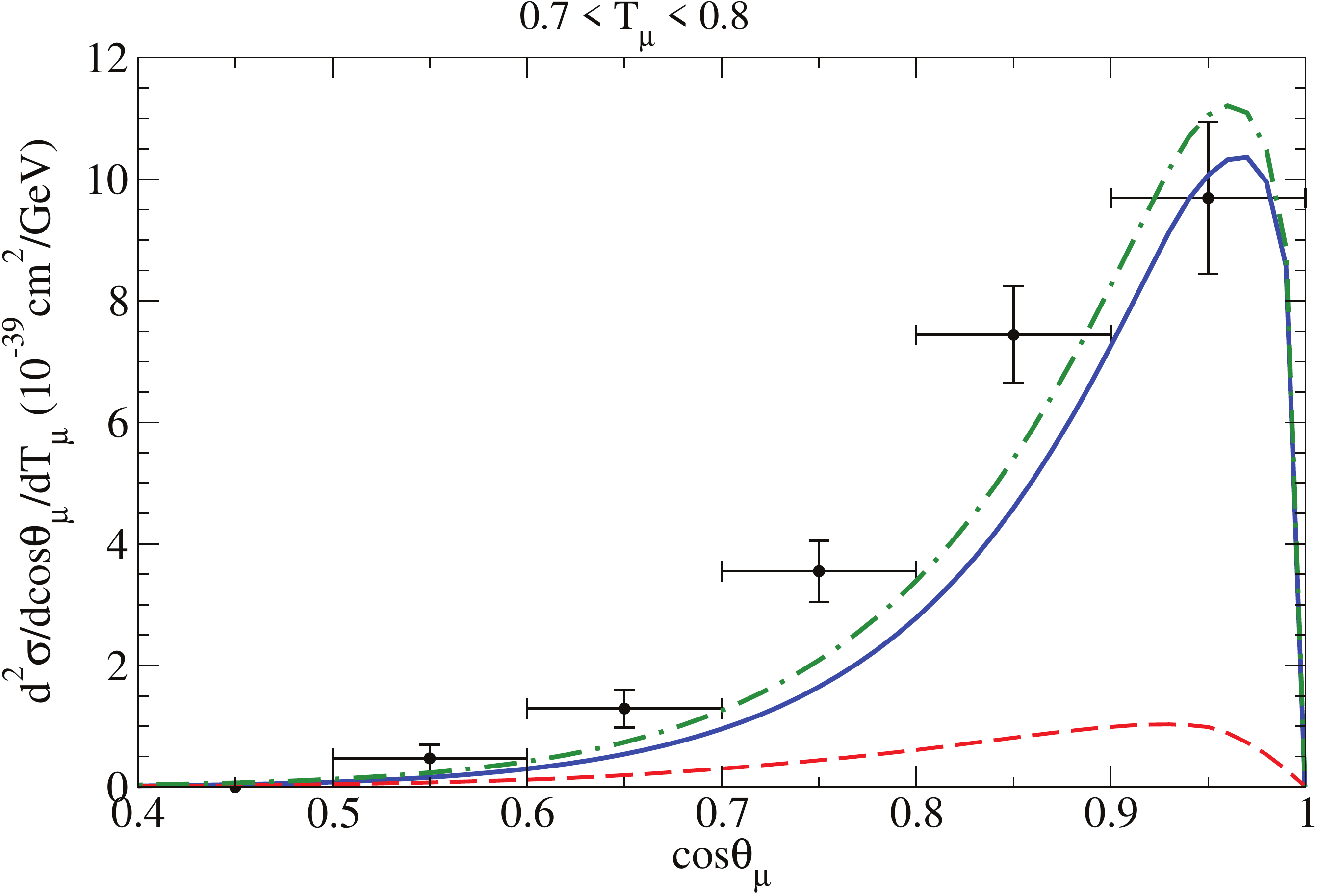}
\includegraphics[scale=0.21, bb=-20 0 784 557, clip]{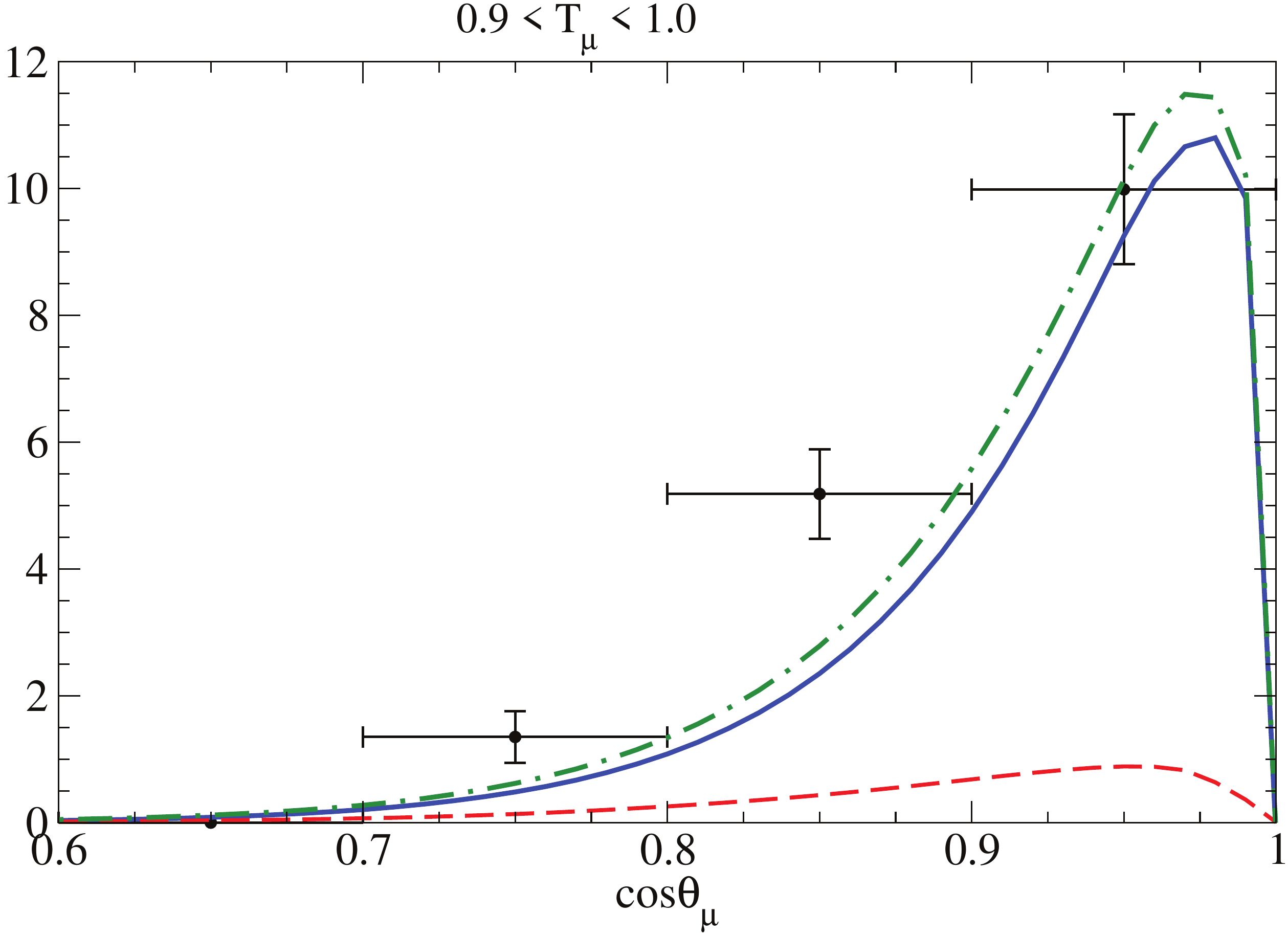} \hspace{-0.25cm}
\includegraphics[scale=0.21, bb=0 0 784 557, clip]{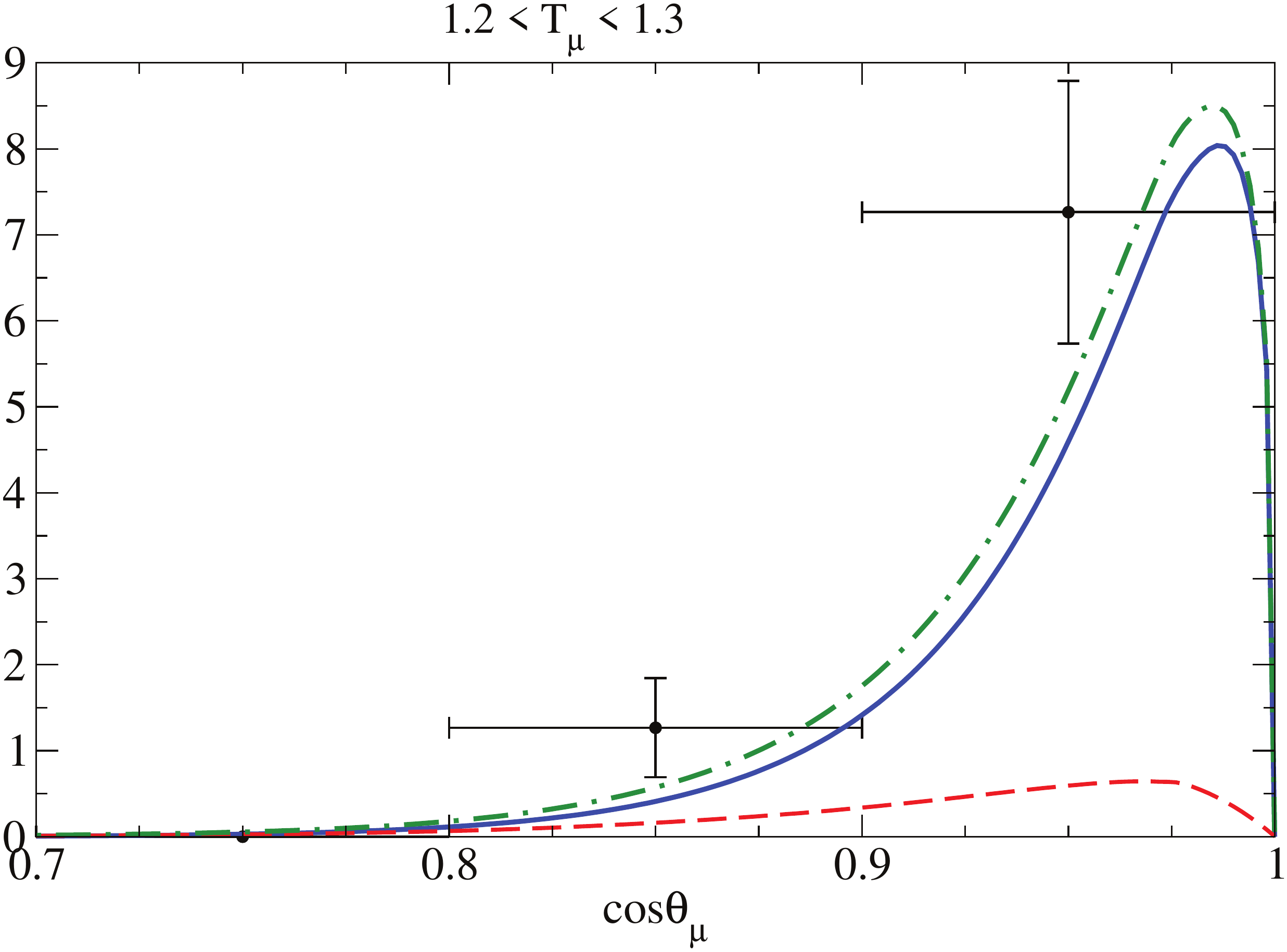}
\caption{(Color online) As for Fig. \ref{fig6}, but for $\bar\nu_\mu$ scattering versus $\cos\theta_\mu$. The data are from Ref. \cite{MiniBooNECC13}.  \label{fig7}}
\end{center}
\end{figure}

As can be seen in Figs. \ref{fig4} and \ref{fig5}, the inclusion of MEC results in an increase of the cross sections, yielding reasonable
agreement with the MiniBooNE data for low angles, up to $\cos\theta_\mu\simeq 0.7$. At larger scattering angles the disagreement with the experiment 
becomes more significant, and the vector-vector transverse MEC do not seem to be sufficient to account for the discrepancy. The same conclusion can 
be drawn by plotting the cross section versus the scattering angle (see Figs. \ref{fig6} and \ref{fig7}) at fixed muon momentum; the inclusion of 
MEC improves the agreement with the data at low scattering angles, but some strength is missing at higher angles, especially for low muon momenta, as observed in \cite{Ivanov14}.

The size of the MEC contribution to the cross section reported here --- of the order of 10$\%$ --- 
corresponds to the average value found within our particular RFG model. Our results show that
processes involving MEC are responsible for a sizable enhancement of
the response in the transverse channel. The extent to which this
enhancement affects the cross section, however, strongly depends on the
kinematics (see discussion in previous section).

We remark that axial-axial and axial-vector transverse MEC responses, $R_{T,\:AA}^{MEC}$ and $R_{T',\:VA}^{MEC}$, are not considered in this work and could partially explain the discrepancy with the data. Furthermore, additional nuclear correlations could contribute to the 2p-2h excitations as the ones induced by MEC; however, since the longitudinal vector contributions come directly from experimental data and hence have all the correlations built in, such contributions would need to break zeroth-kind scaling which has not been demonstrated. Note that extended RFG or RMF models with 2p-2h, as well as 1p-1h, correlations are actually required to preserve gauge invariance, but their inclusion would call for consistent treatments to avoid double-counting.

\begin{figure}[htbp]
\begin{center}
\includegraphics[scale=0.31, bb=0 11 794 529]{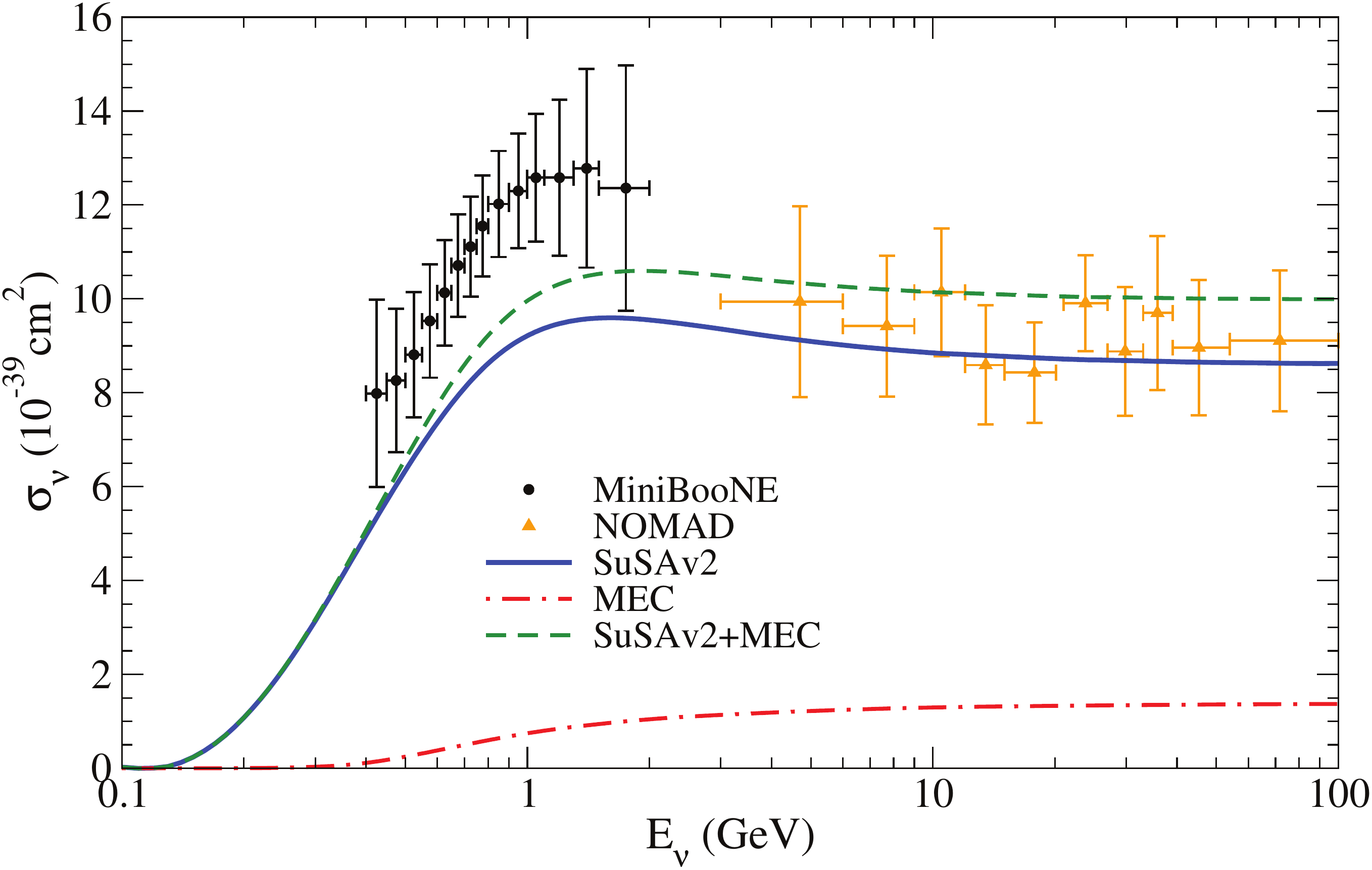} \\
\vspace{0.2cm}
\includegraphics[scale=0.31, bb=0 11 794 529]{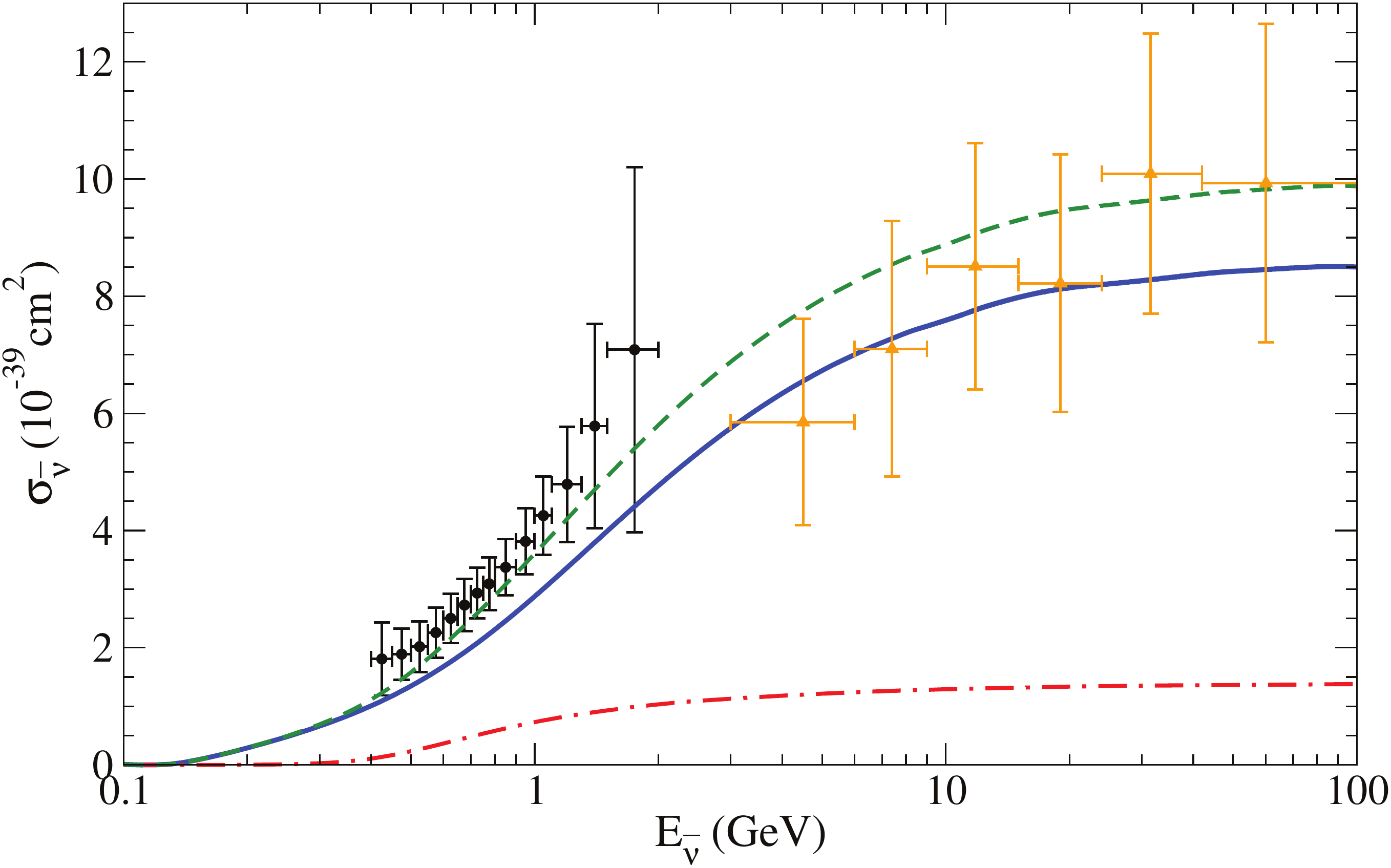}
\caption{(Color online) CCQE $\nu_\mu$ cross section per nucleon displayed versus neutrino energy $E_\nu$ and evaluated using the SuSAv2 and the SuSAv2+MEC approaches (top panel). CCQE $\bar\nu_\mu$ cross section is also shown (bottom panel). Results are compared with the MiniBooNE \cite{MiniBooNECC10, MiniBooNECC13} and NOMAD \cite{NOMAD09} experimental data. Also presented for reference are the results for the MEC contributions.  \label{fig9}}
\end{center}
\end{figure}

\begin{figure}[htbp]
\begin{center}
\includegraphics[scale=0.31, bb=0 11 794 529]{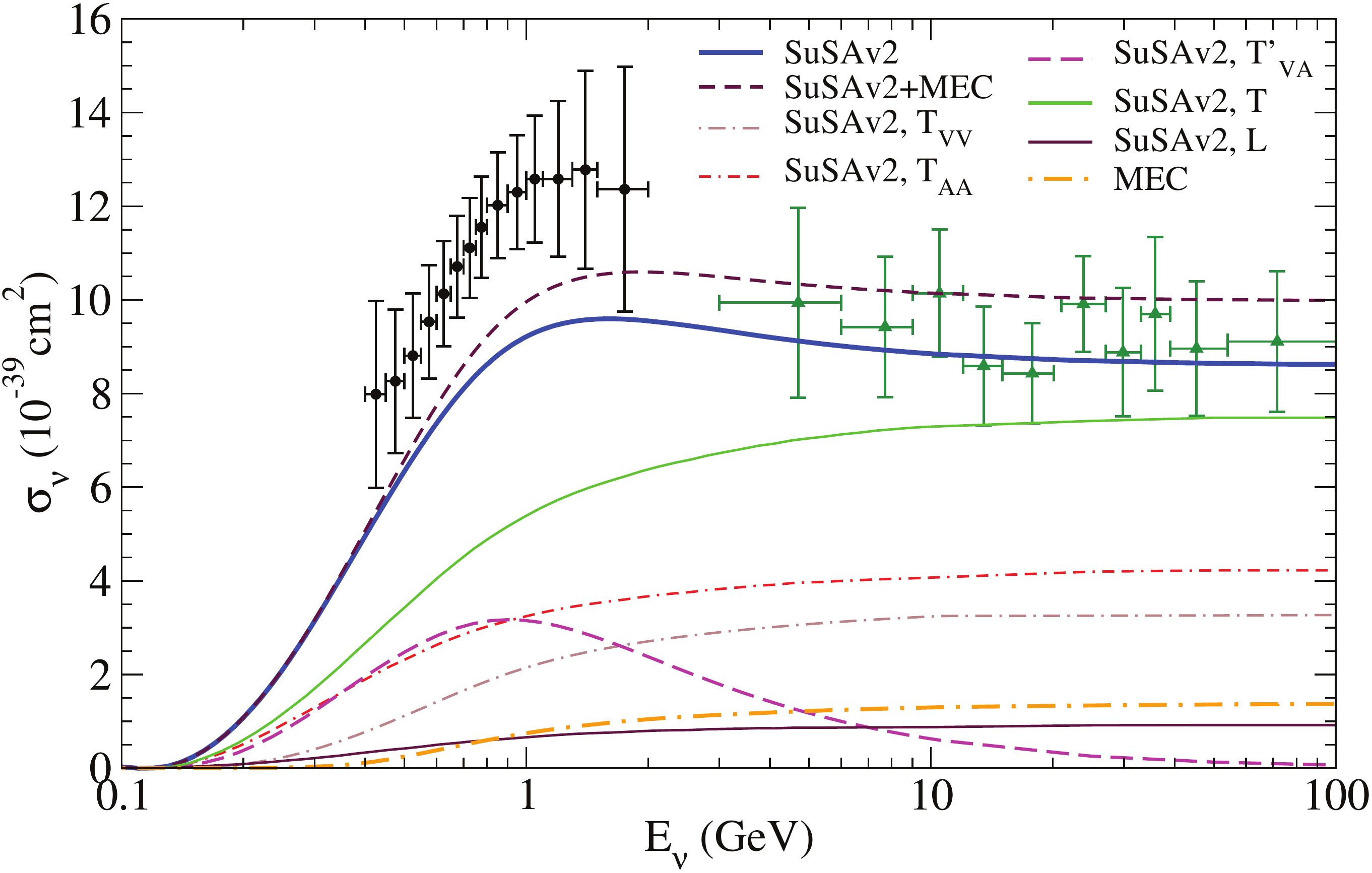}
\caption{(Color online) Separation into components of the CCQE $\nu_\mu$ cross section per nucleon displayed versus neutrino energy $E_\nu$ within the SuSAv2 approach. The MEC and SuSAv2+MEC curves are shown. The MiniBooNE \cite{MiniBooNECC10} and NOMAD \cite{NOMAD09} data are also shown for reference. \label{fig13} }
\end{center}
\end{figure}

\begin{figure}[htbp]
\begin{center}
\includegraphics[scale=0.31, bb=0 11 794 529]{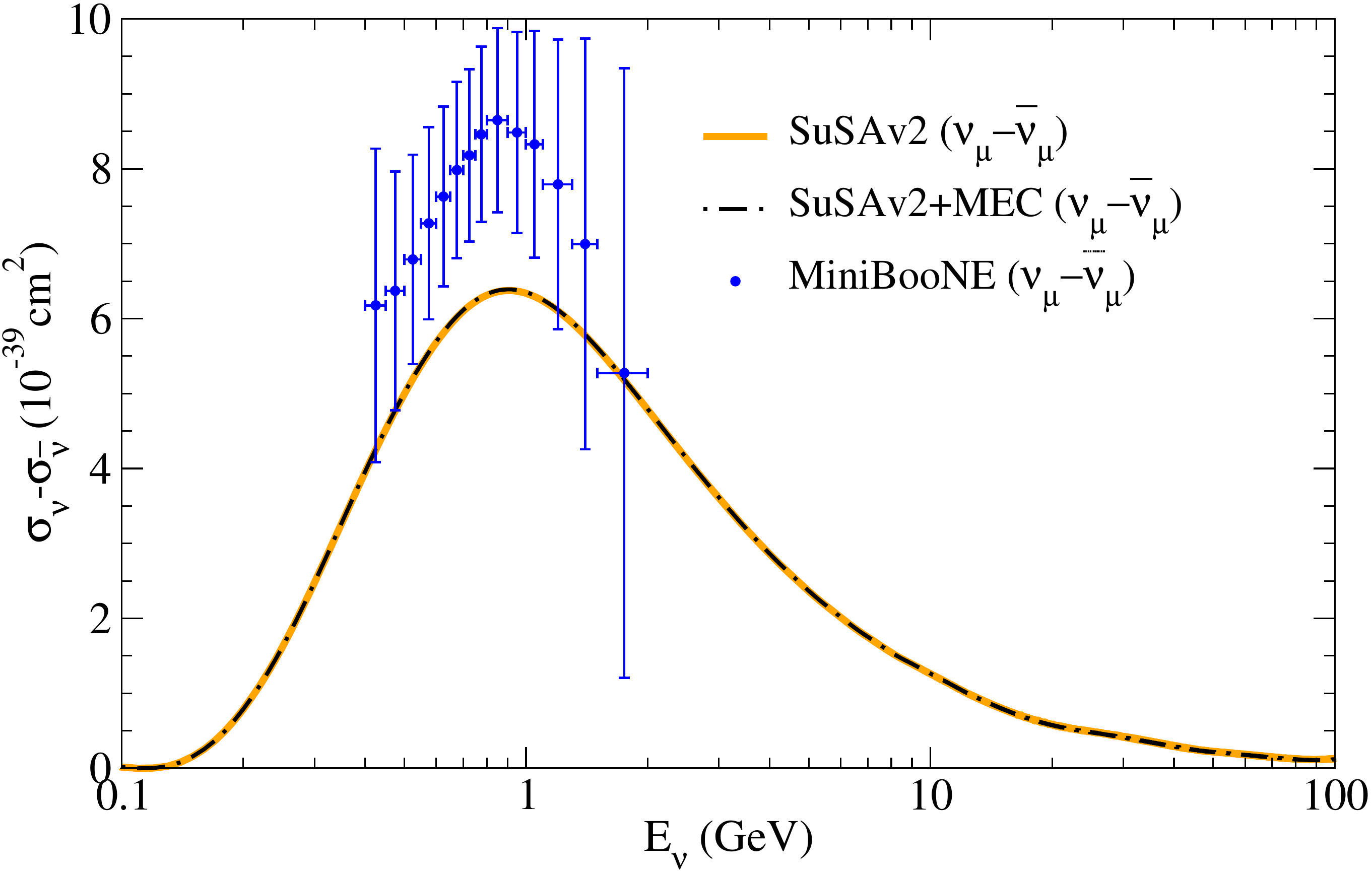}
\caption{(Color online) Experimental difference between neutrino and antineutrino cross sections ($\sigma_{\nu_\mu}-\sigma_{\bar\nu_\mu}$) from MiniBooNE, together with the corresponding theoretical
prediction from SuSAv2+MEC, whose difference with the SuSAv2 prediction is negligible.  \label{fig10}}
\end{center}
\end{figure}

When comparing our theoretical results with the MiniBooNE data one can observe a better agreement for antineutrinos than for neutrinos (see Fig. \ref{fig9}). This is due to the fact that, in the neutrino case, the two missing MEC responses in our calculation are constructively combined, $R_{T,\:AA}^{MEC}$ + $R_{T',\:VA}^{MEC}$, whereas they are destructively combined in the antineutrino case, $R_{T,\:AA}^{MEC}$ - $R_{T',\:VA}^{MEC}$. In other words, we expect a larger strength missing in our calculation in the neutrino case than in the antineutrino case, whose origin possibly can be attributed to the missing MEC pieces. Furthermore, one can see in the total neutrino cross section (Fig. \ref{fig9}) that some strength is missing at intermediate energies, 0.4-1.5 GeV, which is the region where the VA QE component is peaked (Fig. \ref{fig13}); an extra contribution in this channel via 2p-2h MEC would thus improve the agreement with MiniBooNE data.
We can observe in Fig. \ref{fig13} that below 1 GeV the SuSAv2 VA response is higher than the VV one and of the same order as the AA one. Other contributions to the VA response, apart from the QE one (SuSAv2), can be estimated as follows
\begin{equation}
(\sigma_{\nu_\mu})_{T',\:VA}^{other} \simeq \frac{\left( \sigma_{\nu_\mu} - \sigma_{\bar\nu_\mu} \right)_{exp}}{2} - \frac{\left( \sigma_{\nu_\mu} - \sigma_{\bar{\nu}_\mu} \right)_{SuSAv2}}{2} \:,
\end{equation}
as long as one assumes no quenching of the axial current within the nuclear medium with respect to the vector current, as is the case in the superscaling approach. If one considers $(\sigma_{\nu_\mu})_{T',\:VA}^{other}$ as mainly due to MEC, it is found that a VA MEC response as large as the computed VV MEC response would be needed to reproduce the data.
In Fig. \ref{fig10} we show the experimental difference between neutrino and antineutrino cross sections $(\sigma_{\nu_\mu}-\sigma_{\bar\nu_\mu})_{exp}$ from MiniBooNE, together with the corresponding theoretical prediction from SuSAv2, which is approximately equal to $2\:(\sigma_{\nu_\mu})_{T'\:VA}^{SuSAv2}$. The theoretical result from SuSAv2
with VV MEC contributions is also shown in the figure, but
is almost indistinguishable from the SuSAv2 result due
to the VV character of the MEC used.
Apart from the opposite sign in the $VA$ response, some minor differences between neutrino and antineutrino cross sections arise from the different Coulomb distortions of the emitted lepton \cite{Amaro05a} and the final nuclei involved in the CC neutrino (Nitrogen) and antineutrino (Boron) scattering processes.

\begin{figure}[htbp]
\begin{center}
\includegraphics[scale=0.279, bb=0 0 784 557, clip]{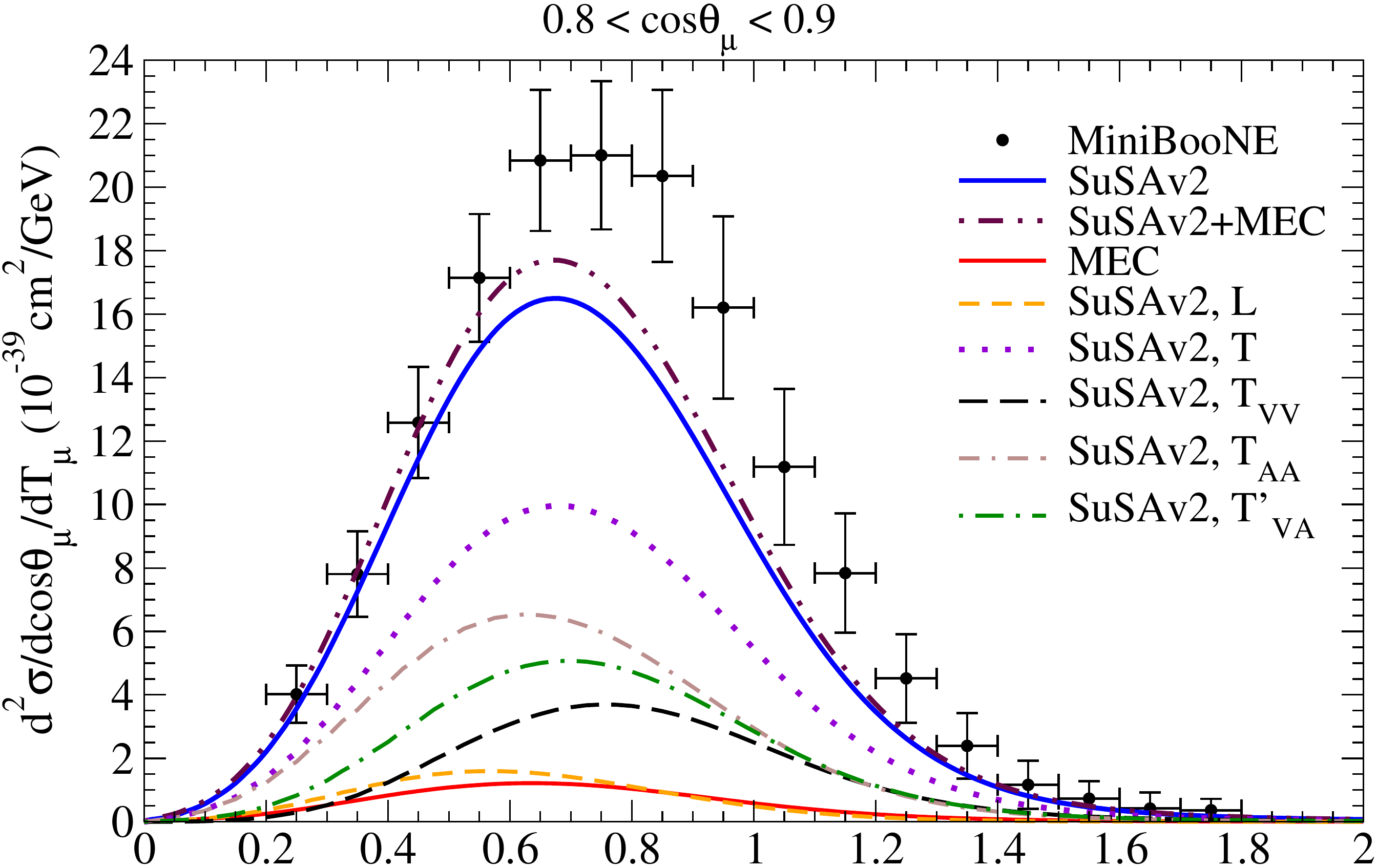}
\includegraphics[scale=0.279, bb=0 0 784 557, clip]{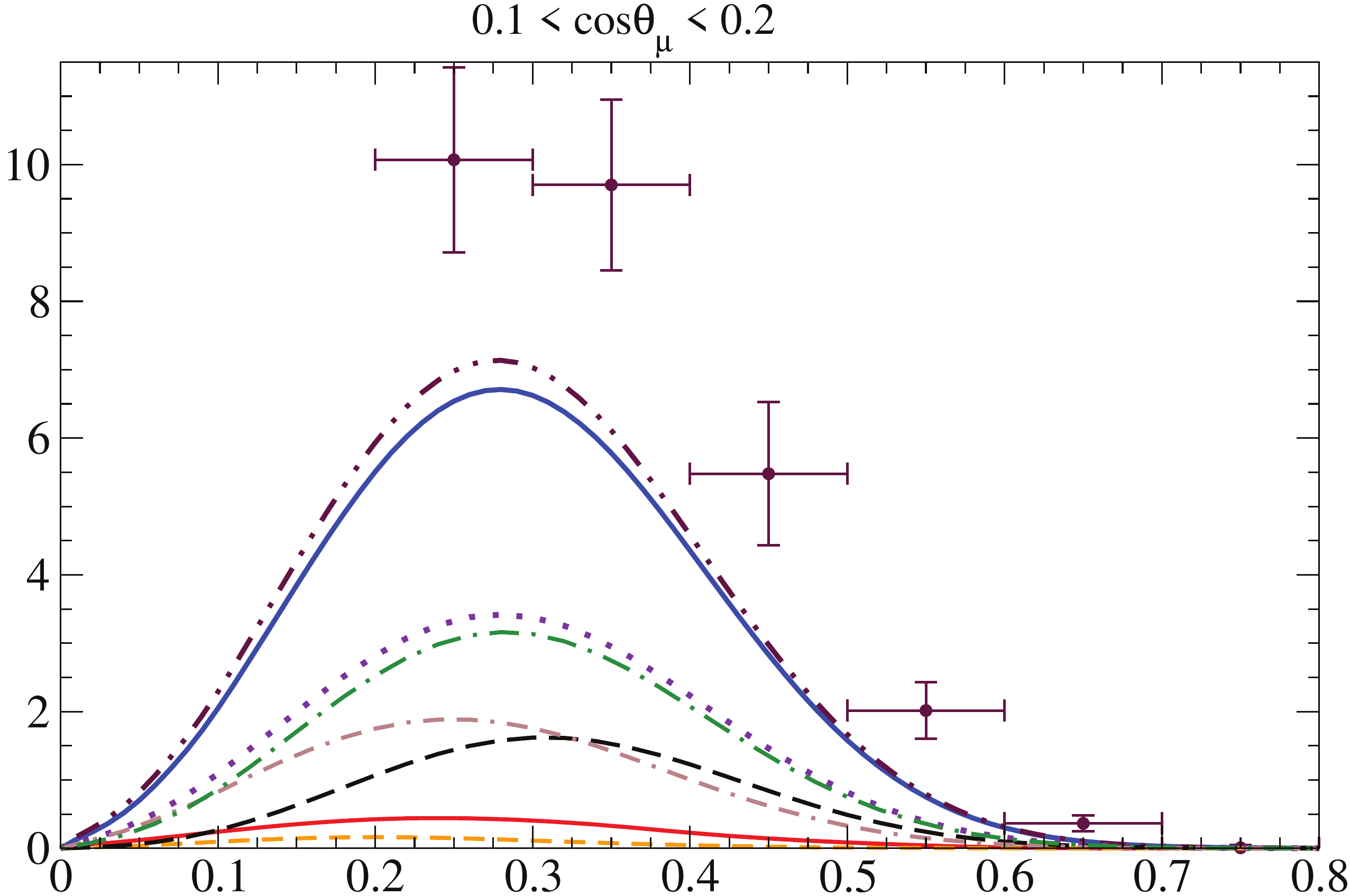} \\
\includegraphics[scale=0.279, bb=0 0 784 557, clip]{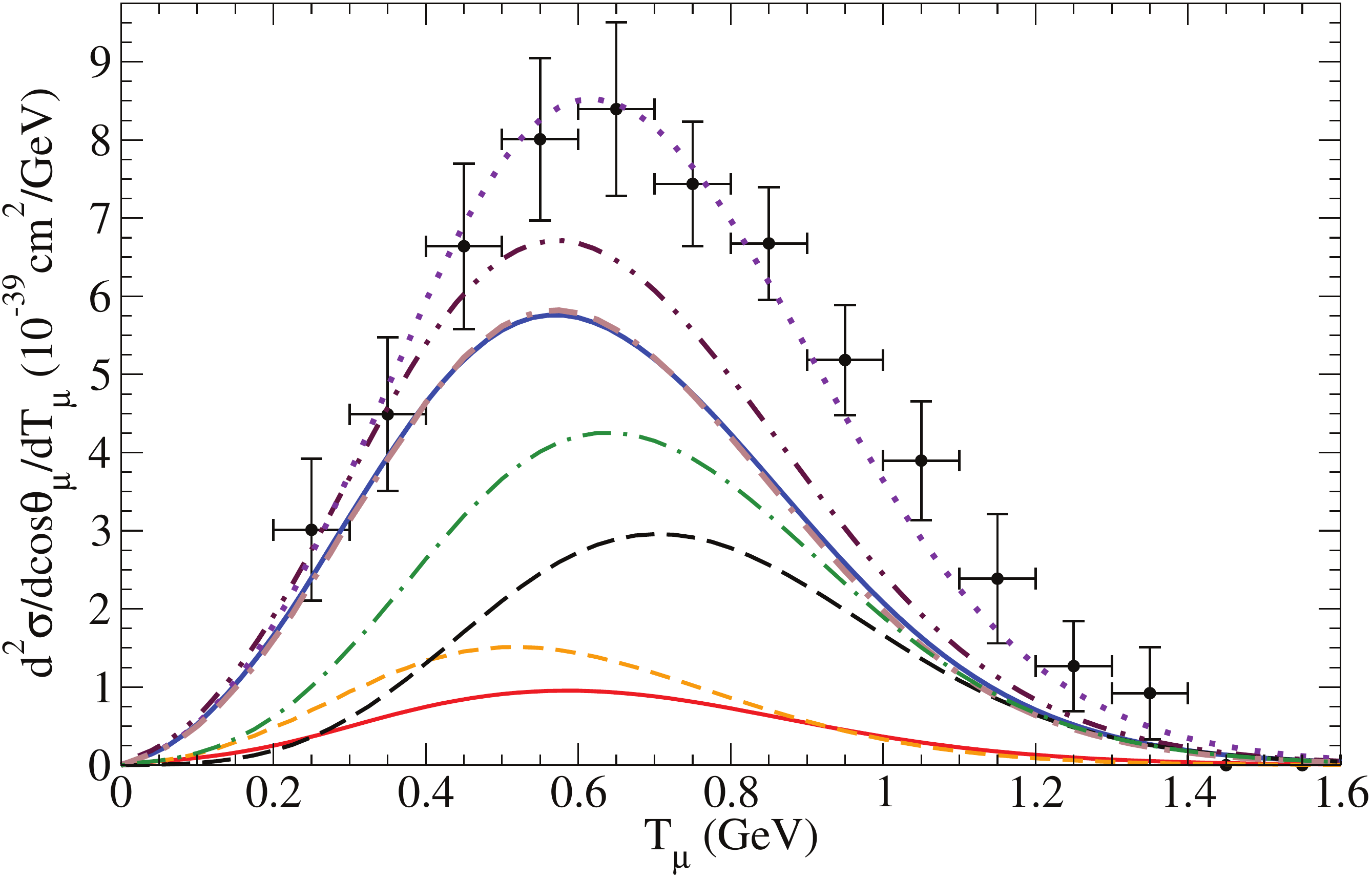}
\includegraphics[scale=0.279, bb=0 0 784 557, clip]{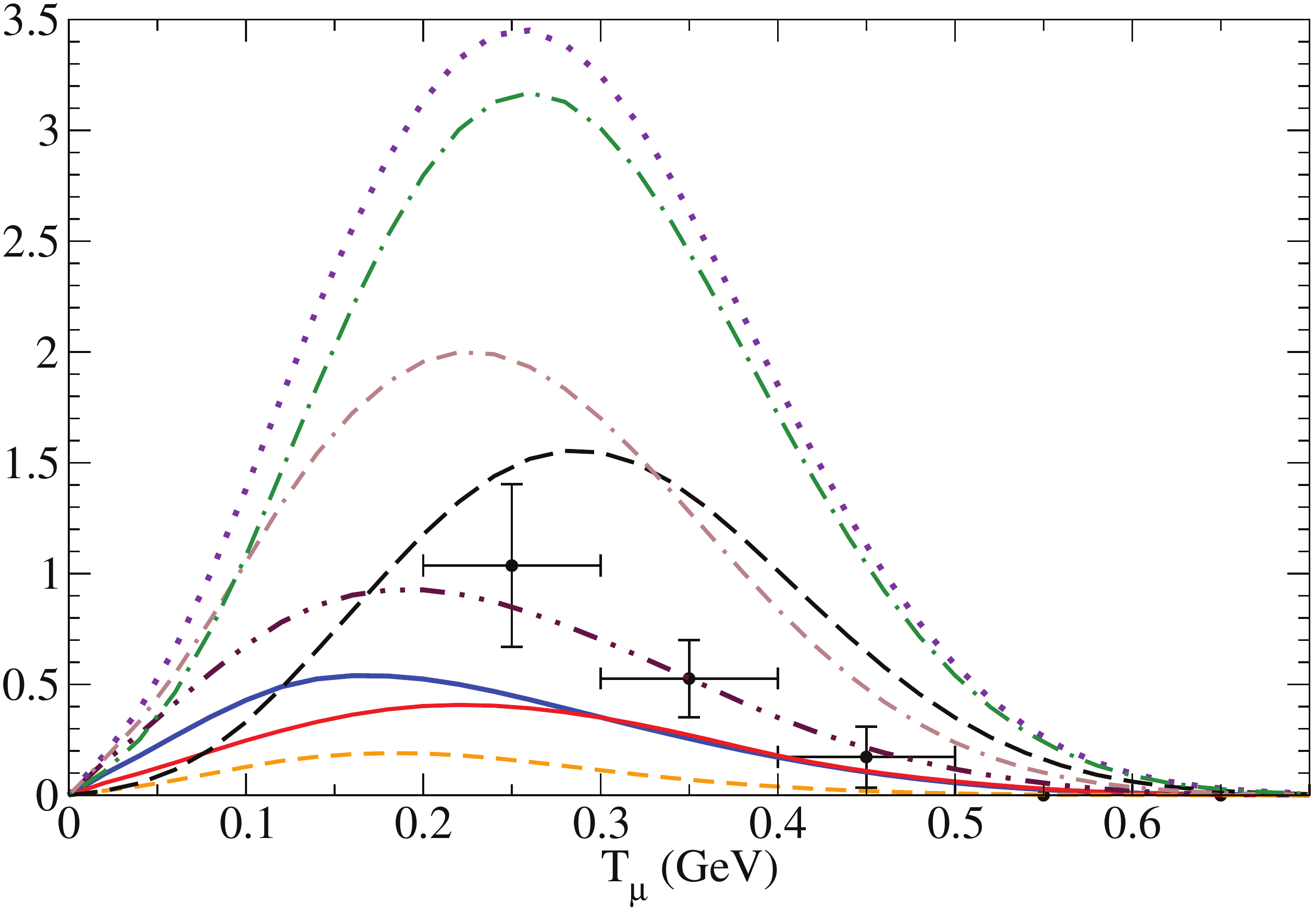}
\caption{(Color online) Separation into components of the MiniBooNE CCQE $\nu_\mu$ (top panel) and $\bar\nu_\mu$ (botoom panel) double-differential cross section per nucleon displayed versus $T_\mu$ for various bins of $\cos\theta_\mu$ within the SuSAv2 approach. The MEC and SuSAv2+MEC curves are shown. The MiniBooNE \cite{MiniBooNECC10, MiniBooNECC13} data are also shown for reference.  \label{fig14}}
\end{center}
\end{figure}

\begin{figure}[htbp]
\begin{center}
\includegraphics[scale=0.279, bb=0 0 784 557, clip]{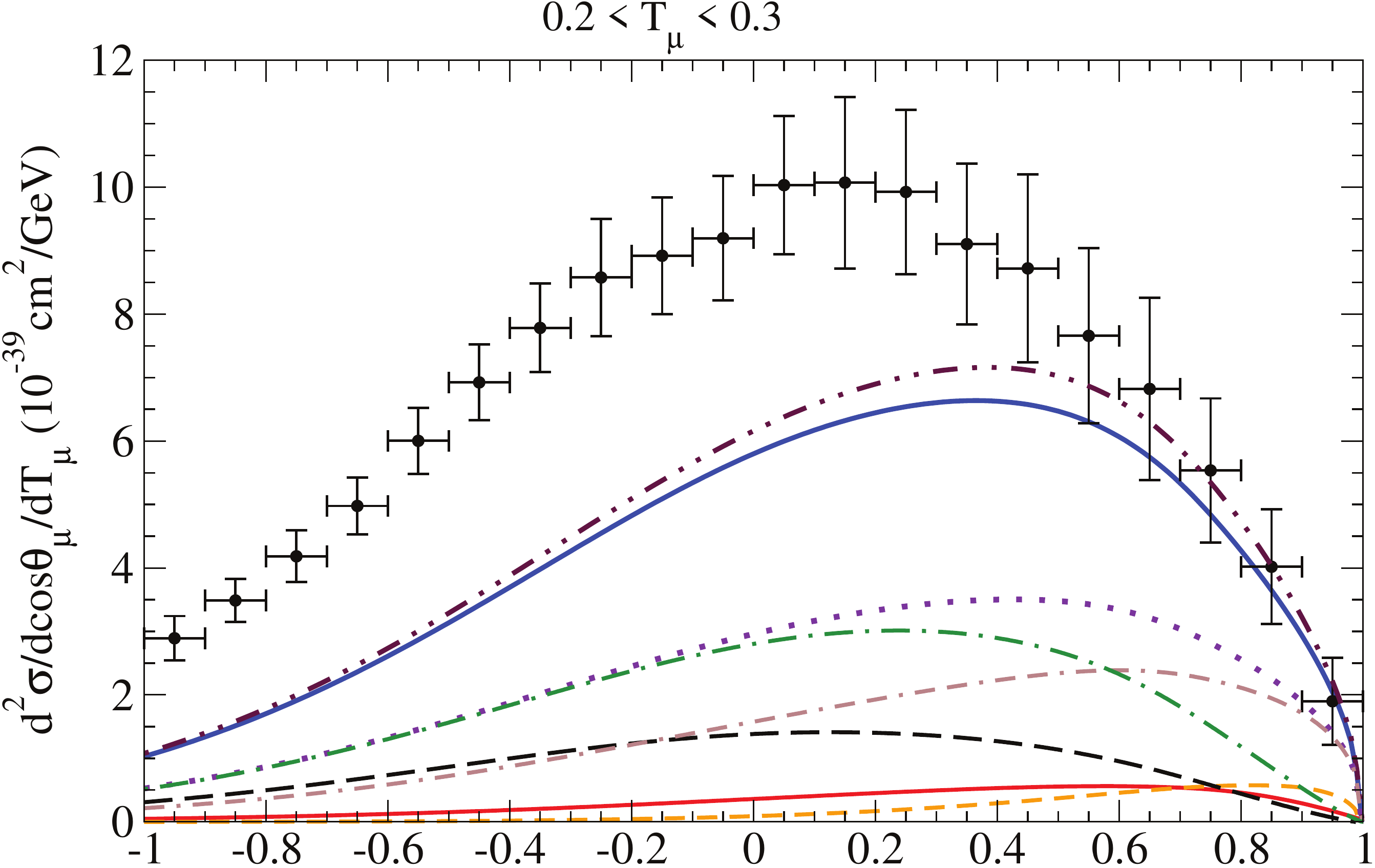}
\includegraphics[scale=0.279, bb=0 0 784 557, clip]{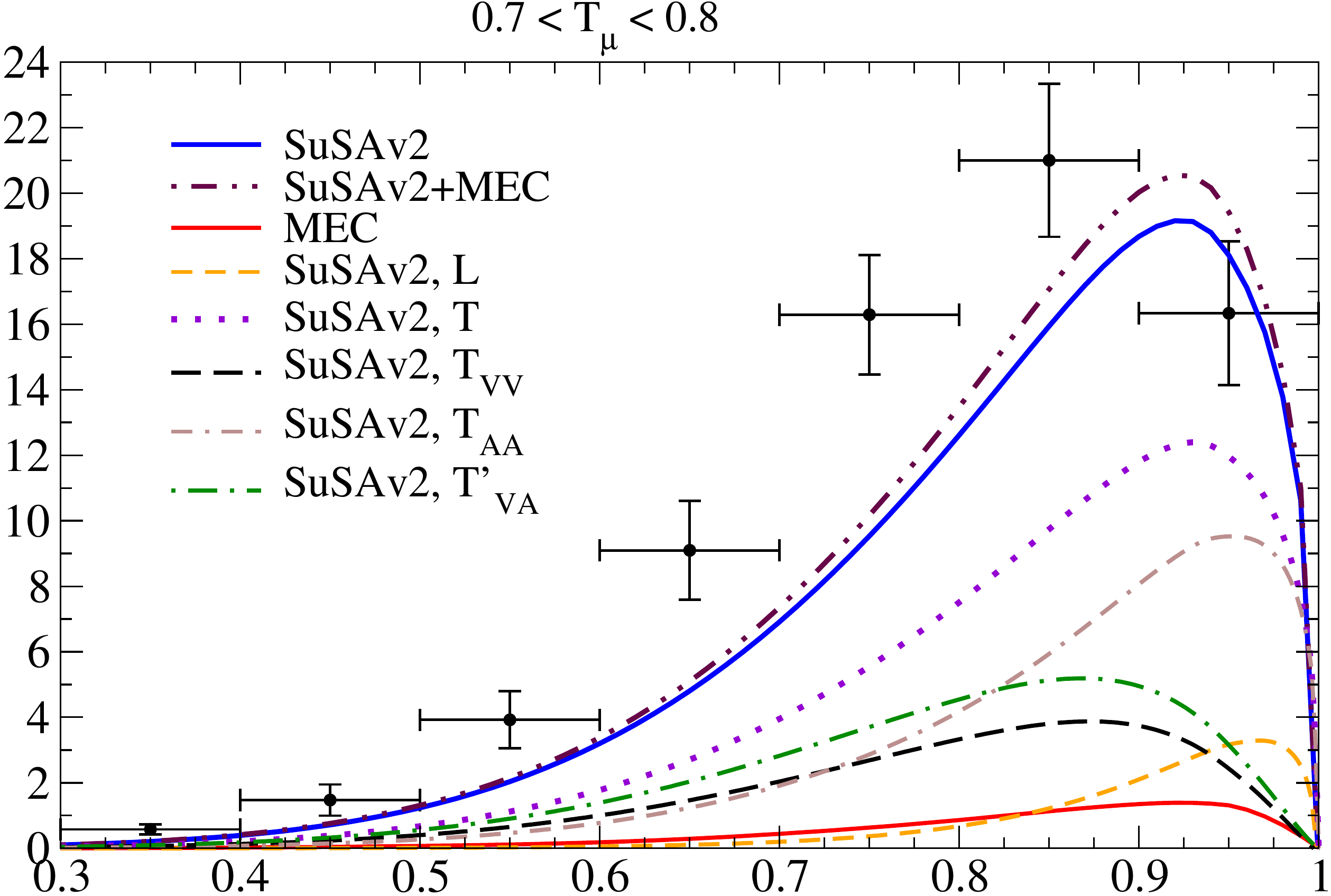} \\
\includegraphics[scale=0.279, bb=0 0 784 557, clip]{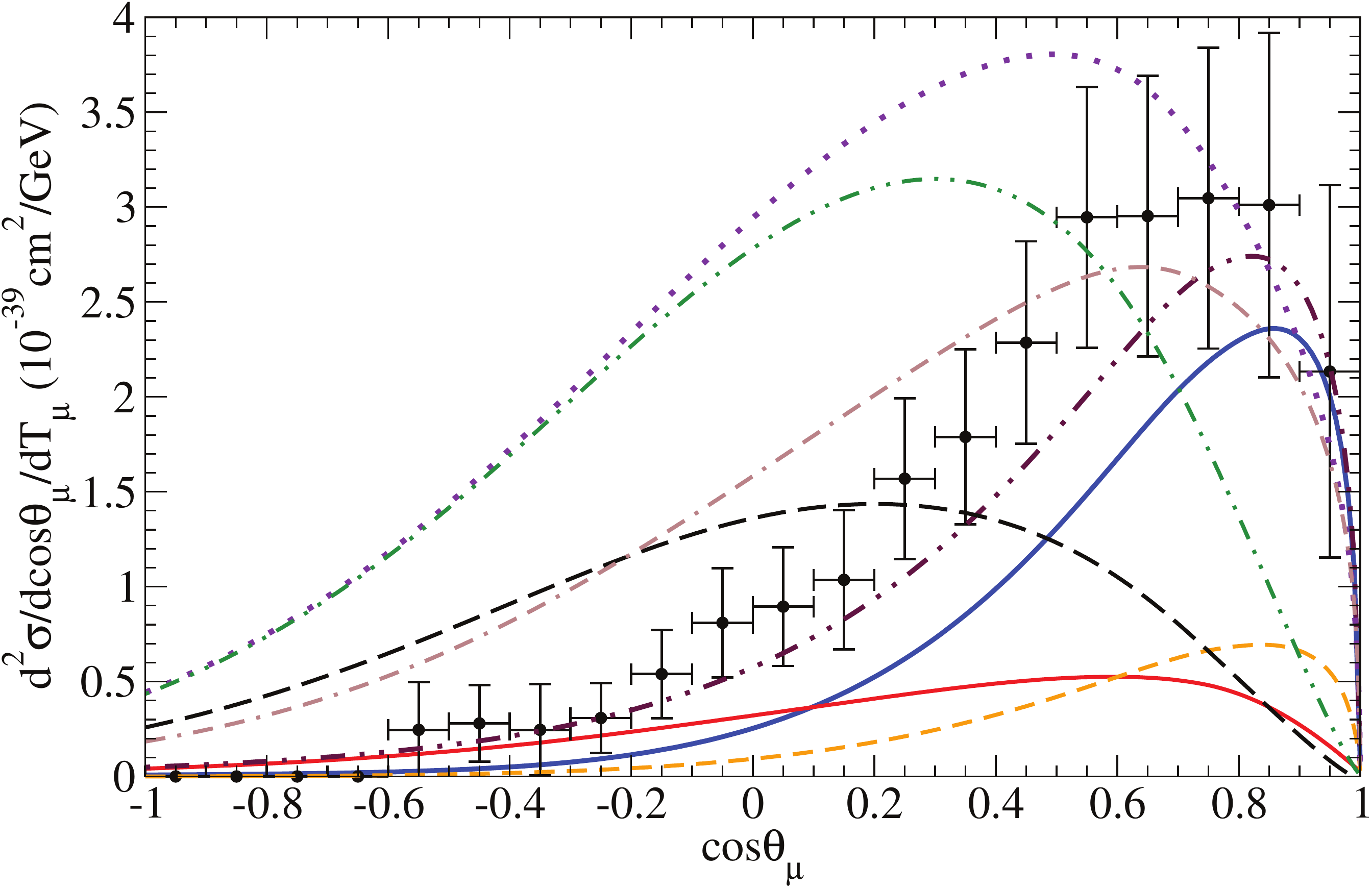} 
\includegraphics[scale=0.279, bb=0 0 784 557, clip]{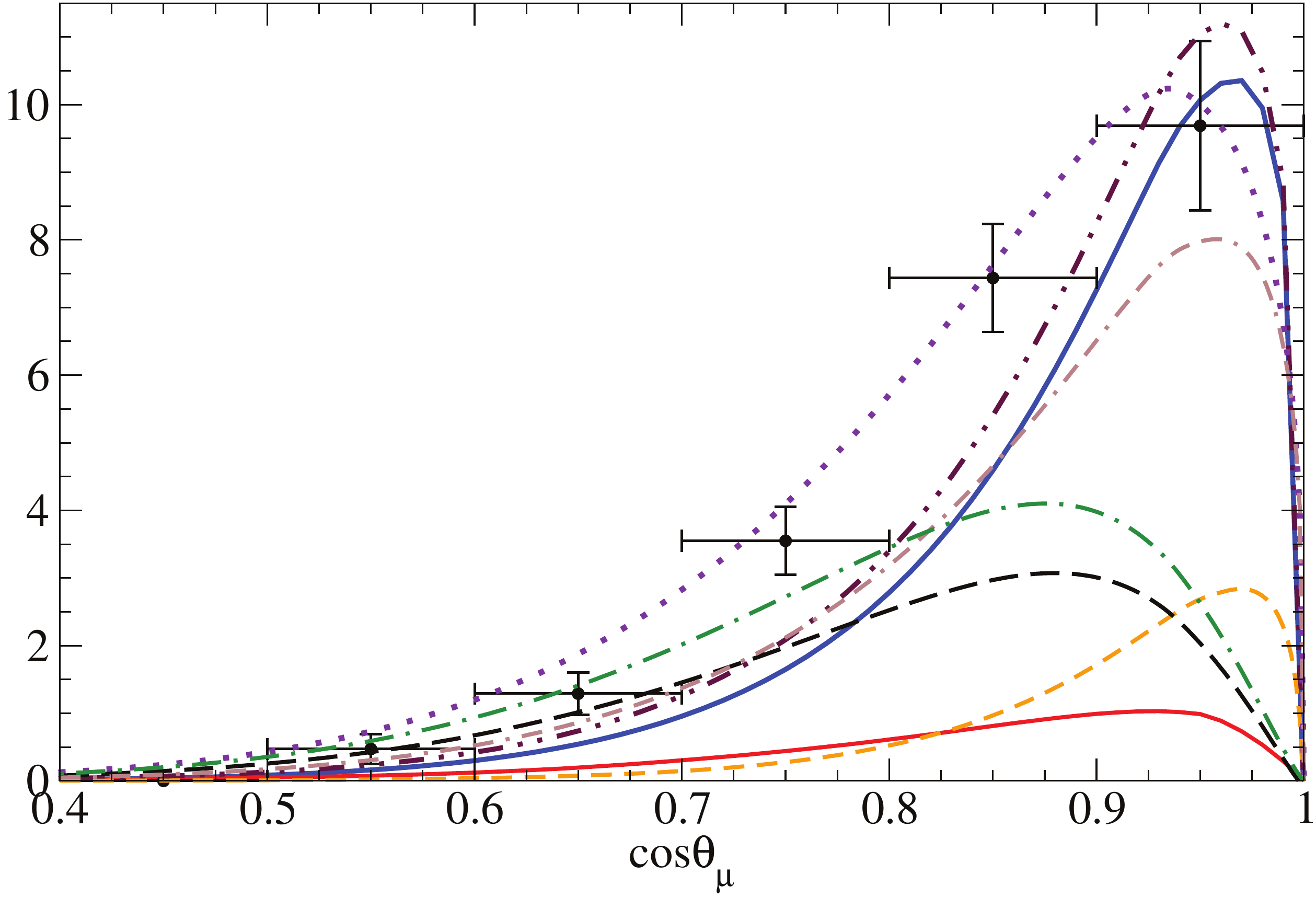}
\caption{(Color online) Separation into components of the MiniBooNE CCQE $\nu_\mu$ (top panel) and $\bar\nu_\mu$ (botoom panel) double-differential cross section per nucleon displayed versus $\cos\theta_\mu$ for various bins of $T_\mu$ within the SuSAv2 approach. The MEC and SuSAv2+MEC curves are shown. The MiniBooNE \cite{MiniBooNECC10, MiniBooNECC13} data are also shown for reference.  \label{fig15}}
\end{center}
\end{figure}

It can be seen that an extra contribution to the VA response from MEC would improve the agreement with the data for the difference between neutrino and antineutrino total cross sections of MiniBooNE, as was noted above for just the neutrino case. In the same way, one could deduce the suitability of extra AA and VA contributions via MEC in the double-differential MiniBooNE cross section by analyzing Figs. \ref{fig14} and \ref{fig15}. At NOMAD kinematics, Fig. \ref{fig9}, we observe a good agreement of the SuSAv2+MEC results, partly due to the negligible contribution of the VA response, whose MEC part is missing in our calculation, in such high-energy processes ($E_{\nu}$ between 5 and 100 GeV). From Fig. \ref{fig13} one sees that the VA interference becomes very small for $E_{\nu}>$ 5 GeV; this arises because the scattering at NOMAD kinematics is very forward-peaked and as $\theta_{\mu} \rightarrow 0$ the factor $v_{T'} \rightarrow 0$ (see Ref. \cite{Amaro05b}). This is also in agreement with some previous QE 
results \cite{Megias13}. 

\begin{figure}[htbp]
\begin{center}
\includegraphics[scale=0.31, bb=0 11 794 529]{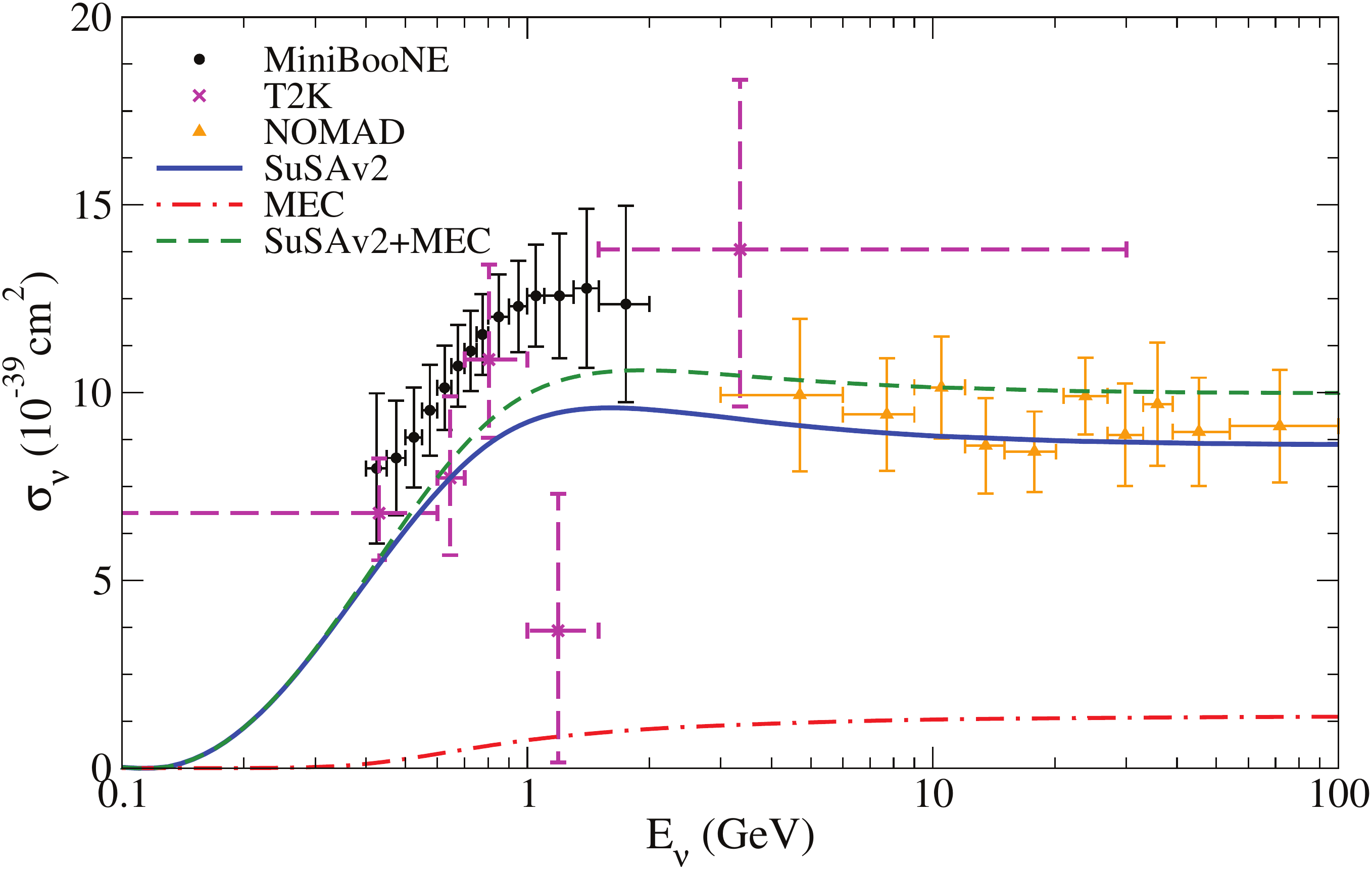}
\caption{CCQE $\nu_\mu$ cross section per nucleon displayed versus neutrino energy $E_\nu$ and evaluated using the SuSAv2 and the SuSAv2+MEC approaches. Results are compared with the MiniBooNE \cite{MiniBooNECC10, MiniBooNECC13}, NOMAD \cite{NOMAD09} and T2K \cite{T2KtotalQE} experimental data. Also presented for reference are the results for the MEC contributions. \label{figt2k}}
\end{center}
\end{figure}

While work is in progress to compute the weak responses with all the V and A contributions, we have found 
that assuming the transverse vector 2p-2h MEC scaling function, $f_{T,VV}^{MEC}$, 
to equal the axial-axial ($f_{T,AA}^{MEC}$) and vector-axial ($f_{T',VA}^{MEC}$) ones - as done for instance in \cite{Martini09} - 
a final result in 
agreement with MiniBooNE data is found. On the contrary, the calculation slightly overstimates NOMAD data. However, such results
cannot be fully justified until a proper 2p-2h MEC calculation for the axial-axial and vector-axial responses is completed. 
Moreover, one should take note of the different ways to analyze the QE-like events in MiniBooNE and NOMAD, where in the latter \cite{NOMAD09} the combination of 1-track and 2-track samples in the case of $\nu_\mu n\rightarrow\nu^-p$ can help to reduce 
some uncertainties as well as some contributions beyond the Impulse Approximation, such as from MEC or correlations that eject two nucleons. For completeness we also show in Fig. \ref{figt2k} recent results from the T2K Collaboration \cite{T2KtotalQE}. One should notice that, as they state, ``there is consistency between the experiments within the current statistical and systematic uncertainties.''

\begin{figure}[htbp]
\begin{center}
\includegraphics[scale=0.31, bb=0 11 794 529]{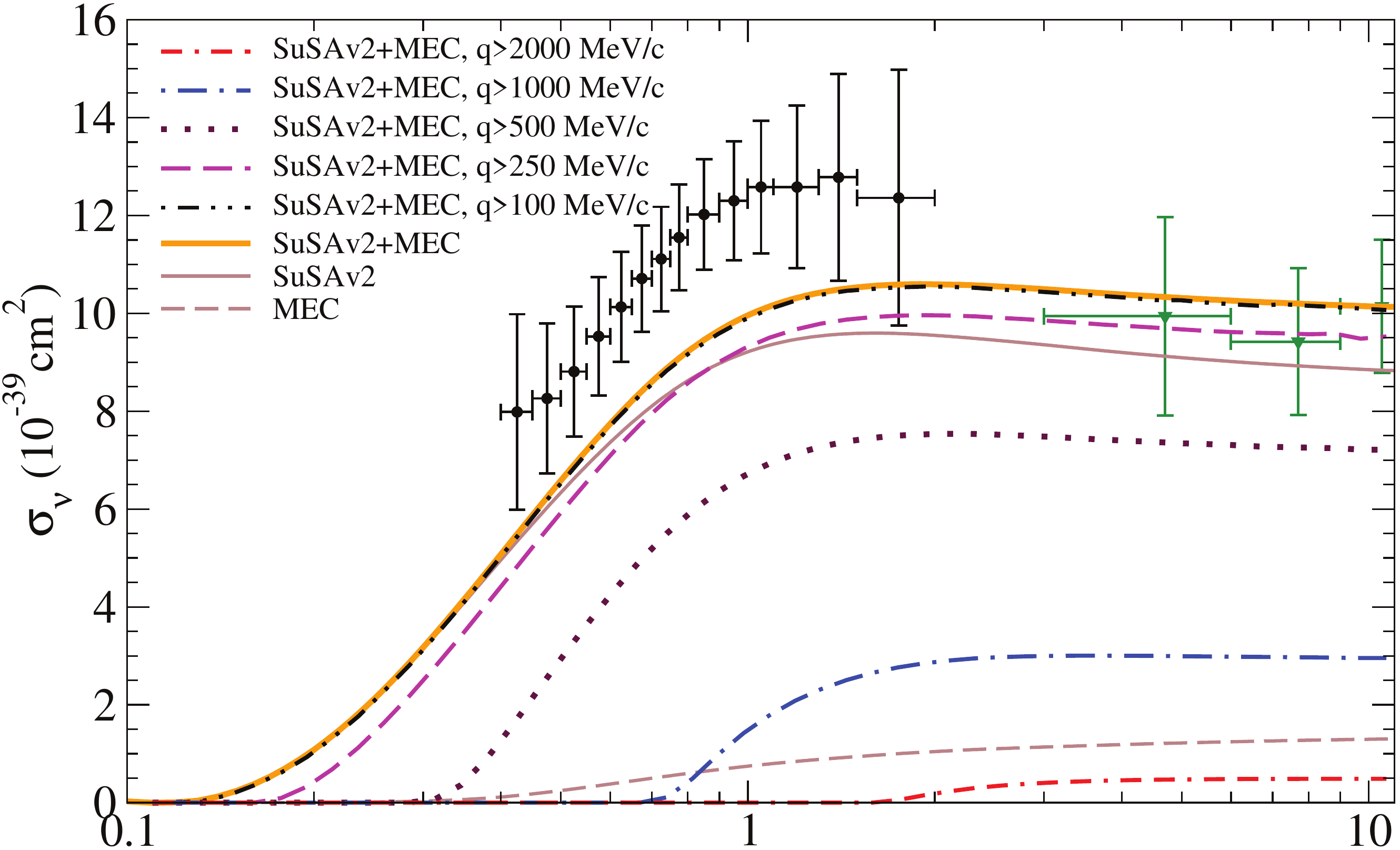}\\
\vspace{0.2cm}
\includegraphics[scale=0.31, bb=0 11 794 529]{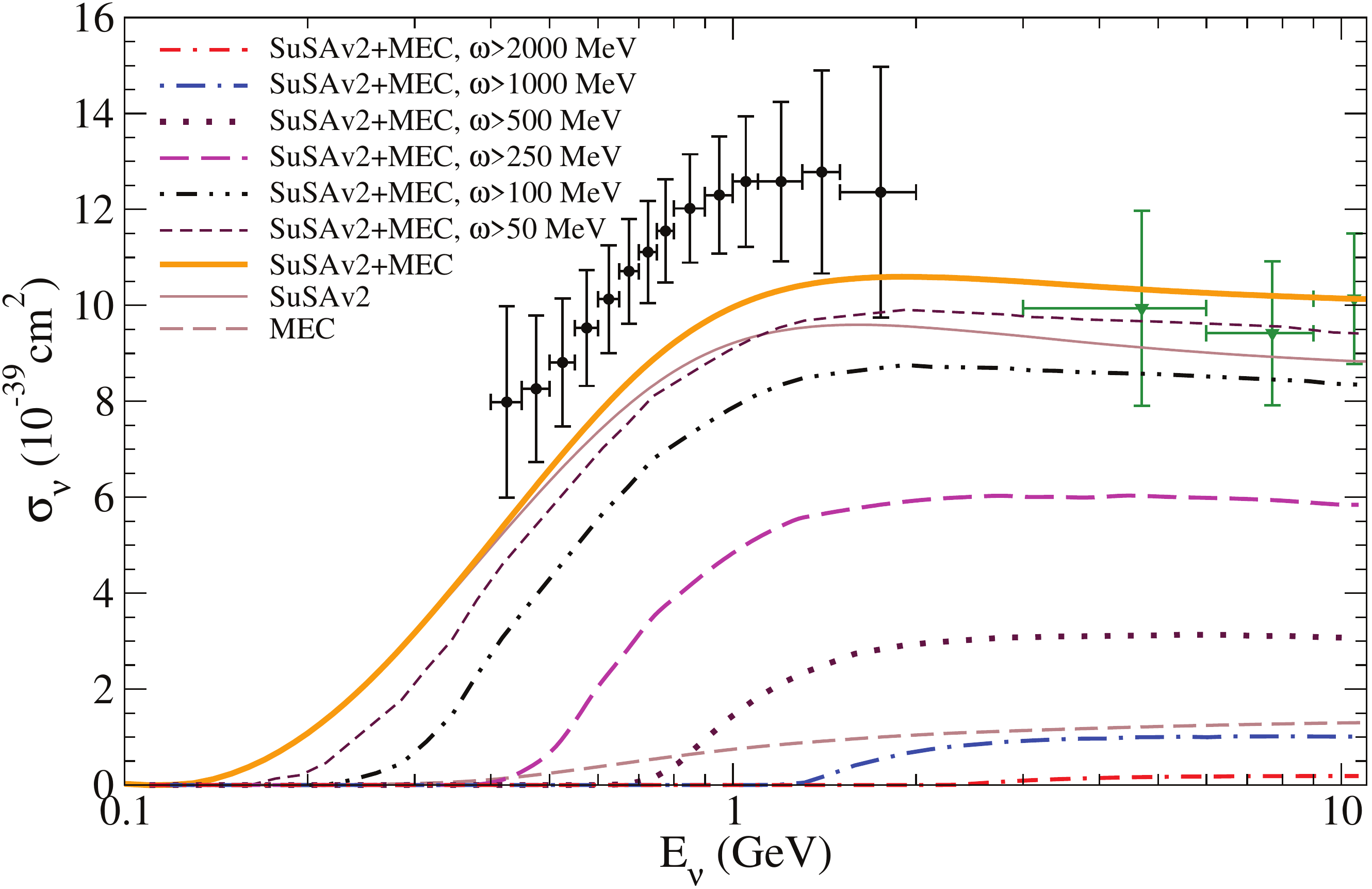}
\caption{(Color online) As for Fig. \ref{fig11}, but for the SuSAv2+MEC CCQE cross section. The MiniBooNE \cite{MiniBooNECC10} and NOMAD \cite{NOMAD09} data are also shown for reference. \label{fig12}}
\end{center}
\end{figure}

Moreover, an analysis of the relevant kinematic regions in the SuSAv2+MEC cross section is shown in Fig. \ref{fig12}, where it is observed that the main contribution to the total cross section comes from $\omega< 1000$ MeV and $q\lesssim 1000$ MeV/c whereas the region of $\omega< 50$ MeV and $q< 250$ MeV/c is not too significant for the cross section (less than 10$\%$). This is in accordance with some previous works \cite{Megias13, Megias14}. The same conclusion can be drawn by analyzing the different kinematics in the total MEC cross section (Fig. \ref{fig11}), where the low kinematic region ($\omega< 50$ MeV, $q< 250$ MeV/c) is even less important ($<2\%$).

\begin{figure}[htbp]
\begin{center}
\includegraphics[scale=0.25,angle=0]{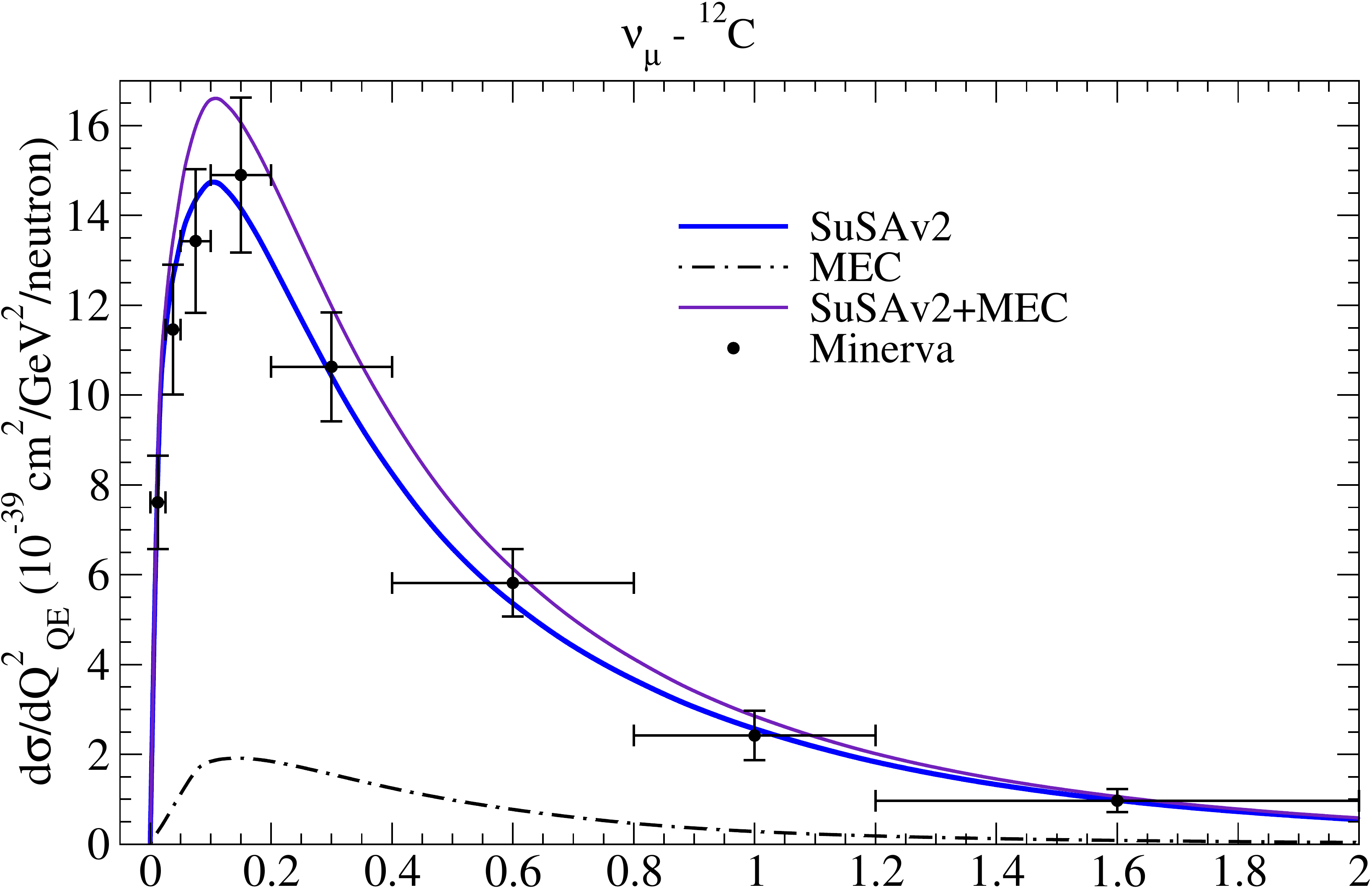} \\
\vspace{0.2cm}
\includegraphics[scale=0.25,angle=0]{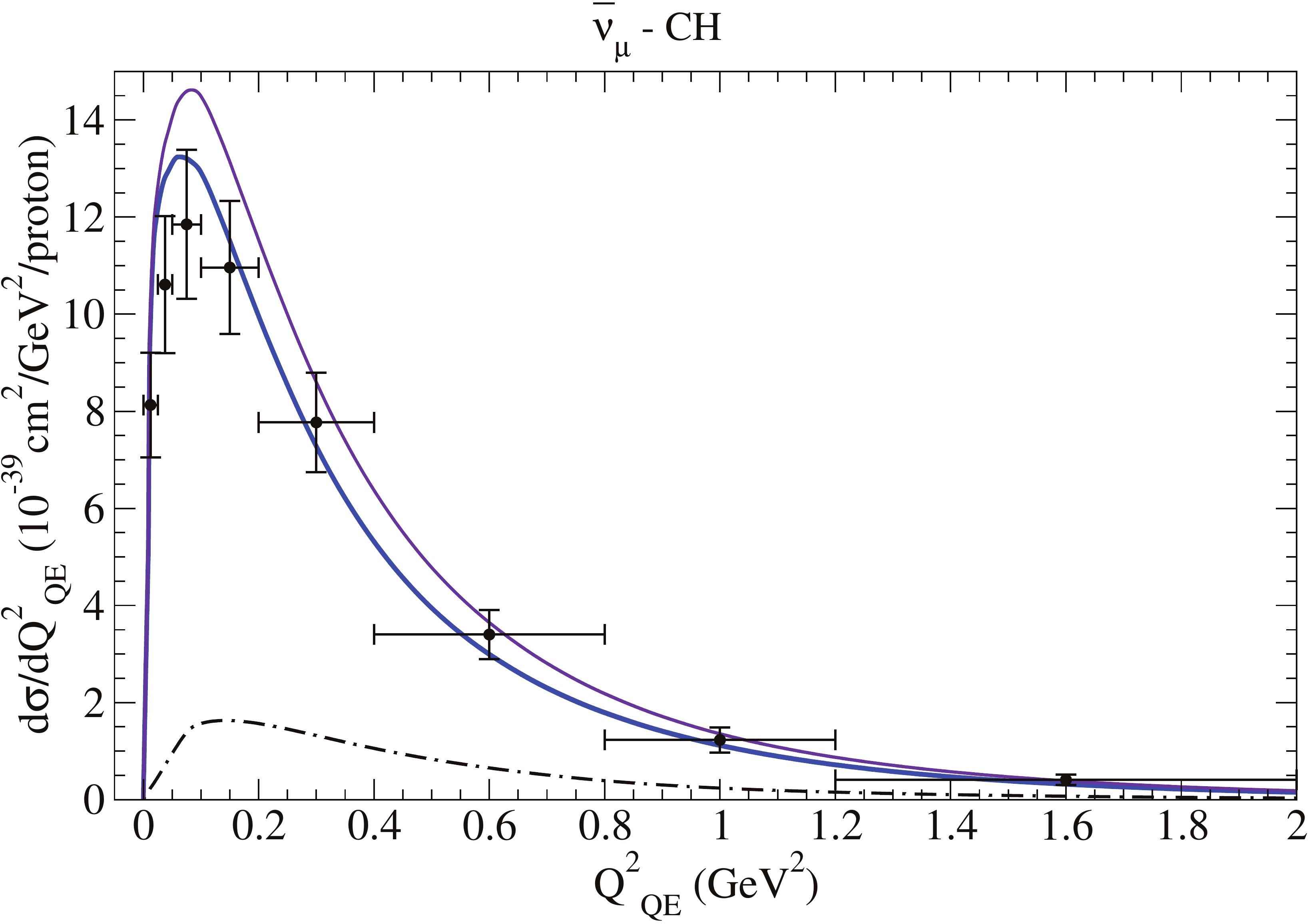}
\caption{(Color online) Flux-folded CCQE $\nu_\mu-^{12}$C (upper panel) and $\bar\nu_\mu-$CH (lower panel) scattering cross section per target nucleon as a function of $Q^2_{QE}$ and evaluated in the SuSAv2 and SuSAv2+MEC models. MINER$\nu$A data are from \cite{MINERVAnu13, MINERVAnub13}.  \label{fig16}}
\end{center}
\end{figure}

\begin{figure}[htbp]
\begin{center}
\includegraphics[scale=0.2924,angle=0]{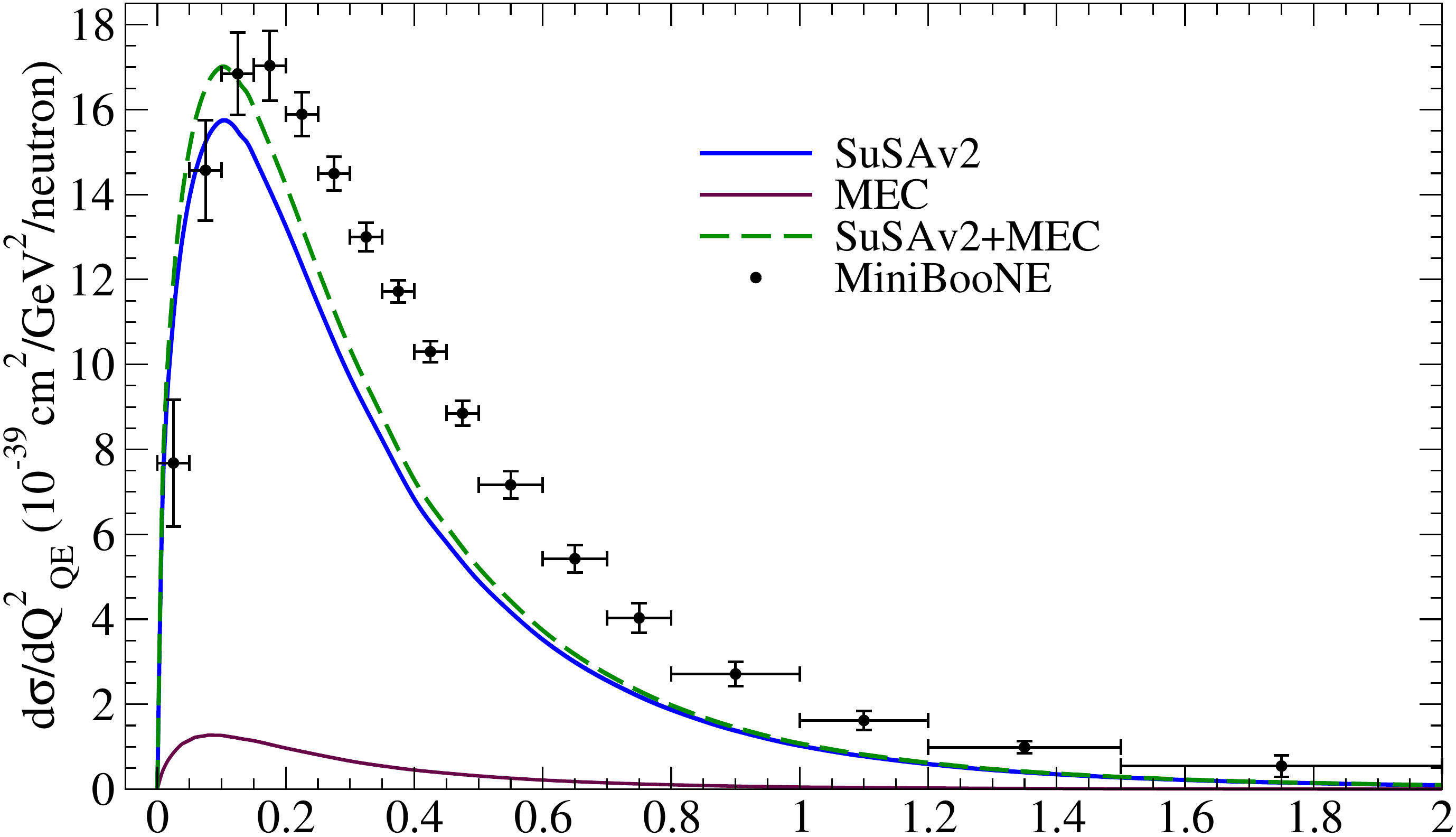} \\
\vspace{0.5cm}
\includegraphics[scale=0.2924,angle=0]{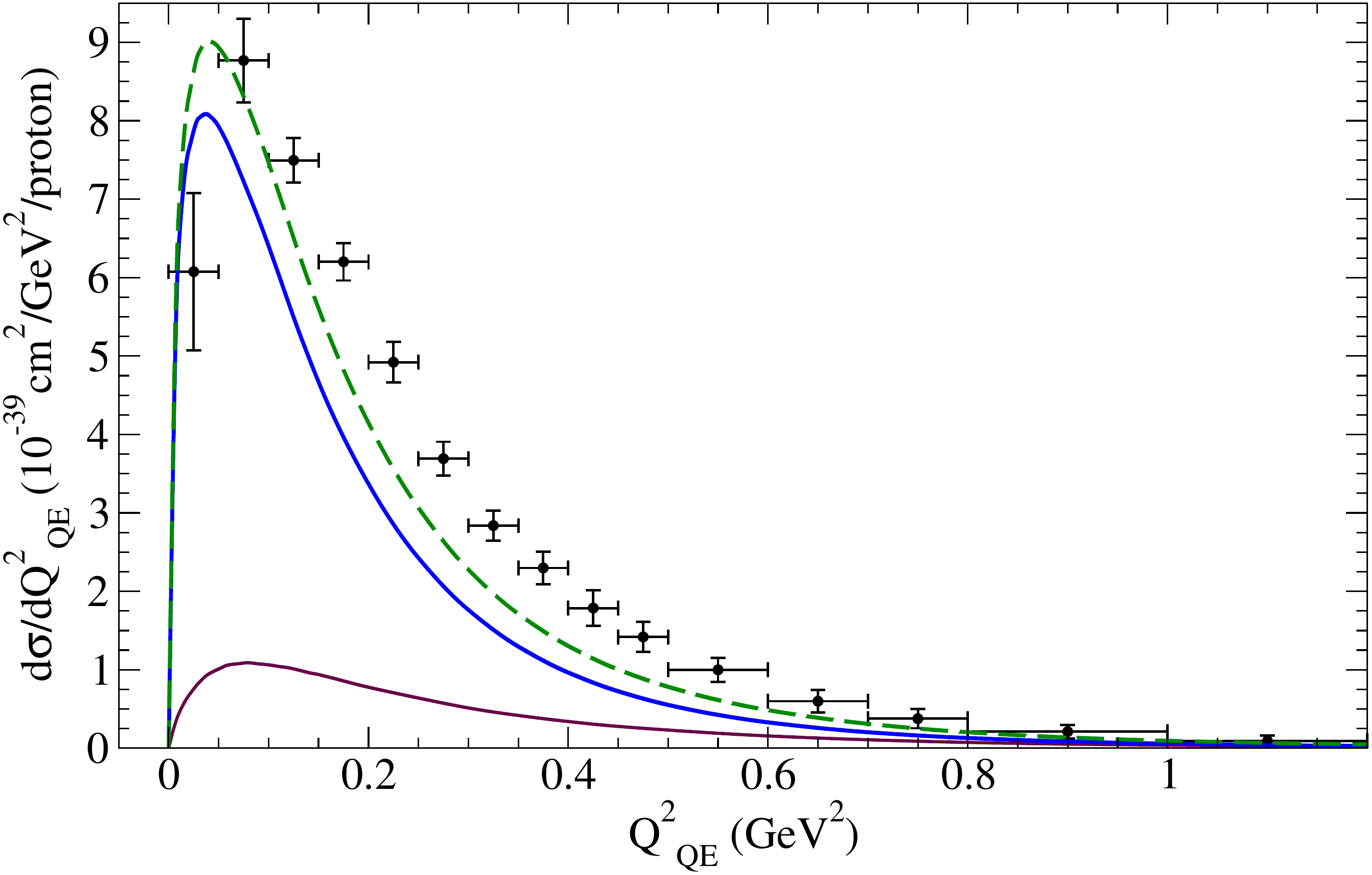}
\caption{(Color online) Flux-folded CCQE $\nu_\mu-^{12}$C (upper panel) and $\bar\nu_\mu-^{12}$C (lower panel) scattering cross section per target nucleon as a function of $Q^2_{QE}$ and evaluated in the SuSAv2 and SuSAv2+MEC models. MiniBooNE data are from \cite{MiniBooNECC10, MiniBooNECC13}.  \label{fig17}}
\end{center}
\end{figure}

At MINER$\nu$A kinematics, a good agreement arises for the purely QE SuSAv2 model with the $d\sigma/dQ^2_{QE}$ data without additional assumptions, Fig. \ref{fig16}, as observed in \cite{Megias14} for other impulse-approximation based models. An overestimation of the data shows up at low $Q^2_{QE}$ when adding 2p-2h MEC contributions. On the contrary, this effect is not observed in the same differential cross sections of MiniBooNE, Fig. \ref{fig17}, which is an example of the discrepancies between the two experiments and their different ways to proceed in the data analysis.


\section{Conclusions}\label{conclusions}

We have obtained CCQE neutrino-$^{12}$C cross sections using the SuSAv2 scaling procedure and a new parametrization of 2p-2h vector-vector transverse MEC. 
Both ingredients are based on relativistic models (RMF, RFG, RPWIA), as demanded by the kinematics of present and future high-energy neutrino experiments, 
where traditional non-relativistic models are questionable. We do not include in this work axial-axial and vector-axial MEC contributions needed 
for the analysis of neutrino scattering processes, nor correlation diagrams --- the calculation of the axial MEC contributions is currently being 
considered using~\cite{AmaroMEC14, AmaroMEC14b}.

Any model aimed at providing a useful and reliable tool to be employed in the analysis of experimental studies of neutrino oscillations needs their
limits of applicability to be completely understood. This has been the case in our present study where the limits of the approach have been stated clearly and 
discussed at length. Various models rely on different assumptions: non-relativistic expansions, factorization approach, mean field, {\it etc.}, that restrict their reliability.
However, in the absence of a ``fully-unlimited'' description of the reaction mechanism, the use of consistent, even limited, theoretical predictions to be contrasted with
data allows one to get insight into the physics underlying neutrino experiments. Hence, in spite
of the limitations mentioned above, our present model provides results that are in accordance
with $(e,e')$ data in the region around the QE peak. This is of great importance, and it gives us confidence in the
consistency and validity of our calculations in order to analyze lepton-nucleus scattering.

By comparing these results with the experimental data of the MiniBooNE, NOMAD and MINER$\nu$A collaborations we have shown that 2p-2h MEC play an important role in CCQE neutrino scattering and may help to resolve the controversy between theory and experiment. The main merit of the parametrization provided here is that it translates a sophisticated and computationally demanding microscopic calculation of MEC into a smooth parametrization which is dependent on the values of the transfer variables of the process. The economy of this MEC parametrization together with the one inherent in a scaling approach might be of interest to Monte Carlo neutrino event simulations used in the analysis of experiments.

%
%

\begin{acknowledgments}
This work was partially supported by Spanish DGI FIS2011-28738-C02-01, Junta de Andaluc\'ia FQM-160, INFN, Spanish Consolider-Ingenio 2000 Program CPAN, U.S. Department of Energy under cooperative agreement DE-SC0011090 (T.W.D), 7th European Community Framework Program Marie Curie IOF ELECTROWEAK (O.M.). G. D. M. acknowledges support from a fellowship from the Junta de Andaluc\'ia (FQM-7632, Proyectos de Excelencia 2011) and financial help from VPPI-US (Universidad de Sevilla). INFN under project MANYBODY (M.B.B. and A.D.P). DGI FIS2011-24149 and Junta de Andaluc\'ia FQM225 (J.E.A.). R.G.J. acknowledges financial help from VPPI-US (Universidad de Sevilla). 
\end{acknowledgments}

\bibliographystyle{apsrev4-1}

\bibliography{bibliography}

\begin{thebibliography}{44}
\bibitem [{foo()}]{footnote1}%
  \BibitemOpen
  \href@noop {} {\ }\bibinfo {note} {Note that, as is common in discussions of
  electron scattering, we define QE to mean the part of the cross section
  arising from nucleon knockout via one-body operators. This is to be
  distinguished from contributions that arise through the action of two-body
  operators such as the MEC effects discussed in the present work. The latter
  can eject single nucleons or two nucleons (or in fact no nucleons at all, as
  in elastic scattering). In contrast, what is referred to as ``quasielastic''
  in the neutrino community really means the ``no-pion cross section'' and that
  should contain both one- and two-body current operators, but should have no
  pions produced in the final state. Indeed, one common concern is how much
  model dependence occurs in defining this no-pion cross section, since
  corrections must be made for events where a pion is actually produced, but is
  absorbed before being detected and hence mistaken as a ``quasielastic''
  event.}\BibitemShut {Stop}%
\end{thebibliography}

\twocolumngrid

\onecolumngrid


\twocolumngrid

\onecolumngrid


\twocolumngrid

\onecolumngrid


\twocolumngrid

\newpage
\clearpage 

%

\end{document}